\documentclass[floatfix,prc,aps,twocolumn]{revtex4}
\usepackage{amssymb}
\usepackage{graphicx,psfrag}

\begin{document}

\newcommand{\note}[1]{\marginpar{\tiny {#1}}}   
\newcommand{\bld}[1]{\mbox{\boldmath $#1$}}     
\newcommand{\code}{$\mathtt{FREYA}$}
\newcommand{\Punf}{$^{239}$Pu(n$_{\rm th}$,f)}
\newcommand{\Pusf}{$^{240}$Pu(sf)}
\newcommand{\Cfsf}{$^{252}$Cf(sf)}
\newcommand{\Cmsf}{$^{244}$Cm(sf)}
\newcommand{\Unf}{$^{235}$U(n$_{\rm th}$,f)}
\newcommand{\Usf}{$^{238}$U(sf)}
\newcommand{\Punfnonf}{$^{239}$Pu}
\newcommand{\Pusfnosf}{$^{240}$Pu}
\newcommand{\Cfsfnosf}{$^{252}$Cf}
\newcommand{\Cmsfnosf}{$^{244}$Cm}
\newcommand{\Unfnonf}{$^{235}$U}
\newcommand{\Usfnosf}{$^{238}$U}
\newcommand{\SKIP}[1]{}
\newcommand{\nubar}{{\overline{\nu}}}
\newcommand{\half}{\mbox{$1\over2$}}

\def\black{
}
\def\red{
}


\title{Improved modeling of photon observables with \code}

\author{R.~Vogt$^{1,2}$ and J.~Randrup$^3$}

\affiliation{
  $^1$Nuclear and Chemical Sciences Division,
  Lawrence Livermore National Laboratory, Livermore, CA 94551, USA\break
$^2$Physics Department, University of California, Davis, CA 95616, USA\break
$^3$Nuclear Science Division, Lawrence Berkeley National Laboratory, 
Berkeley, CA 94720, USA}

\date{\today}

\begin{abstract}
  The event-by-event fission model \code\ has been improved, in
  particular to address deficiencies in the calculation of photon
  observables.  We discuss the improvements that have been made and introduce
  several new variables, some detector dependent, that affect the photon
  observables.  We show the
  sensitivity of \code\ to these variables.  We then compare the results to the
  available photon data from spontaneous and thermal neutron-induced fission.
\end{abstract}

\maketitle

\section{Introduction}

The computational model \code\ generates complete fission events,
{\em i.\,e.}\ it provides the full kinematic information 
on the two product nuclei as well as all the emitted neutrons and photons.
In its development, an emphasis had been put on speed,
so large event samples can be generated quickly.
\code\ therefore relies on experimental data supplemented by
simple physics-based modeling.

In its standard version, to treat a given fission case,
\code\ needs the fission fragment mass distribution, $Y(A)$,
and the average total kinetic energy for each mass split, TKE($A$),
for the particular excitation energy considered.
$Y(A)$ is taken either directly as the measured yields
or as a five-Gaussian fit to the data 
which makes it possible to parameterize its energy dependence, see
Ref.~\cite{VRBDO} for details on how the energy dependence of neutron-induced
fission is handled in \code.

In order to generate an event,
\code\ selects the mass split based on the provided $Y(A)$.
The fragment charges are then sampled from the normal distributions
suggested by experiment \cite{VRBDO}.
The linear and angular momenta of the two fragments 
and their internal excitations are subsequently sampled.
After their formation, the fully accelerated fragments de-excite first
by neutron evaporation and then by photon emission.
In addition to spontaneous fission, \code\ treats neutron-induced fission 
up to $E_n=20\,{\rm MeV}$.  The possibility of pre-fission evaporation
is considered as well as pre-equilibrium neutron emission.
Both play an increasing role at the higher energies.

This paper is a follow up to our previous paper on prompt photon emission
from fission \cite{RVJR_gamma}, describing several improvements to the
modeling of photon emission.  For detailed information on how to download and
run \code, the first published fission event generator, see Ref.~\cite{CPC}
and the subsequent new version announcement that includes the new work
described here \cite{CPC_NVA}.

This paper is organized as follows.  We first discuss the improvements
made to the photon model in \code.  We then describe the model parameters
of \code\ and which observables they affect most strongly.
We then show how the improvements affect prompt photon observables.
Next, we demonstrate the effect of
modifying the \code\ parameters on the photon results, in particular for
$^{252}$Cf(sf) because a first attempt at fitting the \code\ parameters to
data with this version of \code\ was made for this case \cite{Andrew}.

We then compare our results to photon data from $^{252}$Cf(sf) 
as well as from neutron-induced fission when appropriate.
Because no detailed fits have been made for those cases,
the parameter that governs photon emission has the value
determined in a global fit to available $^{252}$Cf(sf) data \cite{Andrew},
as explained further in the text.
The calculations for those cases then yield reasonable preliminary result.
Finally we conclude.

\section{\code}
\label{code}

General descriptions of the physics of \code\ have been published elsewhere
\cite{VRBDO,RVJR_gamma,RVJR_spont,JRRV_ang,RVJR_corr}.  Therefore,
in this paper we describe only the improvements made for photon emission.

In our previous paper on photon emission \cite{RVJR_gamma},
we included statistical emission of photons with no form factor modulation,
corresponding to black-body radiation.
The resulting photon spectrum was too soft, with too few high energy photons.
In addition, although rotational energy was included,
the total angular momentum was not conserved.
Furthermore, the earlier treatment did not include any discrete low-energy 
photon transitions, but carried the statistical de-excitation through until the
yrast line was reached and then disposed of the remaining rotational energy 
by schematic E2 transitions \cite{RVJR_gamma}.

For each fragment, the magnitude of its angular momentum
was governed by a ``spin temperature'' $T_S$,
$\langle S_{\rm f}^2 \rangle = 2 \mathcal{I}_{\rm f}T_S$,
where $\mathcal{I}_{\rm f}$ is the fragment moment of inertia.
The parameter $T_S$ was varied from 0.35 MeV to 2.75 MeV for a fixed value of
$Q_{\rm min}$, 
the energy above the neutron separation energy at which
the fragment de-excitation cascade switches from neutron to photon emission.  
The value of $Q_{\rm min}$ was set to 0.01 MeV in Ref.~\cite{RVJR_gamma} 
and this remains the default value in \code.
However, it was shown that using a large value of $Q_{\rm min}$ (1 MeV)
in conjunction with a small $T_S$ (0.2 MeV) was equivalent to
using a small $Q_{\rm min}$ (0.01 MeV) with a large $T_S$ (2.75 MeV).

Since then, several improvements have been made.  While some have been
reported in preliminary form elsewhere \cite{Andrew,CPC_NVA},
others are described here for the first time.

In Ref.~\cite{JRRV_ang}, we modified \code\ to conserve total angular
momentum, including fluctuations of the dinuclear wriggling and bending modes, 
where the two fragments rotate in the same or opposite sense 
around an axis perpendicular to the dinuclear axis.
The tilting and twisting modes,
where the rotations are around the dinuclear axis, were neglected.
These modes each contribute to fluctuations in the rotational energy,
$\delta E_{\rm rot} = s_+^2/2{\mathcal I}_+ + s_-^2/2{\mathcal I}_-$ 
where wriggling is denoted by $+$ and bending by $-$.
The moments of inertia for these modes (each of which is doubly degenerate),
${\mathcal I}_+$ and ${\mathcal I}_-$, are given in terms of 
the moments of inertia of the individual light and heavy fragments,
${\mathcal I}_L$ and ${\mathcal I}_H$, respectively,
as well as moment of inertia for their relative motion, ${\mathcal I}_R$,
\begin{eqnarray}
  {\mathcal I}_+ &=& ({\mathcal I}_L 
+ {\mathcal I}_H){\mathcal I}/{\mathcal I}_R\ ,\\
  {\mathcal I}_- &=& {\mathcal I}_L{\mathcal I}_H/({\mathcal I}_L 
+ {\mathcal I}_H)\ .
\end{eqnarray}
where ${\mathcal I}={\mathcal I}_L + {\mathcal I}_H + {\mathcal I}_R$
is the total moment of inertia.
The angular momenta of these rotational modes are sampled from
thermal distributions characterized by the spin temperature $T_S$,
$P_\pm(S_\pm) \propto \exp(-S_\pm^2/2 {\mathcal I}_\pm T_S)$.
The individual fragment angular momenta then follow.
Any overall angular momentum, $S_0$ 
(resulting from the absorption of an incoming neutron
or the recoil from any pre-fission neutron emission) is also 
taken into account but the effect tends to be very small.

We express the spin temperature as $T_S = c_S T_{\rm sc}$,
where the ``scission temperature'' $T_{\rm sc}$ 
is the temperature of the system at scission,
and adopt $c_S$ as a convenient parameter for controlling 
the overall magnitudes of the fragment angular momenta in \code.
The general effect of changing $c_S$ is similar to that of
changing $T_S$ in Ref.~\cite{RVJR_gamma}
and we shall discuss how the results depend on $c_S$.
We note that if $c_S$ is taken to be zero, 
the fragments emerge with only their share of the overall rotation,
$S_{\rm f}=S_0{\cal I}_{\rm f}/{\cal I}$, 
which is usually very small (and is zero for spontaneous fission).

In the refined treatment of the statistical photon emission stage,
we modulate the black-body spectrum by a giant dipole resonance (GDR) 
form factor, so that the prompt fission photon spectrum is
\begin{eqnarray}
{dN_\gamma \over d\epsilon_\gamma}\ \sim\ 
{\Gamma_{\rm GDR}^2 \epsilon_\gamma^2 \over 
(\epsilon_\gamma^2-E_{\rm GDR}^2)^2 + \Gamma_{\rm GDR}^2\epsilon_\gamma^2 }\,
\epsilon_\gamma^2\,{\rm e}^{-\epsilon_\gamma/T_{\rm }}\ ,
\end{eqnarray}
from which the photon energy can readily be sampled.
(Its direction is chosen isotropically,
in the frame of the emitting fragment, as earlier.)
The position of the GDR is 
$E_{\rm GDR}({\rm MeV})=31.2/A^{1/3}+20.6/A^{1/6}$ \cite{BermanRMP47},
while its width is $\Gamma_{\rm GDR}=5\,{\rm MeV}$.
Relative to the earlier treatment \cite{RVJR_gamma},
which employed pure black-body radiation,
the inclusion of the GDR hardens the spectrum

Furthermore, as a significant extension,
we now include evaluated discrete transitions 
from the RIPL-3 database \cite{RIPL-3}, as in Refs.~\cite{FIFRELIN,CGMF}.
The RIPL-3 library tabulates a large number of discrete 
electromagnetic transitions for nuclei throughout the nuclear chart.
Some of these lines may be exploited experimentally
to identify the specific fragment species.
Unfortunately, complete information is available 
for only relatively few of the identified transitions, 
so some modeling is required to complement the tabulations (see below).
It is then possible to construct, for each product species,
a table of the possible decays from the included discrete levels.

The RIPL-3 data files are organized by element, 
with one file for each $Z$ value,
and each such file contains similarly structured listings for those isotopes
of that element for which data exist.
For any tabulated nucleus,
we seek to include all levels `in a complete level scheme,
as indicated in the isotope header line on the data file.
Each listed level $\ell$ may decay into a total of $n(\ell)$
lower levels $\{\ell'\}$
and the associated relative transition rate $P(\ell\to\ell')$ 
is indicated for each one, if available.
Often the rate for a listed transition is not given
and the corresponding transition is then ignored.

However, if all the decay rates from a given level $\ell$ are missing
we assign decay rates from from that level to all of the lower levels $\ell'$
based on a simple phase-space consideration,
\begin{eqnarray}
P(\ell\to\ell') \sim 
[\epsilon_\ell-\epsilon_{\ell'}]^2\,{\rm e}^{-(J_\ell-J_{\ell'})^2/2d_J^2}\ .
\end{eqnarray}
Here $\epsilon_\ell$ is the energy of level $\ell$,
$J_\ell$ is its listed spin, and we take $d_J=1$.
It should be recognized that there are many more added transitions (201568)
than tabulated transitions (75809).
That is primarily because a level for which there are tabulated decay rates
tends to decay to only some of the levels below it, 
whereas a level without tabulated decay rates is allowed to decay 
to {\em any} level below it and, furthermore, levels without tabulated decay
rates tend to be high-lying and so have many lower levels.

In the previous de-excitation procedure, the product nucleus 
first made statistical emission until the yrast line was reached
and then proceeded towards the ground state by collective emission.
In order to better emulate the predominantly $E1$ and $M1$ character 
of the statistical transitions, it is assumed that the angular momentum 
of the fragment is reduced by $1\hbar$ for each emission.
This is a somewhat idealized treatment which may need to be refined.
In order to incorporate the subsequent discrete decays,
we follow the earlier procedure until the total excitation has fallen 
below the highest discrete energy $E_\ell$ tabulated for that nucleus.
The energy of the last statistical transition is then increased (slightly) to
ensure that the last statistical decays leads to the closest lower-lying level
and the further de-excitation is then carried out by means of the
discrete rates described above.
The emission of discrete photons is continued until 
the tabulated half-life of a level exceeds a specified value $t_{\rm max}$.

If there are no RIPL-3 transitions available for a given product nucleus, 
the final de-excitation occurs as in Ref.~\cite{RVJR_gamma}
by emission of `collective photons along the yrast line,
with each emitted photon reducing the angular momentum by $2 \hbar$.


We will show how the inclusion of the GDR modulation 
and the RIPL-3 transitions affect the photon observables.
The maximum half-life of the discrete levels, $t_{\rm max}$,
as well as the minimum recorded photon energy, $g_{\rm min}$,
have an impact on the generated photon energy and multiplicity, 
and we will explore how changes in these quantities affect 
the photon observables obtained with \code.
Because $g_{\rm min}$ represents the energy threshold for photon detection 
and $t_{\rm max}$ represents the time gate for the detectors,
the comparison with a particular experiment depends on this information.

\subsection{\code\ parameters}

\code\ contains a number of physics-based parameters
that affect various observables.  
They can be adjusted to available data.
Here we give a brief description of their function and impact.
The following six are code specific parameters:
\begin{description}
\item[$d{\rm TKE}$] is a common, mass-independent, shift of the total
  fragment kinetic energy relative to the input TKE($A$).  
  This shift is tuned to give agreement with the average prompt
  neutron multiplicity $\overline{\nu}$, 
  an adjustment that is typically of the order of one MeV or less.
\item[$x$]   is the relative advantage in excitation energy 
  given to the light fragment over the heavy fragment.
  This parameter significantly affects the neutron multiplicity
  as a function of fragment mass, $\nu(A)$. As shown in Ref.~\cite{RVJR_corr},
  it also affects the shape of the two-neutron angular
  correlation function.  Other codes use systematics of excitation energy
  sharing \cite{GEF} or tune the fragment temperature distribution to the
  available $\nu(A)$ data \cite{FIFRELIN,CGMF}.  
  We have so far kept this parameter single-valued
  since we wish to address cases where 
 $\nu(A)$ is not available for tuning the temperature distribution.
\item[$c_T$]
  is the relative statistical fluctuation in the fragment thermal
  excitation.  Prior to Ref.~\cite{RVJR_corr}, $c_T$ was assumed to be unity by
  default.  In that work, however, it was shown that it had a significant
  effect on the width of the neutron multiplicity distribution $P(\nu)$ and,
  in particular, on the moments of $P(\nu)$ important for some applications.
  In addition, since the value of $c_T$ adjusts the intrinsic excitation energy,
  the extra excitation energy must be taken away from the total kinetic energy
  of the fragments.  Thus $c_T$ effectively governs the width of the TKE
  distribution, $\sigma_{\rm TKE}$, as well.
\item[$e_0$] 
  sets the overall scale of the Fermi-gas level density parameter.  (The
  asymptotic level density parameter is $a \sim A/e_0$.)  It has only a
  negligible effect on the neutron multiplicity distribution, $P(\nu)$;
  the neutron multiplicity as a function of fragment mass, $\nu(A)$;
  the two-neutron angular correlation; and the photon observables.
  It does, however, affect the spectral shape of the prompt fission neutrons.
  We note that while the
  other parameters also affect the spectral shape and normalization to
  $\overline \nu$. the neutron spectrum is effectively the only observable
  dependent on $e_0$.  
\item[$c_S$]
  was defined previously as the ratio of the spin temperature $T_S$ 
  to the scission temperature $T_{\rm sc}$.
  It governs the overall magnitude of the fragment angular momenta.
  It affects the photon observables significantly, as we show here,
  whereas it has only a small effect on the neutron observables.  
  However, there is a strong correlation between $d$TKE and $c_S$ 
  which serves to balance the neutron and photon multiplicities.  
  If $c_S$ is changed to match the photon multiplicity, 
  $d$TKE must also be adjusted to maintain agreement with $\overline \nu$.
  This balance is most important for the multiplicities of prompt emission.
\item[$Q_{\rm min}$]
  is defined as the energy above the neutron separation threshold where
  photon emission takes over from neutron emission.  Since we adjust $c_S$ to
  the photon multiplicity, this parameter is kept fixed at
  $Q_{\rm min} = 0.01$~MeV in the present studies. 
\end{description}
The following two parameters are detector specific
and not internal \code\ parameters:
\begin{description}
\item[$g_{\rm min}$] is the minimum energy of photons that are being recorded
  and it is usually set to the minimum photon energy measurable 
  by the photon detector in the measurement under analysis.
  Photons softer than $g_{\rm min}$ may still be emitted,
  they are just not being recorded in the particular event.
  This parameter is merely introduced for convenience
  and it does not affect the physical process.
\item[$t_{\rm max}$]   is the maximum half-life
  of an energy level in the discrete part of the photon decay chain.
  If the photon cascade leads to a level that has a half-life
  exceeding $t_{\rm max}$, the fragment is effectively stuck at that level
  during the time interval of the measurement and the subsequent history
  is immaterial.
  The effect of changing $t_{\rm max}$ is somewhat subtle and affects
  primarily photons emitted from low-energy levels having relatively 
  low spin which are reached in the later part of the decay cascade.
\end{description}  

\subsection{$^{252}$Cf(sf) analysis}
\label{Andrew_fit}

In Ref.~\cite{Andrew}, we adjusted $d$TKE, $x$, $c_T$, $e_0$ and $c_S$ to
several sets of data.  These included the Mannhart spectral evaluation
\cite{Mannhart}, the $P_n(\nu)$ distribution by Boldeman {\it et al.}\
\cite{Boldeman},
the $\nu(A)$ distribution determined by Dushin {\it et al.}\ \cite{Vorobiev},
the neutron multiplicity as a function of TKE measured by Budtz-J{\o}rgensen
and Knitter \cite{Budtz-Jorgensen},
and the average total photon energy and the average total photon multiplicity
measured by Billnert {\it et al.}\ \cite{BillnertCfgamma}.
While we later compared
to the photon multiplicity distribution measured by DANCE \cite{Czyzhgamma},
we did not include that distribution in the fits.
Indeed, none of the current \code\ parameters
affect the width of the photon multiplicity distribution $P_\gamma(N)$.

A wide range of possible parameter values were considered, 
with some physics bias to guide the fits.  
$d$TKE was varied from -5.0 to 5.0 MeV although large excursions 
from the measured mean would be in strong disagreement with data.
The value of $x$ was assumed to be larger than unity, limiting us to
$1 < x < 1.5$ for the study.  We choose $x > 1$ because the light fragment emits
more neutrons on average than the heavy fragment in spontaneous and
neutron-induced fission, see Ref.~\cite{RVJR_spont} for details.
Also, given the limitations of the single-valued parameter $x$ in describing
the shape of $\nu(A)$ in the low and high fragment mass range, only the mass
region $100 < A < 140$ was used in the fit.
The parameter governing thermal fluctuations,
$c_T$, was also assumed to be larger than unity, between 1 and 2.
The value of the asymptotic level density
parameter, $e_0$, was taken to be in the range $6 < e_0 < 12$ MeV.
Finally, $c_S$ was allowed to vary around unity, $0.5 < c_S < 1.5$.
We will study larger excursions of $c_S$ in Sec.\ \ref{sec:cSdep}
to illustrate the magnitude of the effect.
The comparison to data is made by calculating $\chi^2$ for each data set
individually and summing them to obtain the total $\chi^2$.

In Ref.~\cite{Andrew}, the global $\chi^2$ was minimized using a
particle swarm algorithm.  The best fit value using this method was found to
give $d$TKE = 0.5~MeV, $x = 1.27$, $c_T = 1.08$, $e_0 = 10.37$~MeV, and
$c_S = 0.87$.  
We note that some of these values are not far from those suggested in
Ref.~\cite{RVJR_spont}, where photon observables were not yet included
and $c_T \equiv 1$ by default: $e_0 \sim 10$ and $x = 1.3$.
We are currently working on an approach that will give a better global
$\chi^2$, with well defined uncertainties for $^{252}$Cf(sf), and will apply
the same method to other spontaneously fissioning nuclei as well
as neutron-induced fission.  We note that this must be an ongoing process
as new data are taken and become available.

For the results presented here, we use the best fit values of $e_0$ and $c_S$
found in Ref.~\cite{Andrew} for $^{252}$Cf(sf) 
and keep these same values for our calculations of other cases.
One might expect $e_0$ to be universal since it is independent of the
fissioning nucleus.  We have chosen to leave $c_S$ fixed because it is
strongly correlated with
$d$TKE and its optimal value for a particular nucleus	   
is therefore best determined on the basis of a global analysis of each
isotope.
To calculate results for neutron-induced fission,
we adjust $x$ based on $\nu(A)$ measurements, $c_T$ based on $P(\nu)$ data,
and fix $d$TKE to the measured $\overline \nu$.  These parameter values, while
still preliminary to some degree, provide benchmarks as to how well we can
expect to describe the prompt fission photon data.

\section{Effects of the model refinements}

We begin by showing how the \code\ photon emission results 
have evolved since the publication of Ref.~\cite{RVJR_gamma}.  
That work considered unmodulated (black-body) statistical photon radiation,
\begin{eqnarray}
{dN_\gamma \over d\epsilon_\gamma}\ \sim\ 
	      \epsilon_\gamma^2\,{\rm e}^{-\epsilon_\gamma/T_{\rm }}\ ,
\label{spect:noGDR}
\end{eqnarray}
until the yrast line was reached
and the rotational energy was then dissipated by schematic 
photon emission along the yrast line.
Thus no specific transitions were considered.
In this section we compare three different model scenarios: 
1) no GDR and no RIPL 
corresponding to the earlier treatment \cite{RVJR_gamma};
2) GDR without RIPL, 
showing the effect of the modulation of the statistical decays 
without including the specific tabulated transitions; and
3) GDR with RIPL, 
corresponding to the improved treatment presented here.

All the calculations shown in this section are for $^{252}$Cf(sf).  
The three different model scenarios all use the same parameter values, 
given in Sec.~\ref{Andrew_fit} above with
$g_{\rm min} = 0.10$~MeV and $t_{\rm max} = 1.5$~ns.
Each calculation is based on one million \code\ events
which suffices to ensure negligible statistical uncertainties
aside from regions near the edges of phase space, either in the
low and high $A$ tails of the yields or close to symmetry for observables
given as a function of fragment mass or at extreme values of total fragment
kinetic energy where there are few events for observables given as functions
of TKE.

\begin{figure}[tbh]	  
\includegraphics[angle=0,width=0.45\textwidth]{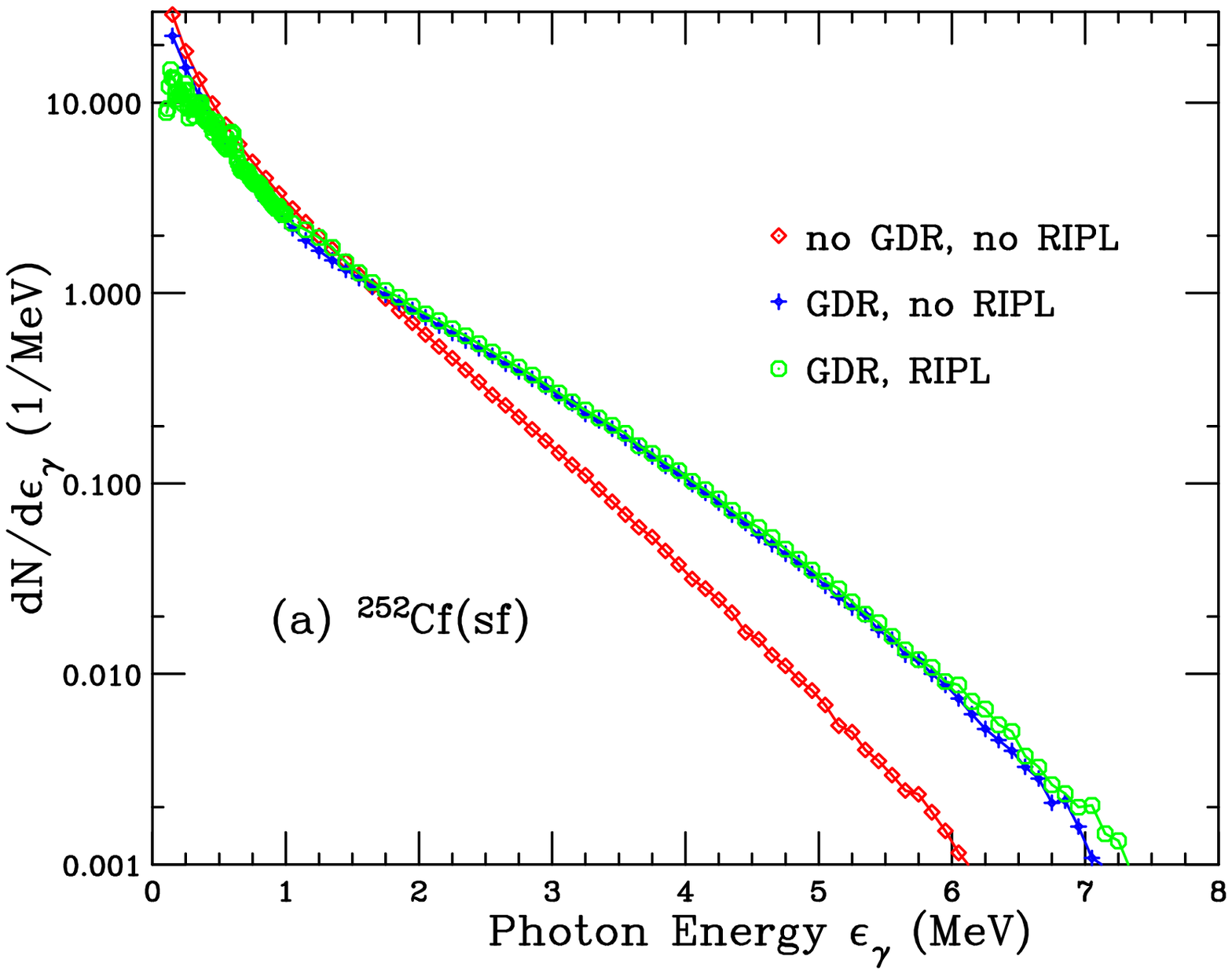}
\includegraphics[angle=0,width=0.45\textwidth]{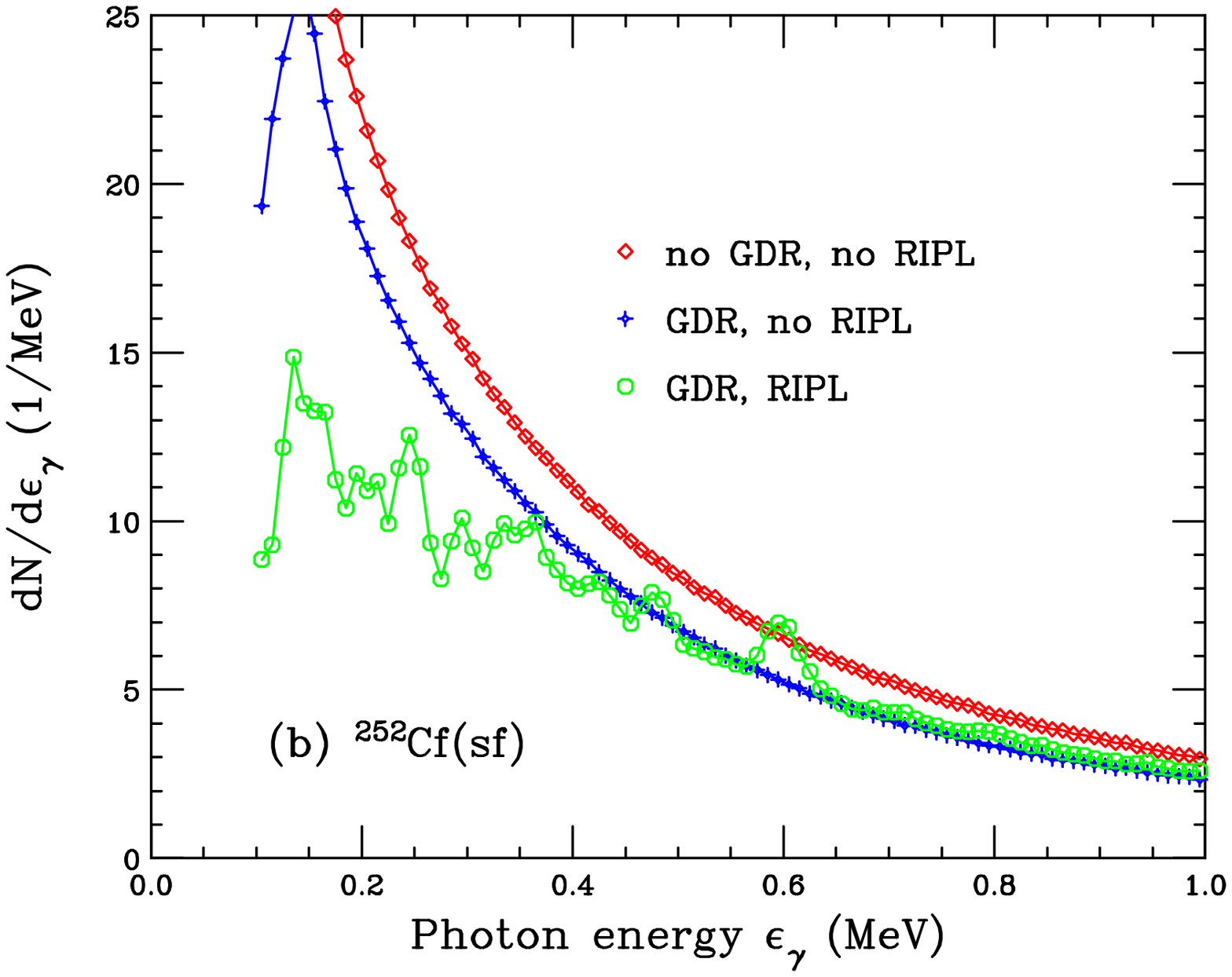}
\caption[]{(Color online) The $^{252}$Cf(sf) photon spectrum from \code\ calculated
  as in Ref.~\protect\cite{RVJR_gamma} without GDR form factor or RIPL-3
  lines, with GDR but without RIPL-3 lines,
  and including both, as in the current version of \code.  
  Panel (a) shows the spectrum over the entire energy range, while
  panel (b) shows the low-energy spectrum for photon energies less than 1 MeV.
  The calculated results in this and all subsequent figures are based on
  one million events and the associated sampling errors are shown.  The spectra
  are normalized to the fission photon multiplicity.
}\label{fig:spect_GDR_RIPL}
\end{figure}

Figure~\ref{fig:spect_GDR_RIPL}(a) shows the prompt photon energy spectrum
over the full energy range.  The calculation without the GDR modulation, 
from Eq.~(\ref{spect:noGDR}), drops exponentially 
and has a negligible yield already for photons of a few MeV.  
When the GDR modulation is included, 
the spectrum broadens significantly above 2 MeV 
and is about an order of magnitude larger in the tail region.
The addition of the RIPL tables does not change the shape 
of the high-energy tail of the photon spectrum further. 
Instead, adding the tabulated RIPL transitions primarily affects
the low-energy end of the spectrum, as shown for the distributions 
at energies less than 1 MeV in Fig.~\ref{fig:spect_GDR_RIPL}(b). 
In this region, the two calculations without the RIPL contributions 
are qualitatively similar, both being smooth and monotonically decreasing.
The effect of including the RIPL tables can be clearly seen:
the emission of low-energy photons is strongly reduced
and a considerable degree of spectral structure appears 
in reflection of specific transitions in fragments with large yields.

\begin{figure}[tbh]   
\includegraphics[angle=0,width=0.45\textwidth]{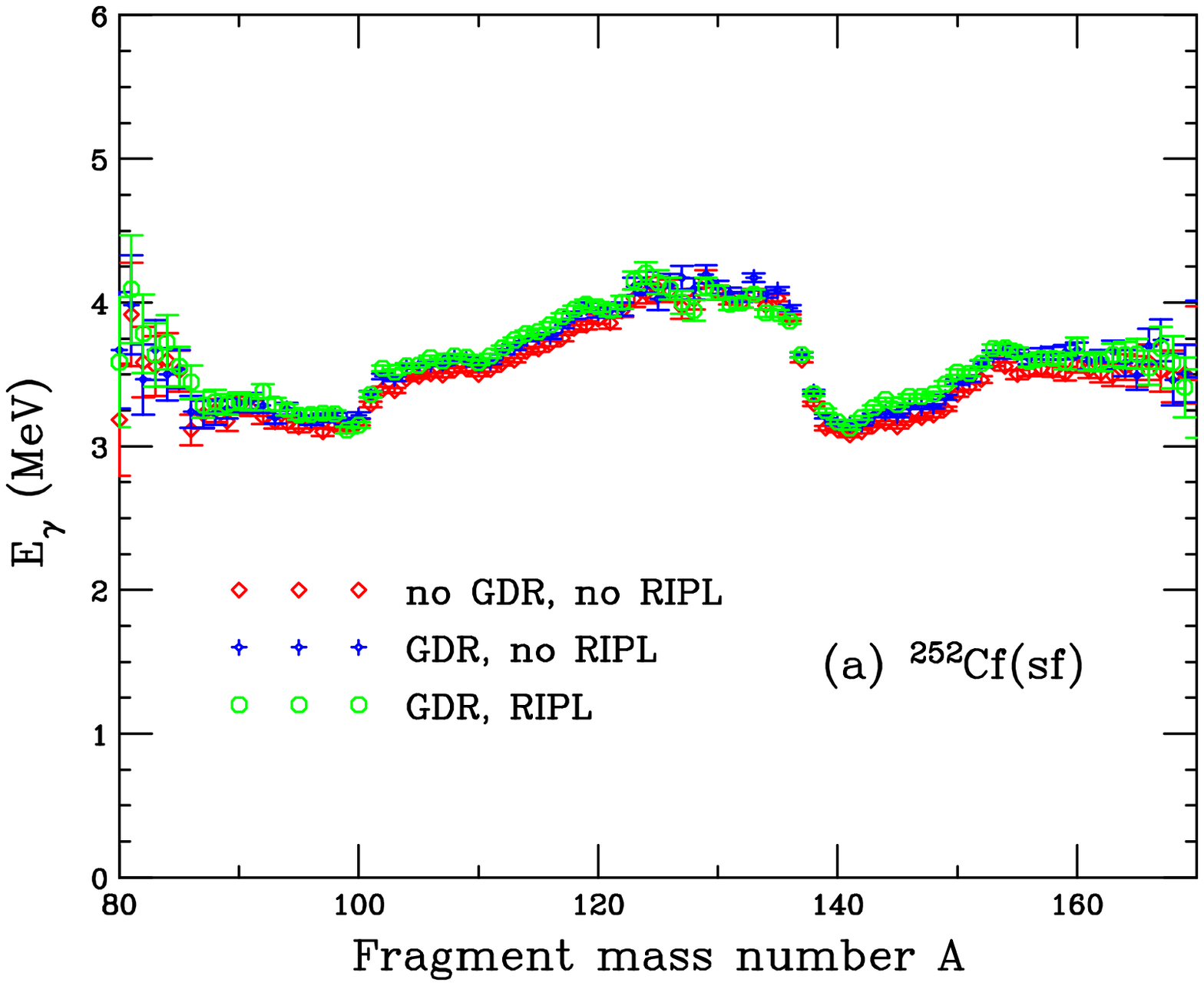}
\includegraphics[angle=0,width=0.45\textwidth]{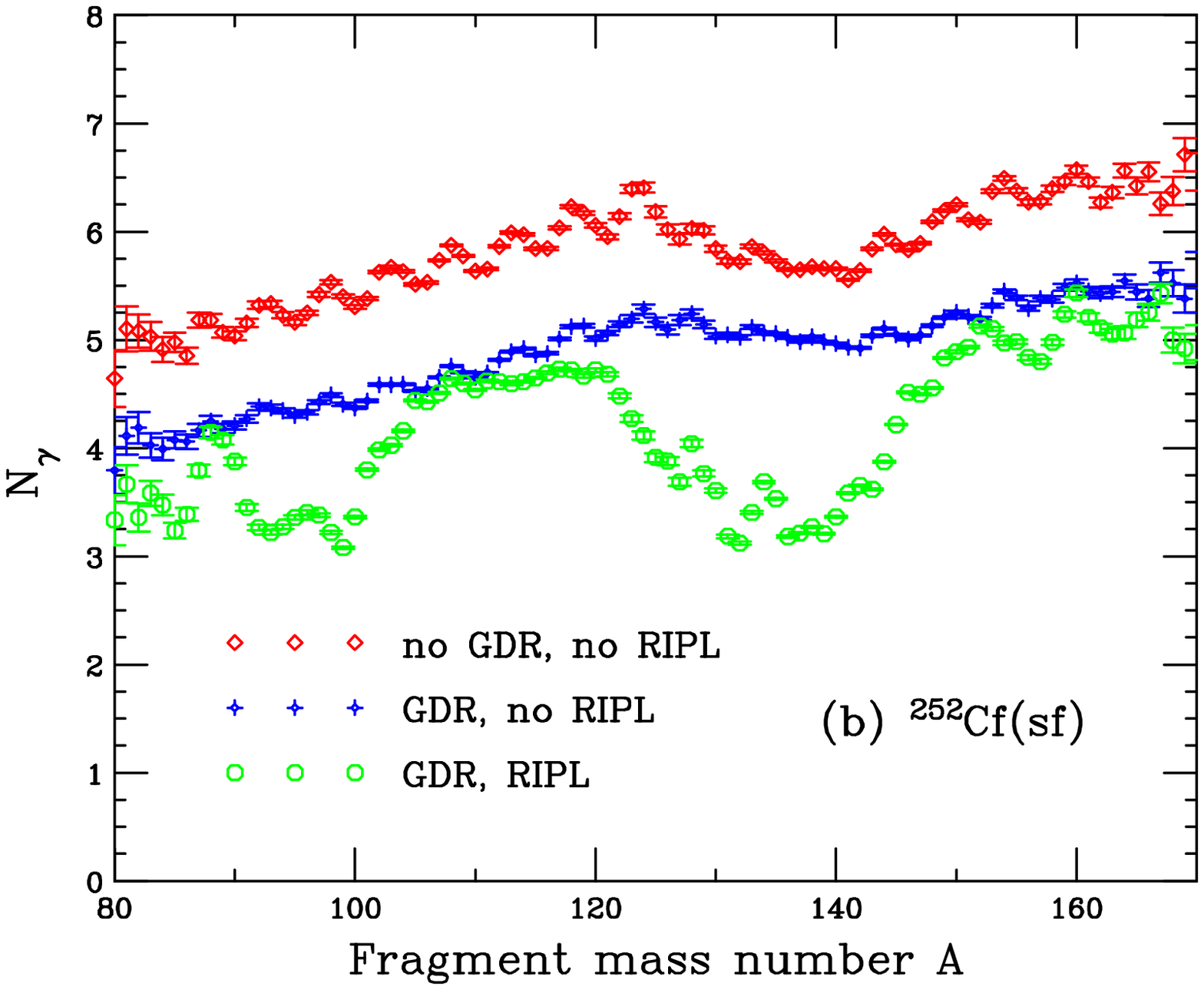}
\caption[]{(Color online) The calculated fragment mass dependence of
(a) the average total photon energy, $E_\gamma(A)$, 
and (b) the average photon multiplicity, $N_\gamma(A)$,
  for $^{252}$Cf(sf) obtained with \code\ for three different model scenarios:
  1) without GDR form factor or RIPL-3 transitions, 
  as in the earlier \code\ \protect\cite{RVJR_gamma};
  2) with the GDR form factor but without RIPL-3 lines;
  and 3) including both, as in the improved \code.
}\label{fig:Adep_Eg_Mg_GDR_RIPL}
\end{figure}

Figure~\ref{fig:Adep_Eg_Mg_GDR_RIPL} shows the total photon energy 
$E_\gamma$ and the photon multiplicity $N_\gamma$ as functions of $A$,
the mass number of the original ({\em i.\,e.}\ pre-evaporation) 
fragment nucleus, for the three different scenarios.
The total energy carried off by photons amounts to the excitation energy
left over after neutron evaporation has ceased
(apart from the small dependence on $t_{\rm max}$).
Consequently, the total photon energy is practically unaffected 
by either the GDR modulation or the inclusion of discrete transitions, 
as seen in Fig.~\ref{fig:Adep_Eg_Mg_GDR_RIPL}(a).
Reflecting the $A$ dependence of the neutron separation energy $S_n$,
the total photon energy is relatively constant for $A<100$ and $A>150$ while, 
in the intermediate region, it increases slowly to a maximum 
near the doubly-closed shell at $A = 132$ 
before dropping and then gradually rising again.  
It varies by around 1 MeV over the full $A$ range.
The shape of $E_\gamma(A)$ is similar to that of Ref.\ \cite{RVJR_gamma}.

On the other hand, the photon multiplicity is affected more strongly
by the inclusion of the GDR and the RIPL transitions,
as seen in Fig.~\ref{fig:Adep_Eg_Mg_GDR_RIPL}(b).  
Relative to the earlier treatment \cite{RVJR_gamma},
the GDR modulation reduces the multiplicity by about one photon
while still yielding a fairly smooth increase with the fragment mass $A$.
In both scenarios, $N_\gamma(A)$ increases linearly
(with a possible odd-even modulation) until symmetry, $A = 126$, where 
the multiplicity decreases and then begins to rise again at $A\approx140$.
By contrast, the inclusion of discrete transitions has a large effect on the
$A$ dependence of $N_\gamma$.  
The transitions introduce more structure, including 
a more pronounced dip near the doubly-closed shell at $A = 132$,
similar to the `sawtooth' pattern in $\nu(A)$.
However, contrary to that behavior,
$N_\gamma(A)$ does not exhibit a pronounced sharp `tooth'
in the light fragment mass region.

\begin{figure}[tbh]	      
\includegraphics[angle=0,width=0.45\textwidth]{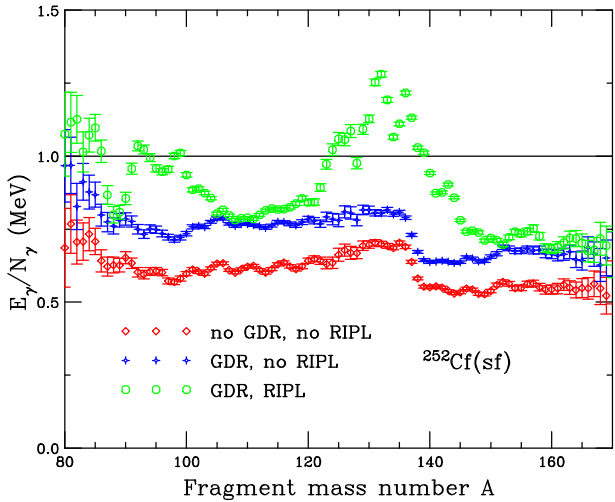}
\caption[]{(Color online) The energy per photon as a function of fragment
  mass for $^{252}$Cf(sf), $E_\gamma(A)/N_\gamma(A)$, obtained
  with \code\ for three different model scenarios:
  1) without GDR form factor or RIPL-3 transitions, 
  as in the earlier \code\ \protect\cite{RVJR_gamma};
  2) with the GDR form factor but without RIPL-3 lines;
  and 3) including both, as done in the improved version of \code.
}\label{fig:Adep_EgoMg_GDR_RIPL}
\end{figure}

The energy per photon, shown in Fig.~\ref{fig:Adep_EgoMg_GDR_RIPL}, 
is the ratio between the mean total photon energy per fragment pair $E_\gamma$, 
shown in Fig.~\ref{fig:Adep_Eg_Mg_GDR_RIPL}(a),
and the mean photon multiplicity $N_\gamma$,
shown in Fig.~\ref{fig:Adep_Eg_Mg_GDR_RIPL}(b).
Because $E_\gamma(A)$ is unaffected by the GDR modulation and the inclusion
of discrete transitions, 
the shape of the ratio is determined by the effect on $N_\gamma(A)$.
All three scenarios show a change in the ratio at $A\approx132$.  
Without the RIPL lines, 
there is simply a shift from a higher plateau for the lighter fragments
to a lower plateau for the heavier fragments.
But when the RIPL lines are included
the pronounced dip in $N_\gamma(A)$ near $A\approx132$ results in 
a peak in $E_\gamma/N_\gamma$ quite different from the other two cases.

\begin{figure}[tbh]	
\includegraphics[angle=0,width=0.45\textwidth]{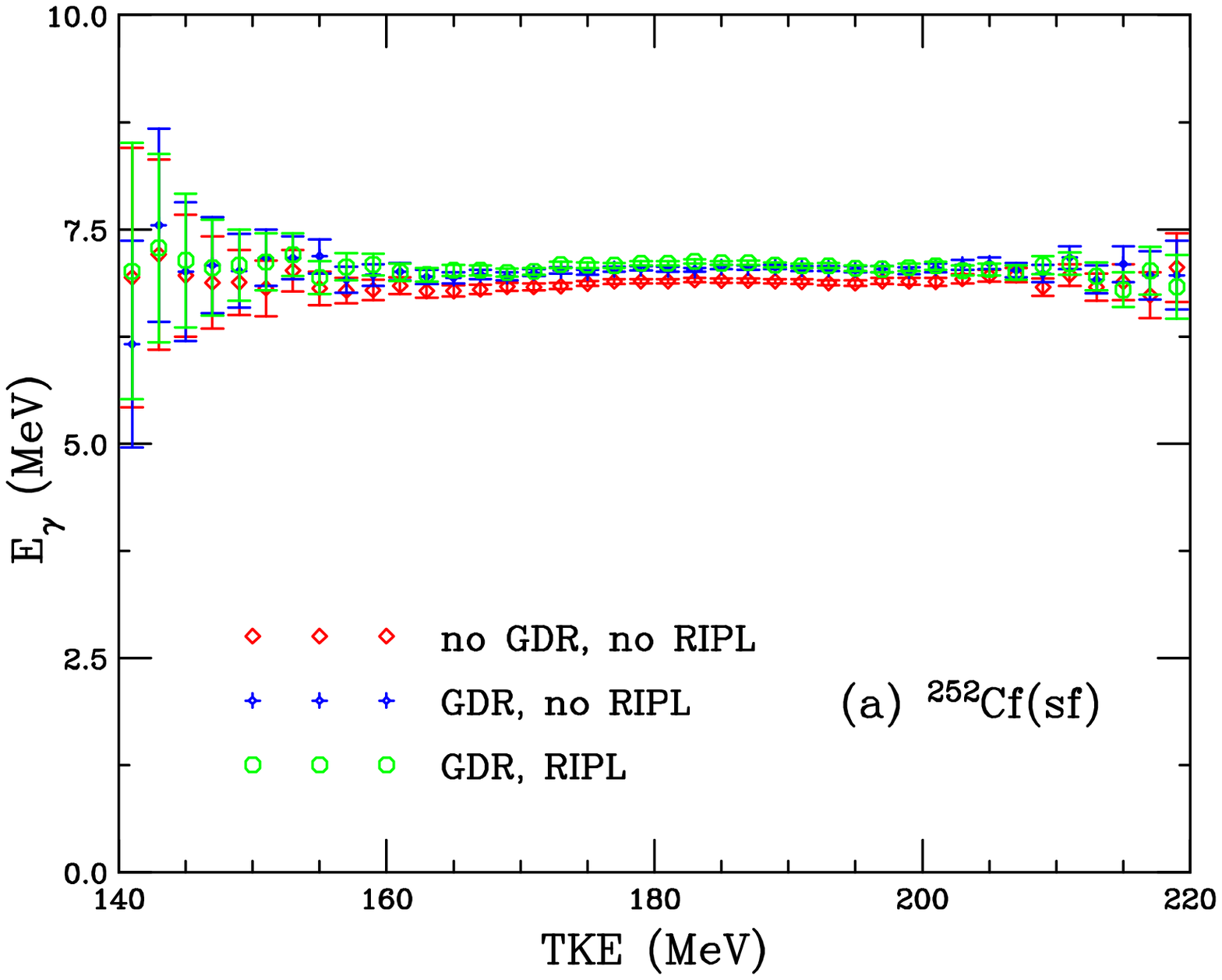}
\includegraphics[angle=0,width=0.45\textwidth]{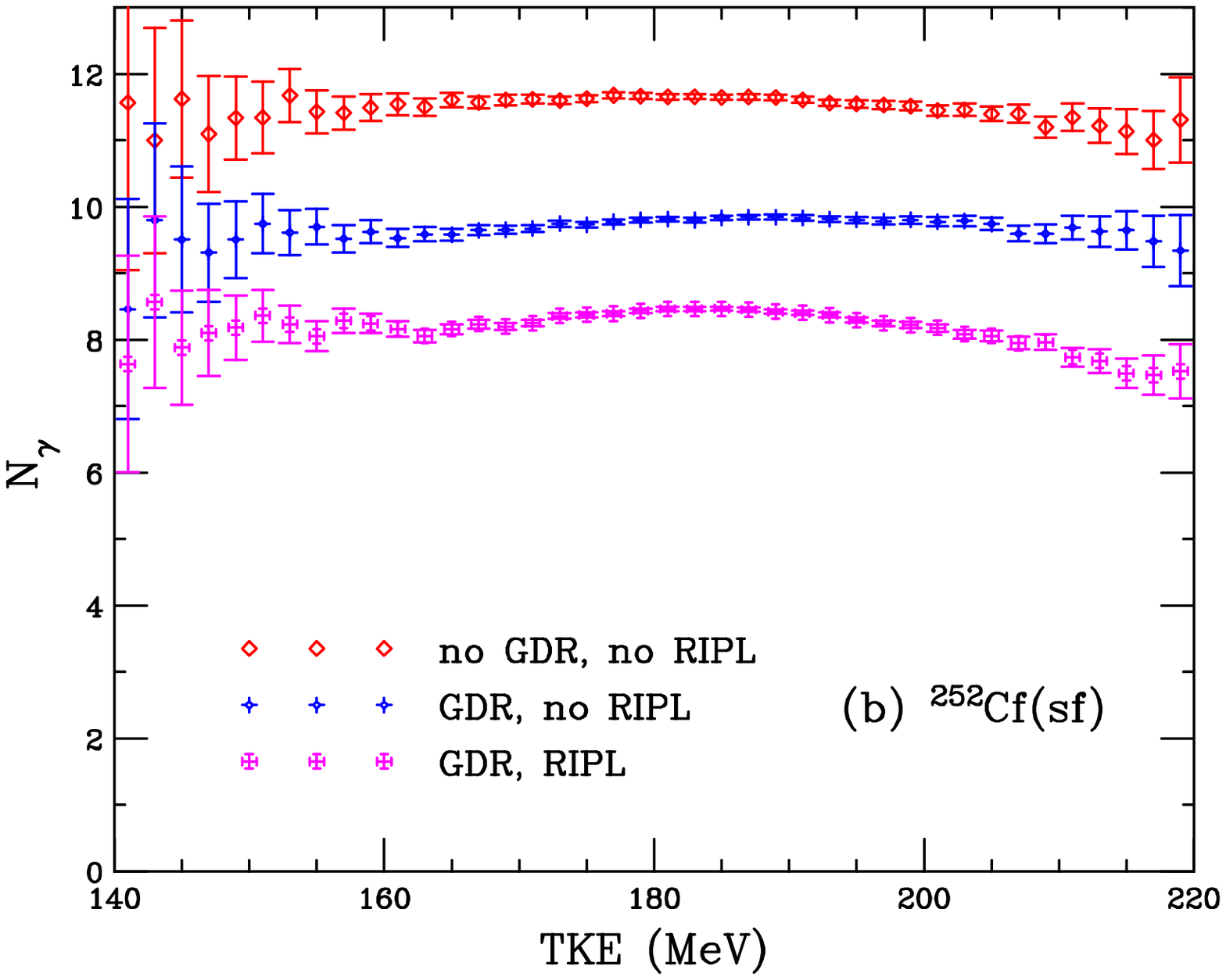}
\caption[]{(Color online) The total photon energy $E_\gamma$ (a) and 
  the photon multiplicity $N_\gamma$ (b) 
  as functions of total fragment kinetic energy TKE 
  calculated for $^{252}$Cf(sf) with \code\ for three different model scenarios:
  1) without GDR form factor or RIPL-3 transitions, 
  as in the earlier \code\ \protect\cite{RVJR_gamma};
  2) with the GDR form factor but without RIPL-3 lines;
  and 3) including both, as done in the improved version of \code.
}\label{fig:TKEdep_Eg_Mg_GDR_RIPL}
\end{figure}

Figure~\ref{fig:TKEdep_Eg_Mg_GDR_RIPL} shows
the total photon energy (emitted from both fragments) $E_\gamma({\rm TKE})$ (a) 
and the total photon multiplicity $N_\gamma({\rm TKE})$ (b)
as functions of the total fragment kinetic energy TKE.
The dependence here mirrors the fragment mass dependence shown
in Fig.~\ref{fig:Adep_Eg_Mg_GDR_RIPL}: $E_\gamma$ is insensitive to
the spectral modulation and the RIPL lines,
while $N_\gamma$ decreases with both the modulation and the RIPL lines.
Each model refinement reduces the multiplicity by nearly one photon.
However, because the total kinetic energy is averaged over mass, 
there is no significant effect on the shape of $N_\gamma({\rm TKE})$.

\begin{figure}[tbh]
\includegraphics[angle=0,width=0.45\textwidth]{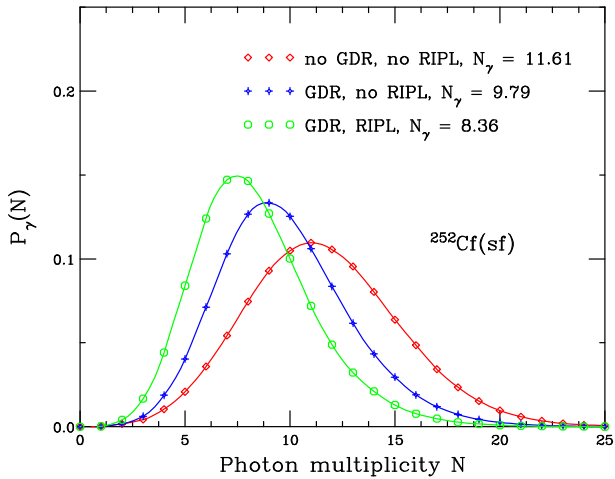}
\caption[]{(Color online) The photon multiplicity distribution 
  $P_\gamma(N_\gamma)$ for $^{252}$Cf(sf) obtained
  with \code\ in three different model scenarios:
  1) without GDR form factor or RIPL-3 transitions, 
  as in the earlier \code\ \protect\cite{RVJR_gamma};
  2) with the GDR form factor but without RIPL-3 lines;
  and 3) including both, as done in the improved version of \code.
}\label{fig:mult_GDR_RIPL}
\end{figure}

Finally, Fig.~\ref{fig:mult_GDR_RIPL} shows
the total photon multiplicity distribution $P_\gamma(N)$
in the three model scenarios.
Unlike the neutron multiplicity distribution, $P_n(\nu)$,
which does not have a Poisson form (primarily because of 
the dominant role played by the separation energy $S_n$),
$P_\gamma(N)$ is more Poisson-like.
The earlier \code\ treatment \cite{RVJR_gamma}
yields the largest average photon multiplicity, 
$N_\gamma = 11.61$, while the GDR spectral modulation
reduces the mean multiplicity to $N_\gamma = 9.79$,
consistent with the fact that it tends to harden the spectrum
(see Fig.\ \ref{fig:spect_GDR_RIPL}).
The inclusion of the discrete transitions reduces the multiplicity even further,
to  $N_\gamma = 8.36$.

After the above illustration of how the two main model refinements
affect various photon observables,
we now move on to discuss how the photon results of the improved \code\
depend on the various model parameters.

\section{Dependence on $c_S$}
\label{sec:cSdep}

In this section we discuss the effect on photon observables caused by changing
the parameter $c_S$ which controls the rotational motion of the fragments.
Recall that the rotational energy is subtracted from the total excitation
energy, leaving the remainder of the energy available for neutron evaporation.

The other parameters are kept fixed while $c_S$ is varied.  
We show results for $c_S = 0.2$, 0.8, 1.4 and 2.0.  
The value of 0.8 is rather close to the best fit value of 0.87.
The value 0.2 is taken as a lower bound,
significantly reducing the degree of rotation removing it altogether.
The two larger values serve to illustrate the range of the effect.

\begin{figure}[tbh]
\includegraphics[angle=0,width=0.45\textwidth]{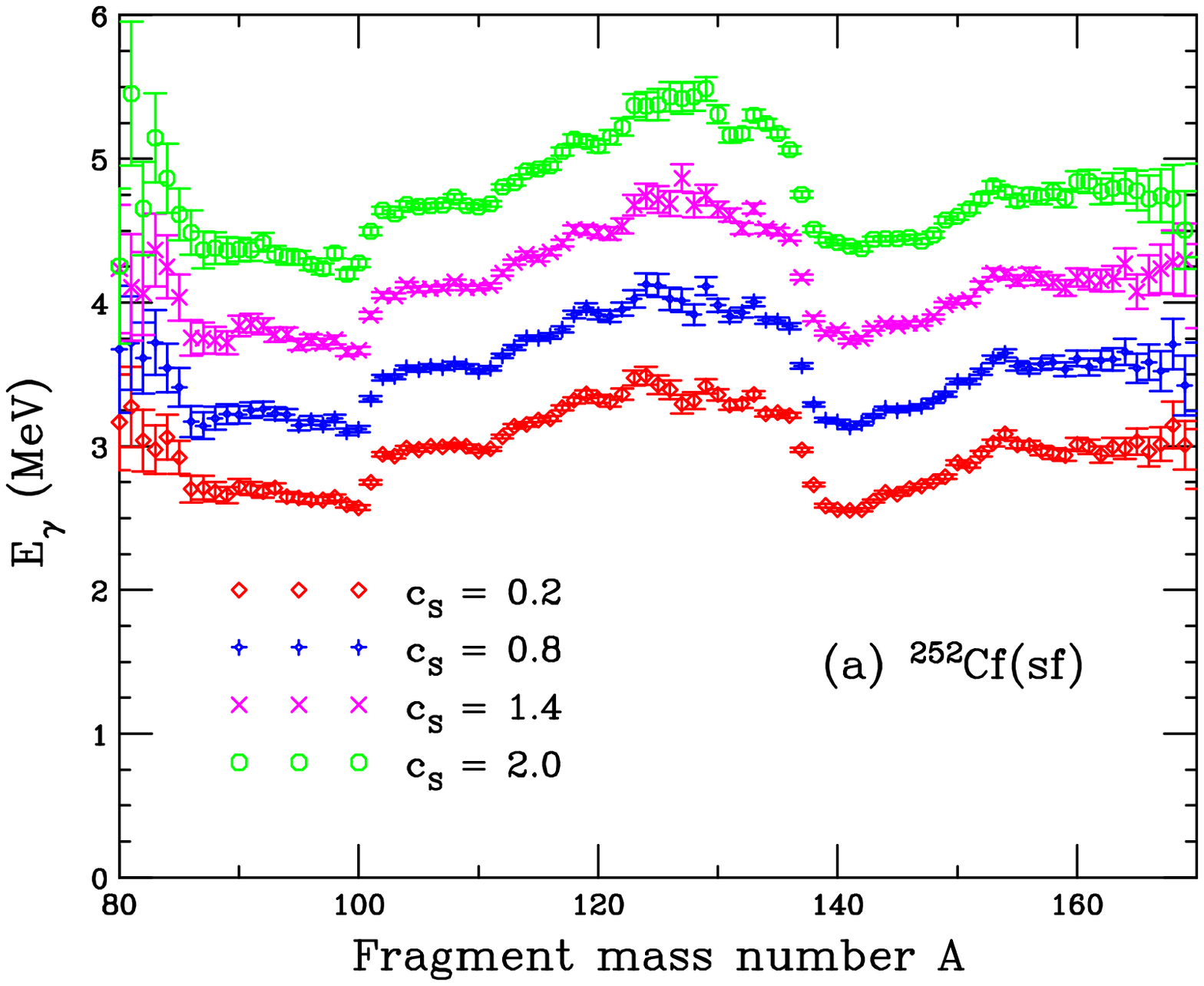}
\includegraphics[angle=0,width=0.45\textwidth]{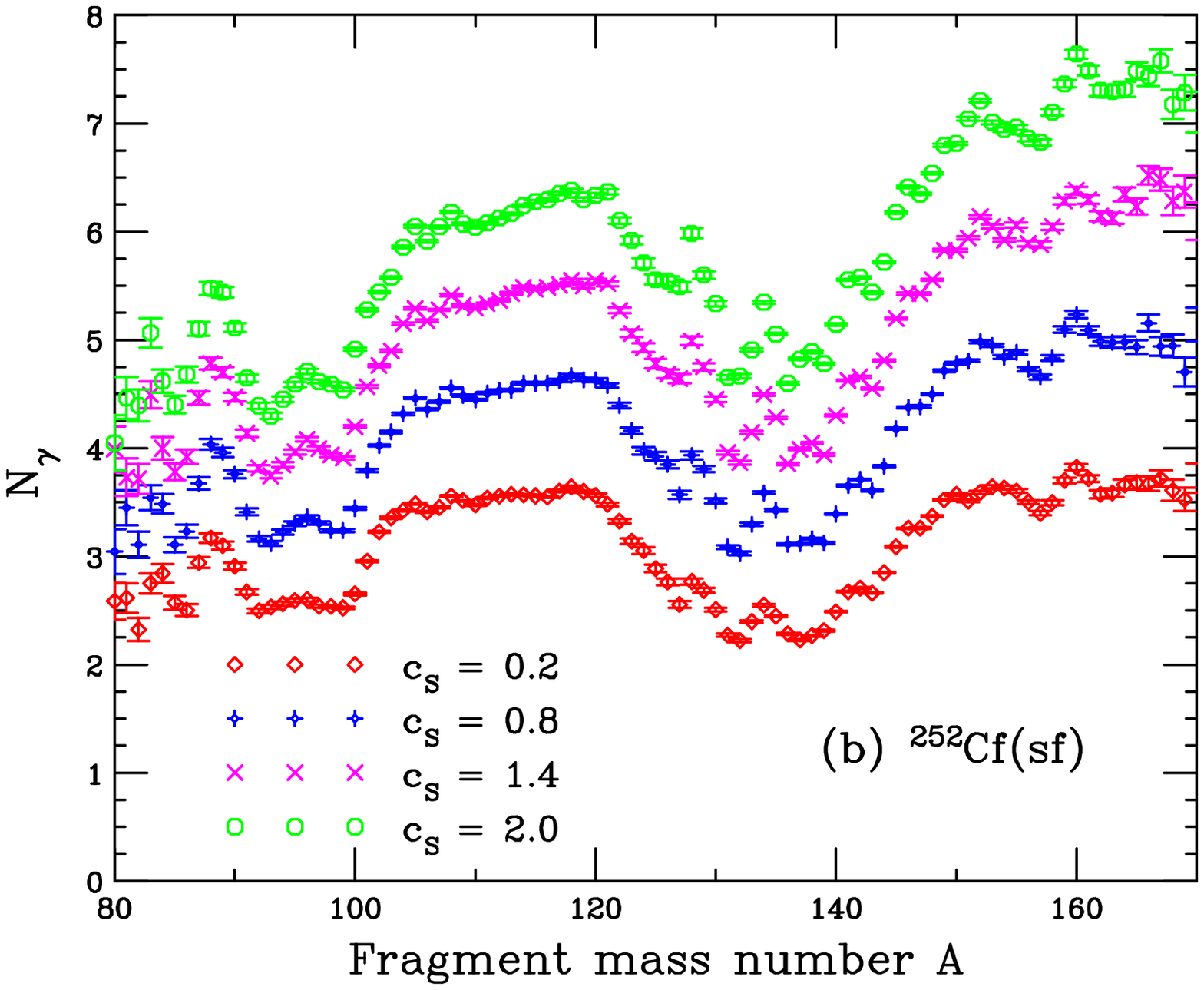}
\caption[]{(Color online) The mean total photon energy $E_\gamma$ (a) 
  and the mean photon multiplicity $N_\gamma$ (b)
  as functions of the fragment mass number $A$ calculated for $^{252}$Cf(sf) 
  for four different values of the parameter $c_S$ 
  which sets the magnitude of the fragment spins.
  Results are shown for $c_S = 0.2$, 0.8, 1.4, and 2.0.
}\label{fig:Adep_Eg_Mg_cs}
\end{figure}

Figure~\ref{fig:Adep_Eg_Mg_cs} presents $E_\gamma(A)$ and $N_\gamma(A)$
for the four illustrative values of $c_S$.
As $c_S$ is increased and the fragments are endowed with ever more rotation,
there is less energy available for the neutrons.
In Fig.~\ref{fig:Adep_Eg_Mg_cs}(a), it can be seen that the photon energy
simply seems to increment by approximately 0.5 MeV when $c_S$ increases by
0.6.  There is no visible mass dependence on changing $c_S$.  However, the
photon multiplicity does seem to show some modification of the fragment mass
dependence with increasing $c_S$, see Fig.~\ref{fig:Adep_Eg_Mg_cs}(b).  While
the general trend is the same for all values of $c_S$, there seems to be a
larger increase in the multiplicity for the heavy fragment while the dip at
$A\approx132$ deepens with increasing $c_S$.  In addition, there appears 
to be a rather flat plateau for the light fragment masses $100 < A < 120$
for $c_S = 0.2$ that acquires a positive slope as $c_S$ increases.

\begin{figure}[tbh]  
  \includegraphics[angle=0,width=0.45\textwidth]{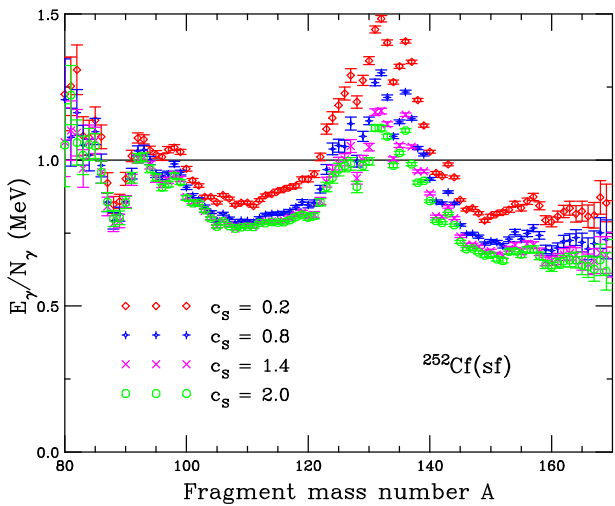}
  \caption[]{(Color online) The energy per photon $E_\gamma/N_\gamma$
  as a function of the fragment mass number $A$ calculated for $^{252}$Cf(sf)
  for four different values of the parameter $c_S$
  which sets the magnitude of the fragment spins.
  Results are shown for $c_S = 0.2$, 0.8, 1.4, and 2.0.
}\label{fig:Adep_EgoMg_cs}
\end{figure}

The changes in the photon multiplicity with fragment mass result in the
observed differences in the energy per photon ratio shown in
Fig.~\ref{fig:Adep_EgoMg_cs}.  The lowest value of $c_S$ actually produces the
most pronounced peak in $E_\gamma/N_\gamma$ for $A\approx132$.  As $c_S$ is
increased, the energy per photon is reduced, particularly for the heavy
fragment.

\begin{figure}[bh]	
\includegraphics[angle=0,width=0.45\textwidth]{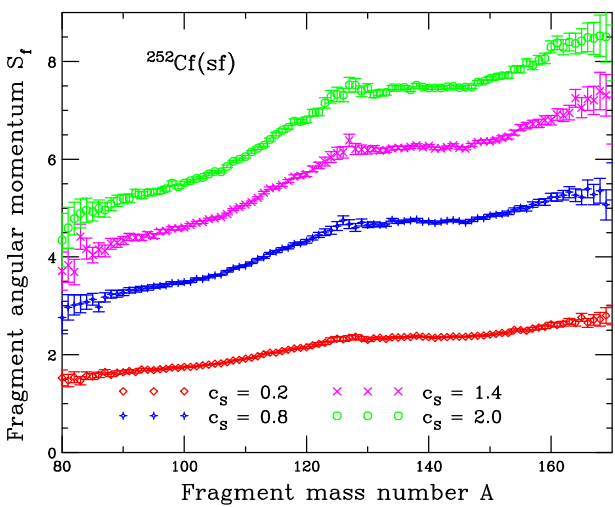}
  \caption[]{(Color online) The average fragment angular momentum $S_{\rm f}$
  as a function of the fragment mass $A$ for $^{252}$Cf(sf)
  calculated for four different values of the parameter $c_S$ 
  (which controls the magnitude of the fragment spins).
  Results are shown for $c_S = 0.2$, 0.8, 1.4, and 2.0.
}\label{fig:Adep_spins_cS}
\end{figure}

Figure~\ref{fig:Adep_spins_cS} shows how the average magnitude 
of the fragment angular momentum, $S_{\rm f}$, grows with the parameter $c_S$.
The value of $S_{\rm f}$ is almost independent of fragment mass for $c_S = 0.2$, 
remaining near $2\hbar$.
As $c_S$ is increased, $S_{\rm f}$ and, thus, the portion of the excitation
energy captured in rotational energy increase as well. \
$S_{\rm f}(A)$ grows nearly linearly in the light mass region
and then remains relatively constant in the heavy region.
The increase in the light fragment spin grows more pronounced 
for larger values of $c_S$.

We recall, however, that these results are calculated assuming that no other
parameter value changes.  Thus, if the total excitation energy is held fixed,
increasing the rotational energy, as is the case for increased $c_S$, then
less energy is available for neutron emission.  Thus increasing $c_S$ while
keeping the other parameters fixed will decrease the average neutron
multiplicity.  For example, if all other parameters remain unchanged, the
average neutron multiplicity can decrease by as much as 12\% when $c_S$ is
increased from 0.2 to 2.0.
Although this could be compensated for by changing the value
of $d$TKE, one has to be careful to adjust it within reasonable physics limits.
Furthermore, when adjusting variables it is important to check the effect 
on other observables to ensure that overall description remains good.
The results here are thus for illustrative purposes only.

\begin{figure}[tbh]  
\includegraphics[angle=0,width=0.45\textwidth]{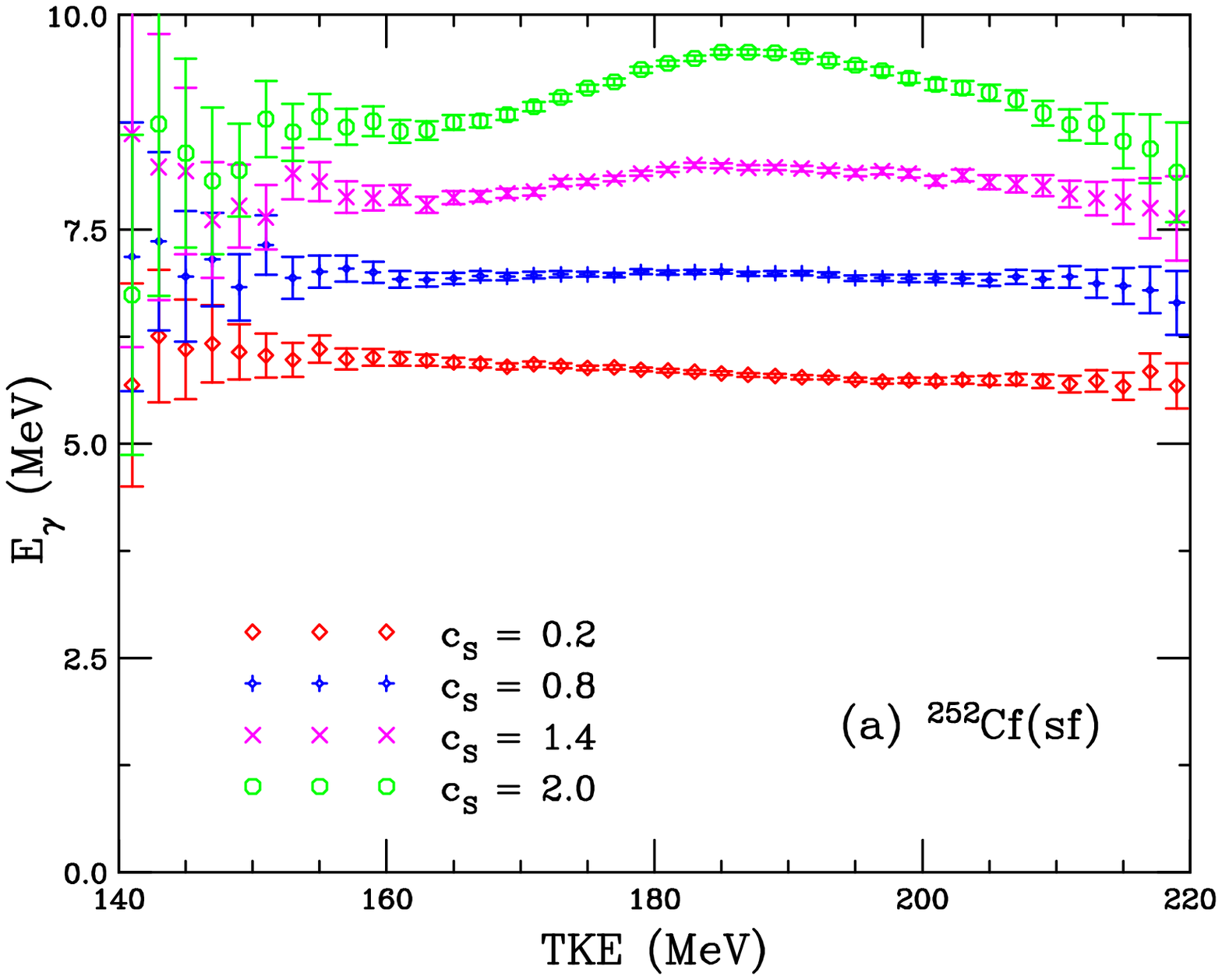}
\includegraphics[angle=0,width=0.45\textwidth]{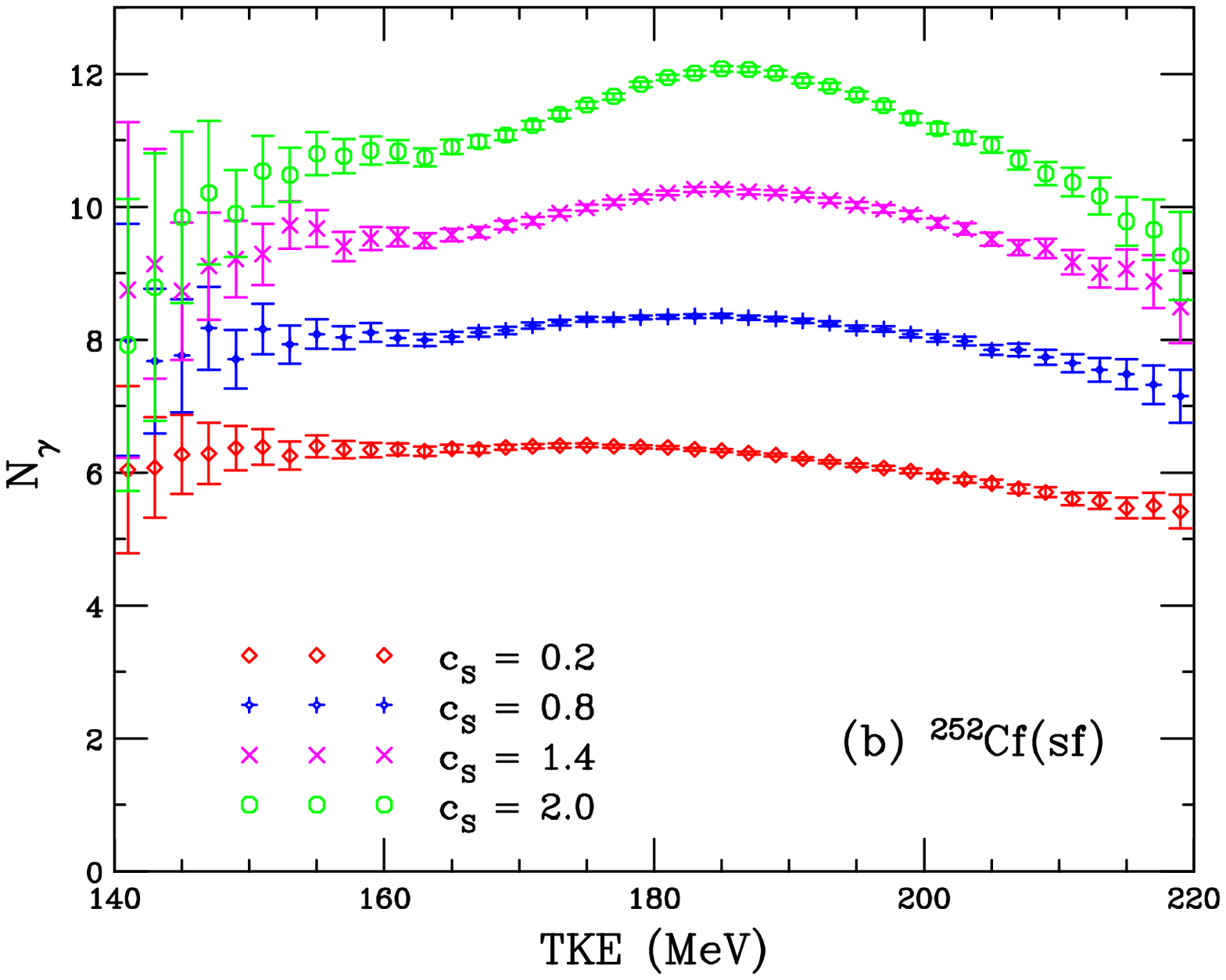}
\caption[]{(Color online) The total photon energy $E_\gamma$ (a)
  and the photon multiplicity $N_\gamma$ (b)
  as functions of total fragment kinetic energy TKE for $^{252}$Cf(sf) 
  calculated for four different values of the parameter $c_S$
  (which controls the magnitude of the fragment spins,
  see Fig.\ \ref{fig:Adep_spins_cS}).
  Results are shown for $c_S = 0.2$, 0.8, 1.4, and 2.0.
}\label{fig:TKEdep_Eg_Mg_cS}
\end{figure}

Figure~\ref{fig:TKEdep_Eg_Mg_cS} shows the variation of $E_\gamma({\rm TKE})$
and $N_\gamma({\rm TKE})$ with respect to $c_S$.  
As $c_S$ is increased, the these functions develop some
curvature with an enhancement appearing around ${\rm TKE}\approx190$ MeV.

\begin{figure}[tbh]	      
\includegraphics[angle=0,width=0.45\textwidth]{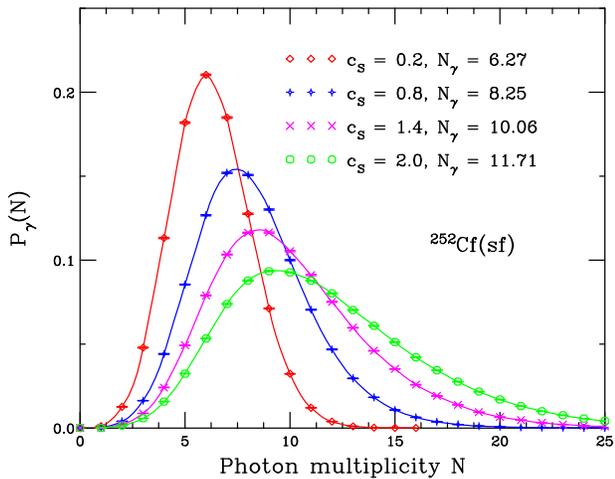}
  \caption[]{(Color online) The photon multiplicity distribution
  $P_\gamma(N)$ for $^{252}$Cf(sf) calculated
  for four different values of the parameter $c_S$ 
  (which sets the magnitude of the fragment spins).
  Results are shown for $c_S = 0.2$, 0.8, 1.4, and 2.0
  using $g_{\rm min}=0.1$ MeV and $t_{\rm max} = 10$ ns.
  The mean multiplicity for each value of $c_S$ is indicated.
}\label{fig:mult_cs}
\end{figure}

Finally, the $c_S$ dependence of the photon multiplicity distribution
$P_\gamma(N)$ is shown in Fig.~\ref{fig:mult_cs}.
As $c_S$ is increased, the fission fragments are formed with 
ever larger angular momenta and because these remain largely unchanged 
during the neutron evaporation chain, the resulting post-evaporation 
fragments tend to have correspondingly higher excitations.
Consequently, a larger number of photons may be emitted,
increasing both the mean multiplicity and the width of $P_\gamma(N)$.

\section{Dependence on $g_{\rm min}$ and $t_{\rm max}$}

Here we discuss the dependence of the \code\ results on the  
detector-related parameters $g_{\rm min}$ and $t_{\rm max}$.  
In our discussion of the dependence on $g_{\rm min}$, 
we show results similar to those in Sec.~\ref{sec:cSdep}.
We refer to photons above $g_{\rm min}$ as detected fission photons since photons
with energies below $g_{\rm min}$, while emitted, will not be detected.
The effect of the detection time window, $t_{\rm max}$, is more subtle, 
however, so we present the $t_{\rm max}$ dependence 
relative to an infinitely wide detection window, $t_{\rm max}\to\infty$.
While most of the results in this section are shown only for $^{252}$Cf(sf), 
some results for $^{235}$U($n_{\rm th}$,f) and $^{239}$Pu($n_{\rm th}$,f) 
are included as well.
It should be recognized that neither of these quantities affects 
the physical photon emission, only the recording of the emission.

\subsection{Dependence on $g_{\rm min}$}

We begin by considering values of $g_{\rm min}$ 
that are in the range of typical photon detectors, $0.05-0.20$ MeV.
We show the dependence of the total photon multiplicity $N_\gamma$
on the total fragment kinetic energy TKE.
We then illustrate the effect of increasing $g_{\rm min}$ up to 2 MeV
which puts the focus ever more on the high-energy 
(and, hence, mostly statistical) photons.

\begin{figure}[tbh]	
\includegraphics[angle=0,width=0.45\textwidth]{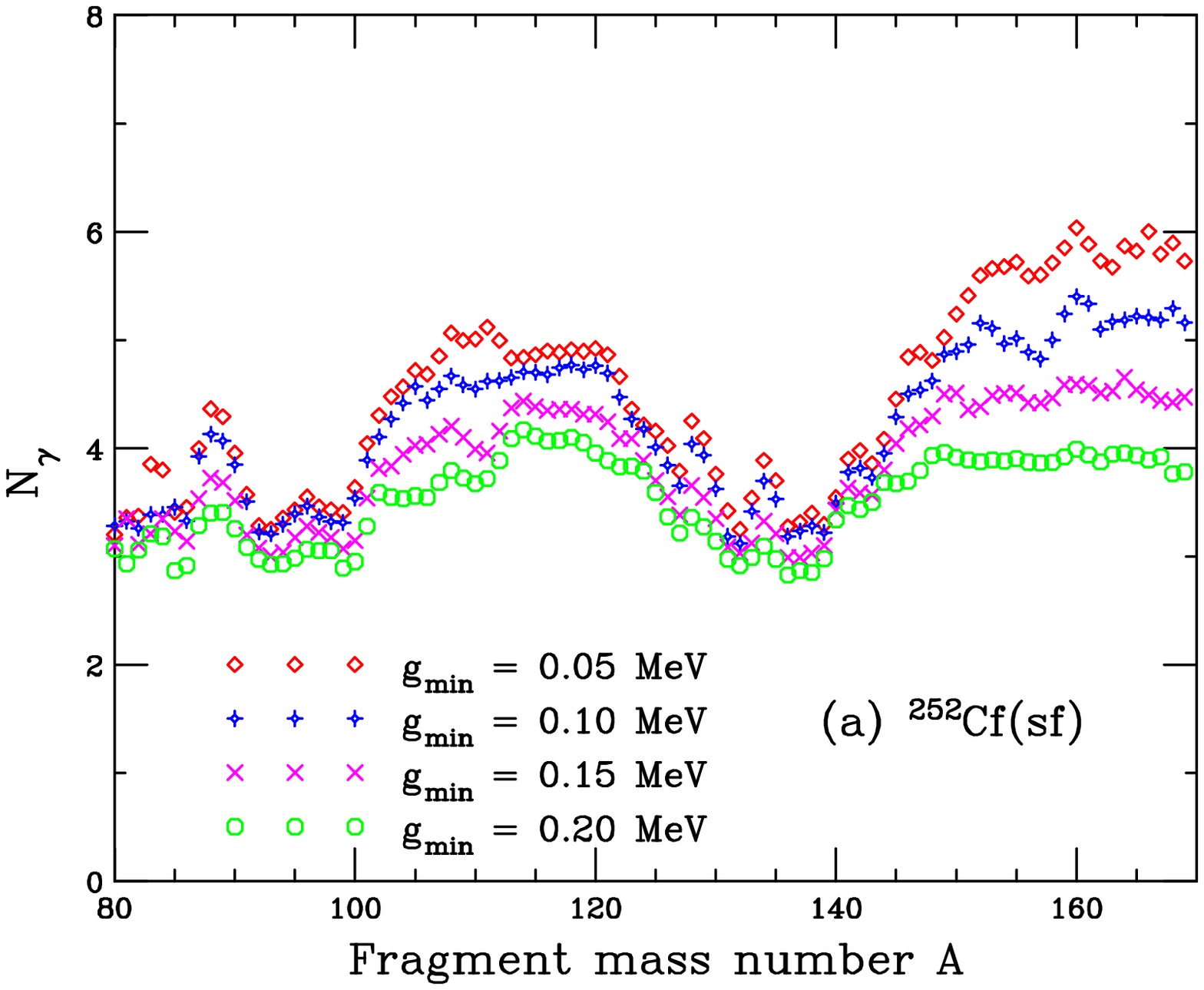}
\includegraphics[angle=0,width=0.45\textwidth]{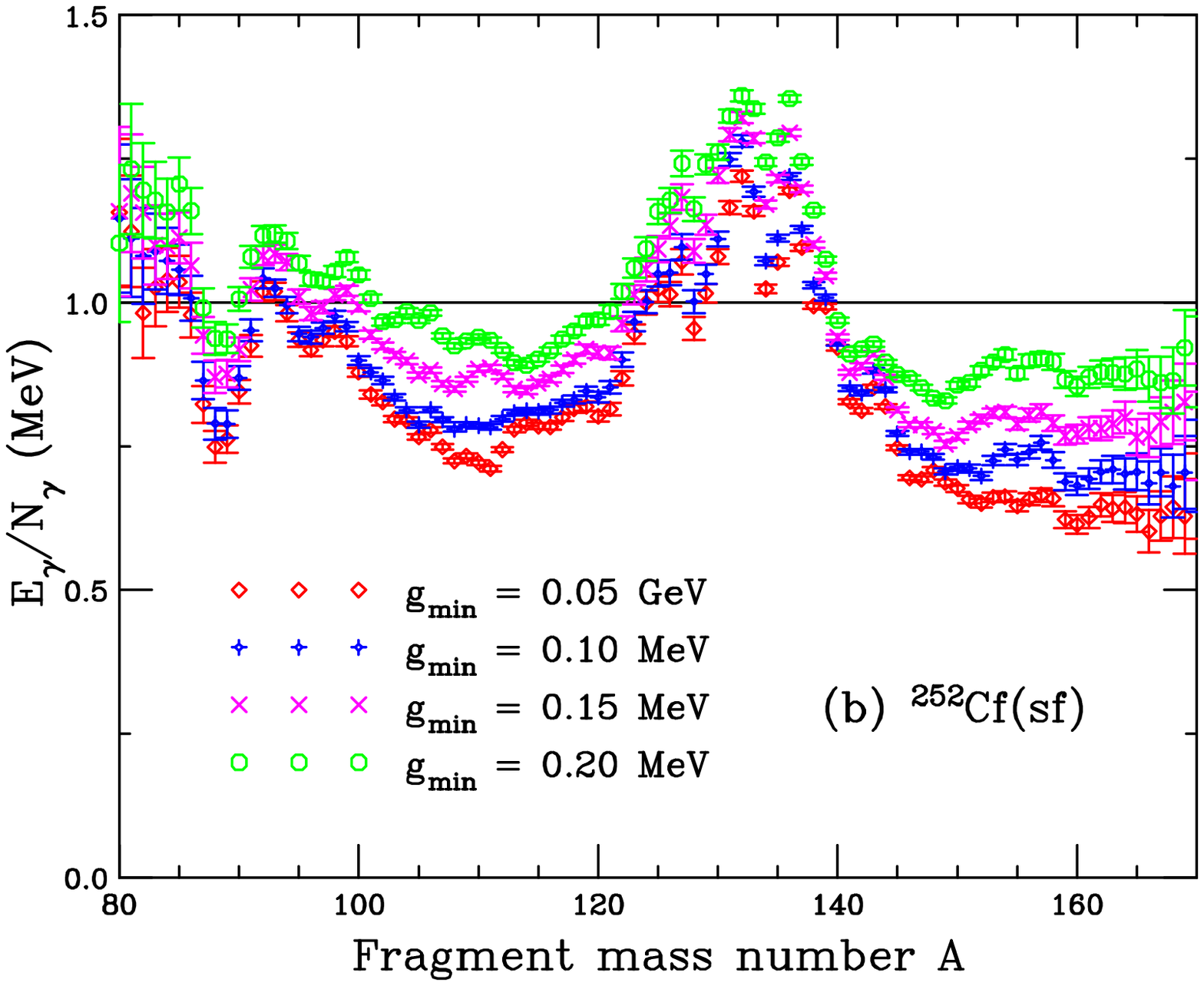}
\caption[]{(Color online) The photon multiplicity $N_\gamma$ (a) 
  and the energy per photon $E_\gamma/N_\gamma$ (b) 
  as functions of fragment mass number $A$ for $^{252}$Cf(sf) calculated
  for four different values of $g_{\rm min}$,
  the minimum photon energy detected.  
  Results are shown for $g_{\rm min}=0.05$, 0.10, 0.15, and 0.20 MeV.
}\label{fig:Adep_Mg_gmin}
\end{figure}

The effect on $E_\gamma(A)$ is very small, 
with a change in $g_{\rm min}$ between 0.05 and 0.20 MeV 
producing a reduction of only $\approx2$\%,
and it is therefore not shown in a separate figure,
whereas the $g_{\rm min}$ dependence of $N_\gamma(A)$ and $E_\gamma(A)/N_\gamma(A)$
are shown in Fig.~\ref{fig:Adep_Mg_gmin} (a) and (b), respectively.  

The effect on the photon multiplicity is significant 
(albeit not as large as changing $c_S$ by a factor of 10, 
between 0.2 and 2, as shown in Fig.~\ref{fig:mult_cs}).
If $g_{\rm min} \rightarrow 0$, ever more soft photons could be emitted, 
increasing the total multiplicity.  
The photon multiplicity from the heavy fragment is affected the most
by a change in $g_{\rm min}$, 
with a $\approx50$\% change in $N_\gamma$ near $A=160$ 
relative to a $\approx20$\% change near $A=110$.  
Because the discrete transitions tend to be relatively soft,
their significance will diminish rapidly as $g_{\rm min}$ is increased.  There are
also soft statistical photons at small $g_{\rm min}$ that will be removed from
the multiplicity count.
As a result, the dependence of $N_\gamma$ on $A$ will grow ever weaker
until it is effectively constant.  

Figure \ref{fig:Adep_Mg_gmin}(b) shows the dependence of 
$E_\gamma(A)/N_\gamma(A)$ on $g_{\rm min}$.  
Also here the $A$-dependence weakens as $g_{\rm min}$ is increased.
However, the characteristic shape shown in the previous sections
remains relatively unchanged.

\begin{figure}[tbh]	
\includegraphics[angle=0,width=0.45\textwidth]{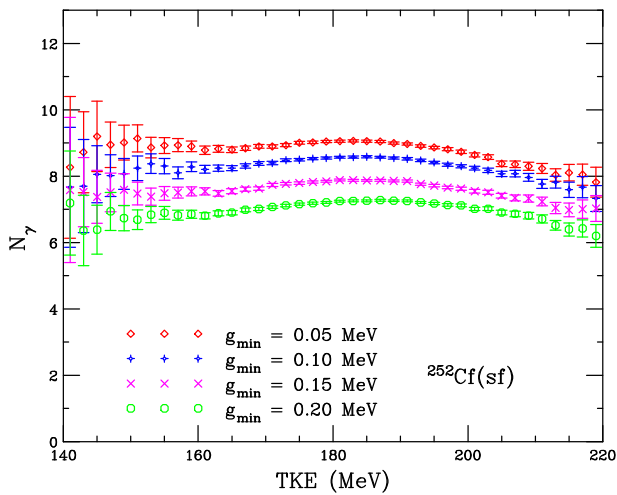}
\caption[]{(Color online) The average photon multiplicity
  $N_\gamma$ as a function of total fragment kinetic energy TKE
  calculated for $^{252}$Cf(sf) 
  for four different values of the parameter $g_{\rm min}$,
  the minimum photon energy detected.
  Results are shown for $g_{\rm min} = 0.05$, 0.10, 0.15, and 0.20~MeV.
}\label{fig:TKEdep_Mg_gmin}
\end{figure}

For the same values of $g_{\rm min}$, Fig.~\ref{fig:TKEdep_Mg_gmin} displays
the multiplicity $N_\gamma$ as a function of the total kinetic energy TKE.  
While the total photon energy is generally independent of TKE for $^{252}$Cf, 
there is a mild TKE dependence of the total photon multiplicity.
The main effect of increasing the detection threshold $g_{\rm min}$
is an overall reduction in $N_\gamma$.

\begin{figure}[tbh]	  
\includegraphics[angle=0,width=0.45\textwidth]{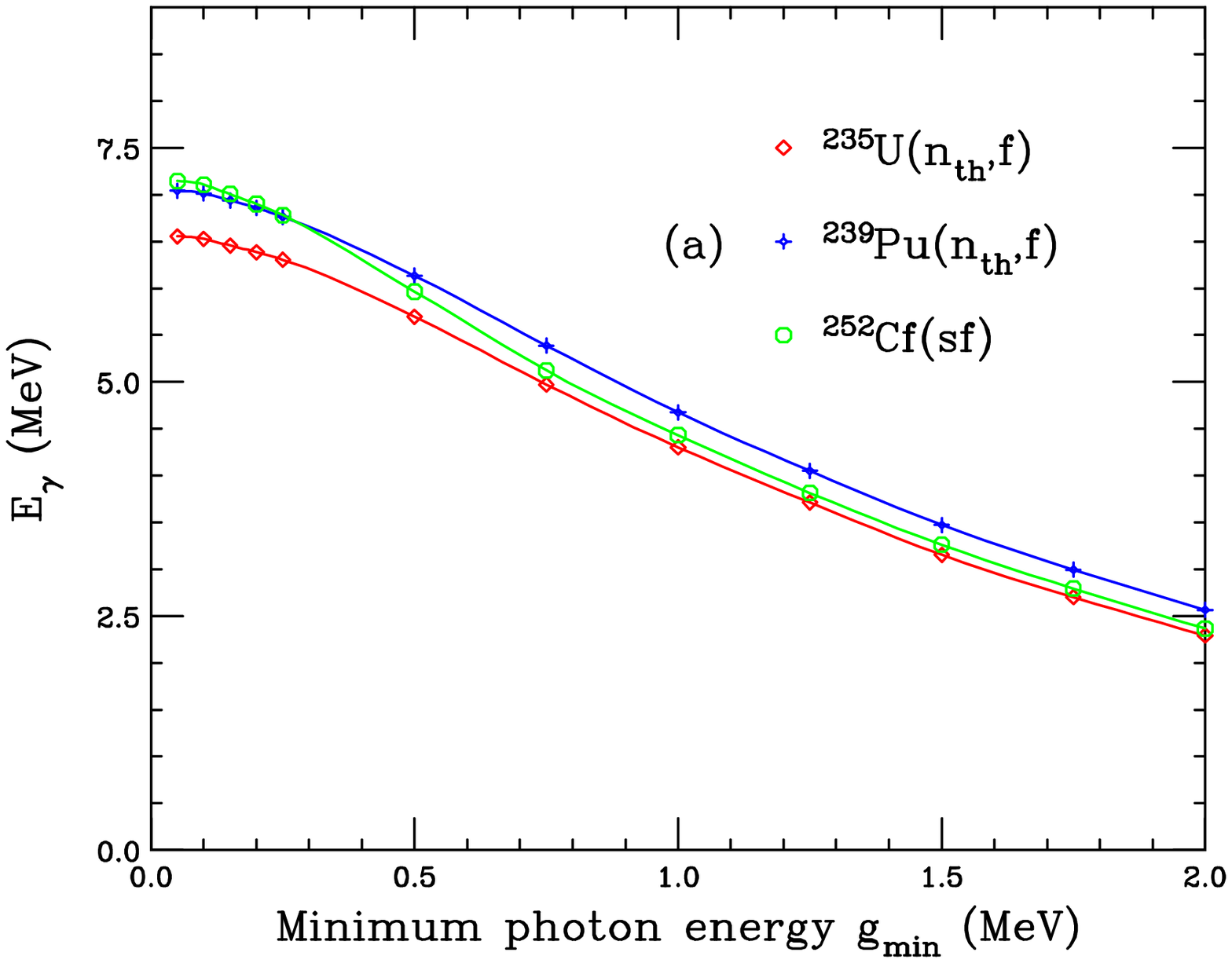}
\includegraphics[angle=0,width=0.45\textwidth]{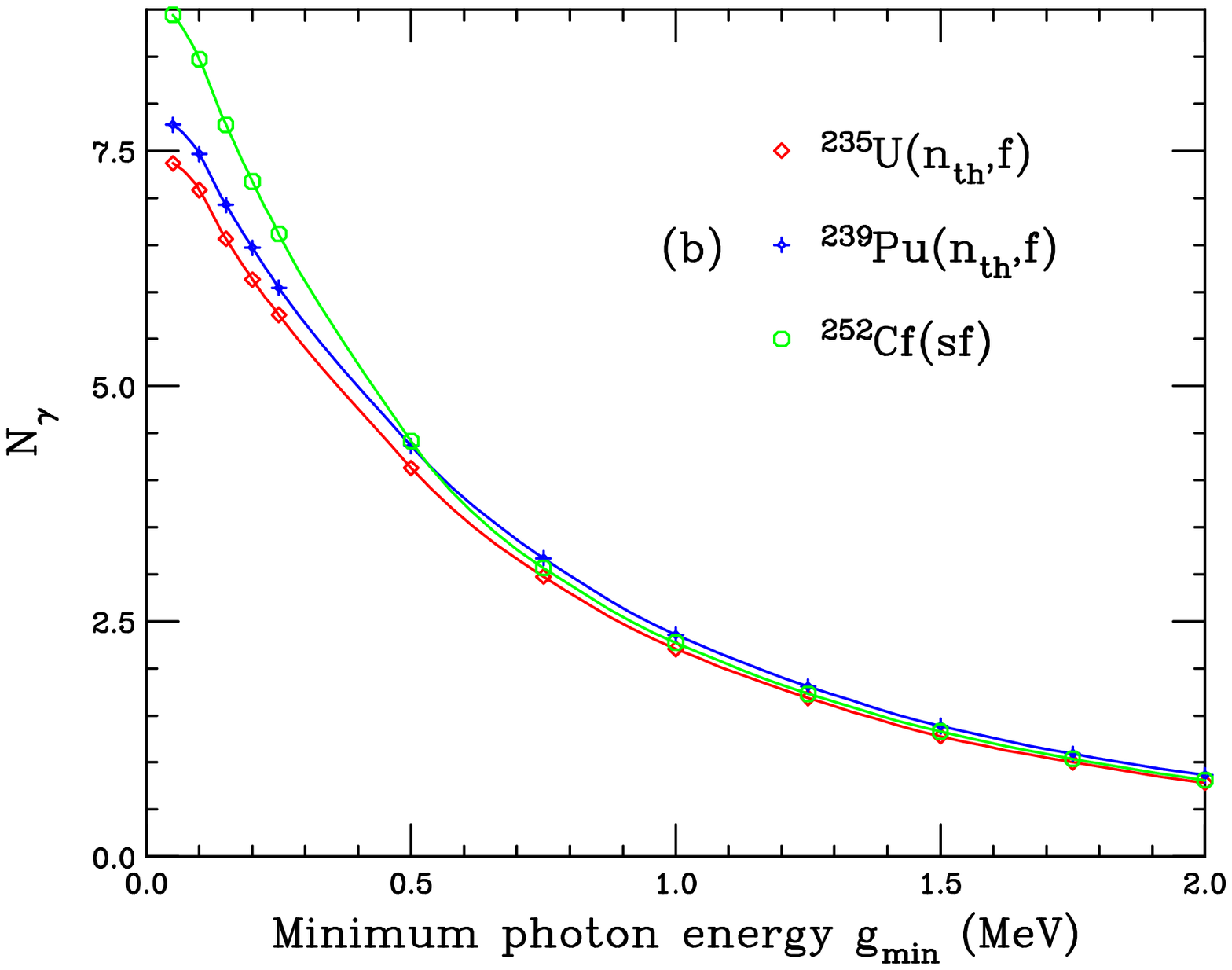}
\caption[]{(Color online) The total photon energy $E_\gamma$ (a)
  and the photon multiplicity $N_\gamma$ (b) 
  calculated for $^{235}$U($n_{\rm th}$,f), $^{230}$Pu($n_{\rm th}$,f), and
  $^{252}$Cf(sf) as a function of the photon energy cutoff $g_{\rm min}$.
  The value of $t_{\rm max}$ was 10~ns for all cases.
}\label{fig:gmindep_CfUPu}
\end{figure}

We now show the dependence of $E_\gamma$ and $N_\gamma$ on $g_{\rm min}$
over a significantly broader range of values, up to 2~MeV.
These results are shown in Fig.~\ref{fig:gmindep_CfUPu} 
for $^{252}$Cf(sf), $^{235}$U($n_{\rm th}$,f) and $^{239}$Pu($n_{\rm th}$,f).
Figure~\ref{fig:gmindep_CfUPu}(a) shows that the slow decrease, noted above
for $g_{\rm min} < 0.2$~MeV, grows stronger for larger values of $g_{\rm min}$.
The dependence is approximately linear for all three systems.
The drop-off of $E_\gamma$ is somewhat steeper for $^{252}$Cf(sf) 
than for $^{235}$U($n_{\rm th}$,f) and $^{239}$Pu($n_{\rm th}$,f)
whose slopes are very similar.

When $g_{\rm min}$ is small, the detected total fission photon energy approaches 
its maximum possible value, namely the total radiated photon energy.
For each product nucleus, the radiated photon energy is given by 
its total excitation energy after neutron evaporation has ceased
which is typically $2-3$ MeV below the threshold at $E^*=S_n\approx6$ MeV.
As $g_{\rm min}$ is increased from 0.05 to 2 MeV,
$E_\gamma$ decreases by a factor of $2.5-3$,
corresponding to a couple of MeV for each product nucleus.

Because most of the radiated photons are relatively soft,
the detected fission photon
multiplicity $N_\gamma$ drops off significantly more rapidly
as the threshold $g_{\rm min}$ is increased,
as seen in Fig.~\ref{fig:gmindep_CfUPu}(b).
Again, the dependence on $g_{\rm min}$ is stronger for $^{252}$Cf(sf) 
which also starts out from a somewhat higher value at $g_{\rm min}\approx0$
than the other two cases displayed.
By $g_{\rm min} > 0.5$~MeV, all three cases show essentially 
the same $N_\gamma(g_{\rm min})$ 
which decreases to slightly less than one by $g_{\rm min} = 2$~MeV.

The similarity in photon multiplicity is likely due to the fact that
(in the current version of \code), 
no photons are emitted until the excitation energy of the fragment has
fallen below the neutron separation energy.  As already mentioned in the
introduction, this has the consequence that neutron observables are insensitive
to the choice of $g_{\rm min}$ and $t_{\rm max}$. 
Due to this, one may also expect
that the residual excitation energy left for photon emission has a weak
dependence on incident neutron energy, as already mentioned in previous work
\cite{RVJR_gamma,RVJR_spont}.

\begin{figure}[tbh]
\includegraphics[angle=0,width=0.45\textwidth]{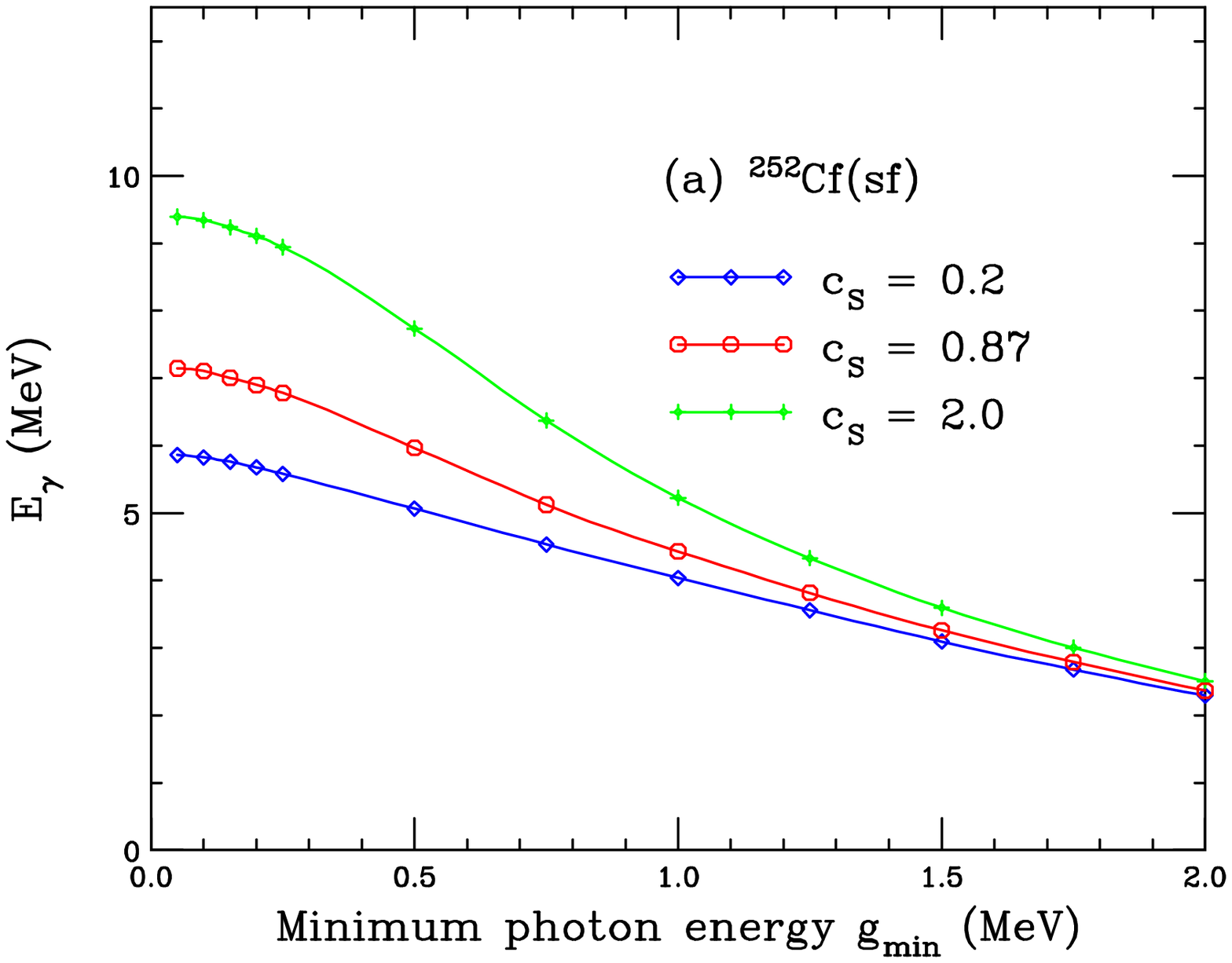}
\includegraphics[angle=0,width=0.45\textwidth]{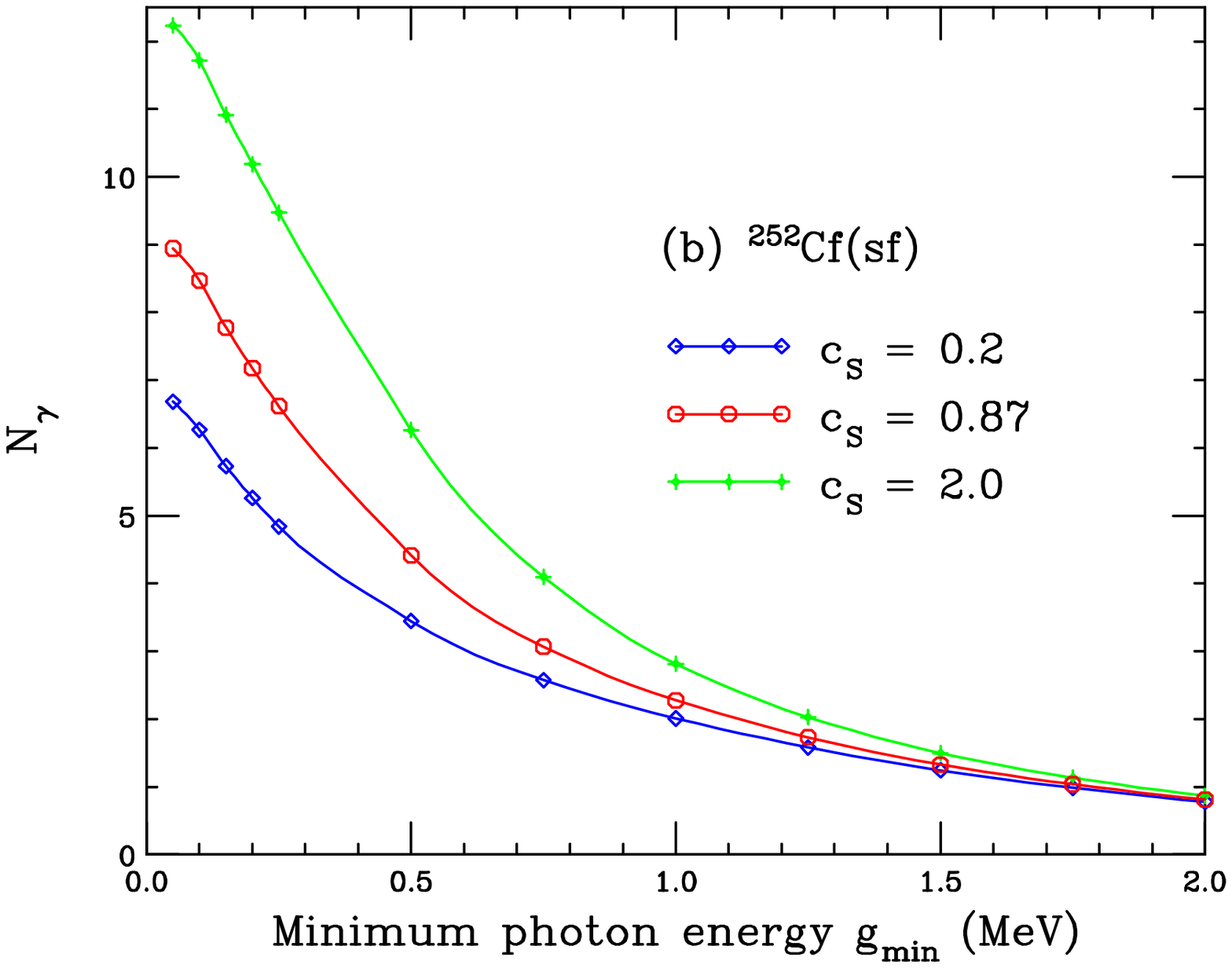}
\caption[]{(Color online) The total photon energy $E_\gamma$ (a)
  and the photon multiplicity $N_\gamma$ (b)
  calculated for $^{252}$Cf(sf) 
  as a function of photon energy cutoff $g_{\rm min}$
  for three values of $c_S$: $c_S=0.2$, 0.87, and 2.0.
  The value of $t_{\rm max}$ was 10~ns.
}\label{fig:gmin_cS_dep_Cf}
\end{figure}

We also note that the similarities between the three cases shown
may be due in part to the use of the same values of $c_S$,
namely the one determined by the preliminary fit to $^{252}$Cf(sf) data.
A fit of $c_S$ to the 
data available for additional cases may result in 
a greater range of $E_\gamma$ and $N_\gamma$ at $g_{\rm min}\approx0$.

The total detected fission photon energy $E_\gamma$ 
is shown in Fig.~\ref{fig:gmin_cS_dep_Cf}(a). For $c_S = 0.2$, 
$E_\gamma$ decreases approximately linearly with increasing $g_{\rm min}$.
As $c_S$ is increased, the fission fragments are created with ever larger
angular momenta and the associated rotational energy is eventually disposed of
by radiation of relatively soft photons.
Consequently, $E_\gamma$ goes up as well. The effect is about 50\% 
in the ideal case when the detection threshold $g_{\rm min}$ vanishes.
When $g_{\rm min}$ is increased, 
an ever larger proportion of these soft photons are not seen.  
Thus the sensitivity to $c_S$ diminishes.

\begin{figure}[tbh]
\includegraphics[angle=0,width=0.45\textwidth]{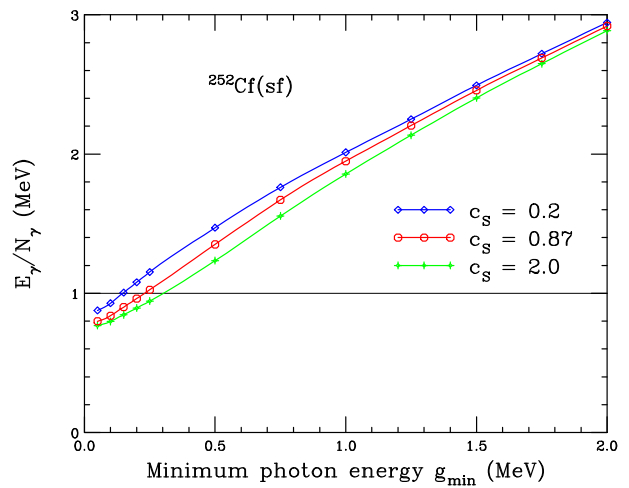}
\caption[]{(Color online) The energy per photon $E_\gamma/N_\gamma$
  calculated for $^{252}$Cf(sf) 
  as a function of the photon energy cutoff $g_{\rm min}$
  for three values of $c_S$: $c_S=0.2$, 0.87, and 2.0.
  The value of $t_{\rm max}$ was 10~ns taken as for all cases.
}\label{fig:gmin_cS_dep_Cf2}
\end{figure}

Figure \ref{fig:gmin_cS_dep_Cf}(b) shows the corresponding results 
for the detected total fission photon multiplicity,
which has a stronger dependence on $g_{\rm min}$,
as noted already in the discussion of Fig.~\ref{fig:gmin_cS_dep_Cf}(a). 
The dependence is particularly strong for the lowest values of $g_{\rm min}$ 
where, for $c_S = 2$, 
increasing $g_{\rm min}$ from 0.1 to 0.2~MeV reduces $N_\gamma$ by $\sim 20$\%,
while increasing $c_S$ from 0.2 to 2 decreases $N_\gamma$ by nearly a factor of two
in the same region of $g_{\rm min}$.  
Thus the dependence of $N_\gamma$ on $g_{\rm min}$ is more power-law like.  
As was the case for $E_\gamma$, when $g_{\rm min}$ is increased,
$N_\gamma$ becomes independent of $c_S$ because, as $g_{\rm min}$ approaches 2~MeV, 
effectively only a single (likely statistical) photon 
has sufficient energy to be detected.

Lastly, we show the dependence of the energy per photon, $E_\gamma/N_\gamma$,
on the detection threshold $g_{\rm min}$ for the same values of $c_S$.
The energy per photon is almost independent of $c_S$,
as shown in Fig.~\ref{fig:gmin_cS_dep_Cf2}.
It exhibits an almost linear increase with $g_{\rm min}$.
Starting out from slightly below 1 MeV for low $g_{\rm min}$,
$E_\gamma/N_\gamma$ becomes greater than 1 MeV at $g_{\rm min}\approx0.15$~MeV
for $c_S = 0.2$ and at $g_{\rm min}\approx0.25$~MeV for $c_S = 2$.  
Thus already for thresholds $g_{\rm min}$ far below 1 MeV, 
each detected fission photon carries over 1~MeV of energy on average.

\subsection{Dependence on $t_{\rm max}$}

With the inclusion of the discrete transitions from the RIPL-3 library 
in \code, it has become possible to study the effect of the time window 
in which the detector operates on the measured photon spectrum
which is expected to be particularly significant at low energies.  
For example, if the photon cascade from a fission product
arrives at a long-lived isomeric state, 
the decay chain may not proceed further during the measurement time 
and no more prompt photon emission can be detected from that nucleus.

Because the fission fragment distributions differ for $^{252}$Cf(sf),
$^{235}$U($n_{\rm th}$,f) and $^{239}$Pu($n_{\rm th}$,f), 
it is instructive to look
at how $N_\gamma$ and $E_\gamma$ change as the detection time window is varied.
The effect for $^{252}$Cf(sf) should be notably different from the effects
for $^{235}$U(n$_{\rm th}$,f) and $^{239}$Pu(n$_{\rm th}$,f)
because the light fragments are shifted upwards in mass
for the former case relative to the latter ones.
In addition, 
the wings of the fragment mass distribution are broader for $^{252}$Cf 
and the dip in the symmetric region is less pronounced.
All these differences could lead to a significantly different
population of the relevant isomeric states.  
Furthermore, $N_\gamma$ should be more affected than $E_\gamma$
because the isomeric states are encountered relatively far down the cascade 
so the photons affected are rather soft 
and will not strongly affect $E_\gamma$.

\begin{figure}[tbh]
\includegraphics[angle=0,width=0.45\textwidth]{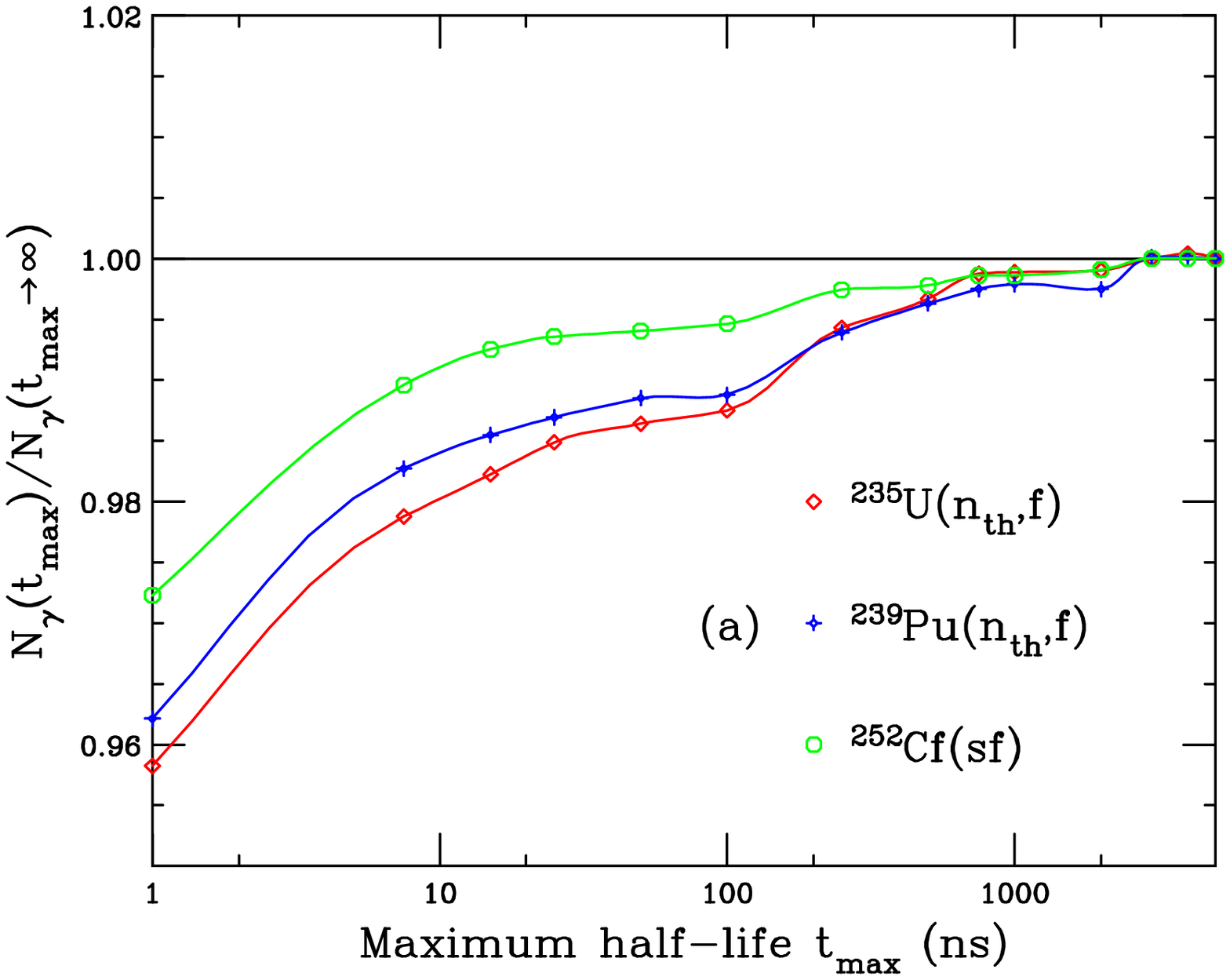}
\includegraphics[angle=0,width=0.45\textwidth]{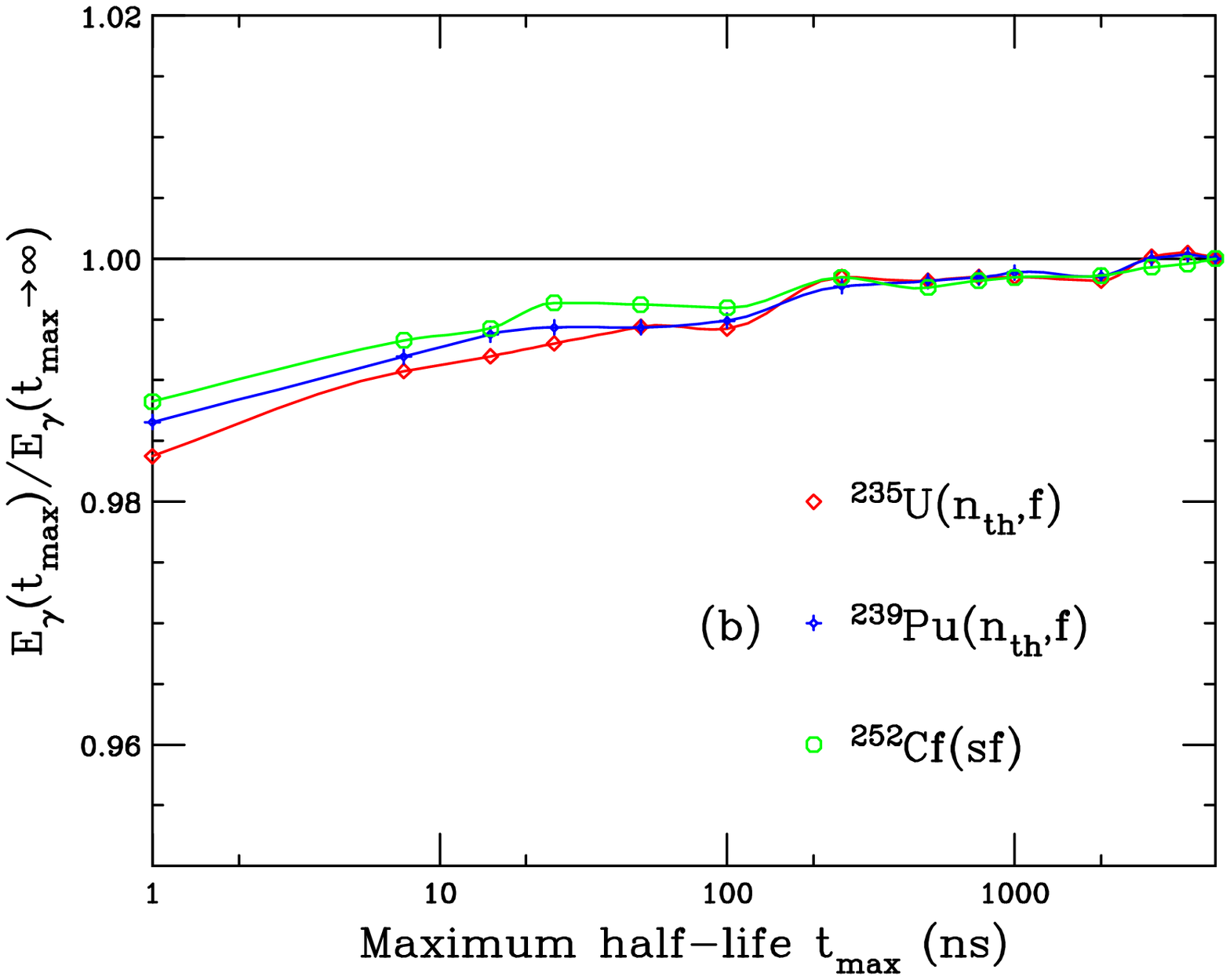}
\caption[]{(Color online) 
  The total photon energy (a) and the photon multiplicity (b)
  as functions of the maximum level half-life $t_{\rm max}$,
  relative to the results for $t_{\rm max} \rightarrow \infty$
  (effectively $t_{\rm max} = 5$ $\mu$s), for $^{235}$U(n$_{\rm th}$,f),
  $^{230}$Pu(n$_{\rm th}$,f), and $^{252}$Cf(sf).
}\label{fig:tmaxdep_CfUPu}
\end{figure}

Figure \ref{fig:tmaxdep_CfUPu} shows the dependence of $N_\gamma$ and $E_\gamma$
on the effective detector time window, $t_{\rm max}$, for all three cases,
employing a detection threshold of $g_{\rm min} = 0.1$~MeV.
The extracted value for a given $t_{\rm max}$ is shown relative to the corresponding 
value obtained with an effectively infinite time window, $t_{\rm max}=5$~$\mu$s.
Thus the ratio represents a cumulative value of the multiplicity or energy 
as a function of the duration of the detector time window.  
As expected, there is a noticeable difference between the three cases
for the multiplicity ratios, see Fig.~\ref{fig:tmaxdep_CfUPu}(a).
The differences are largest for the shortest time windows,
with the ratio being largest for $^{252}$Cf.
All are very close to unity for $t_{\rm max} > 500$~ns.  
On the other hand, as also expected, 
the $t_{\rm max}$ dependence of the ratios for the total photon energy
are very similar for the three cases, see Fig.~\ref{fig:tmaxdep_CfUPu}(b).

\begin{figure}[tbh]
\includegraphics[angle=0,width=0.45\textwidth]{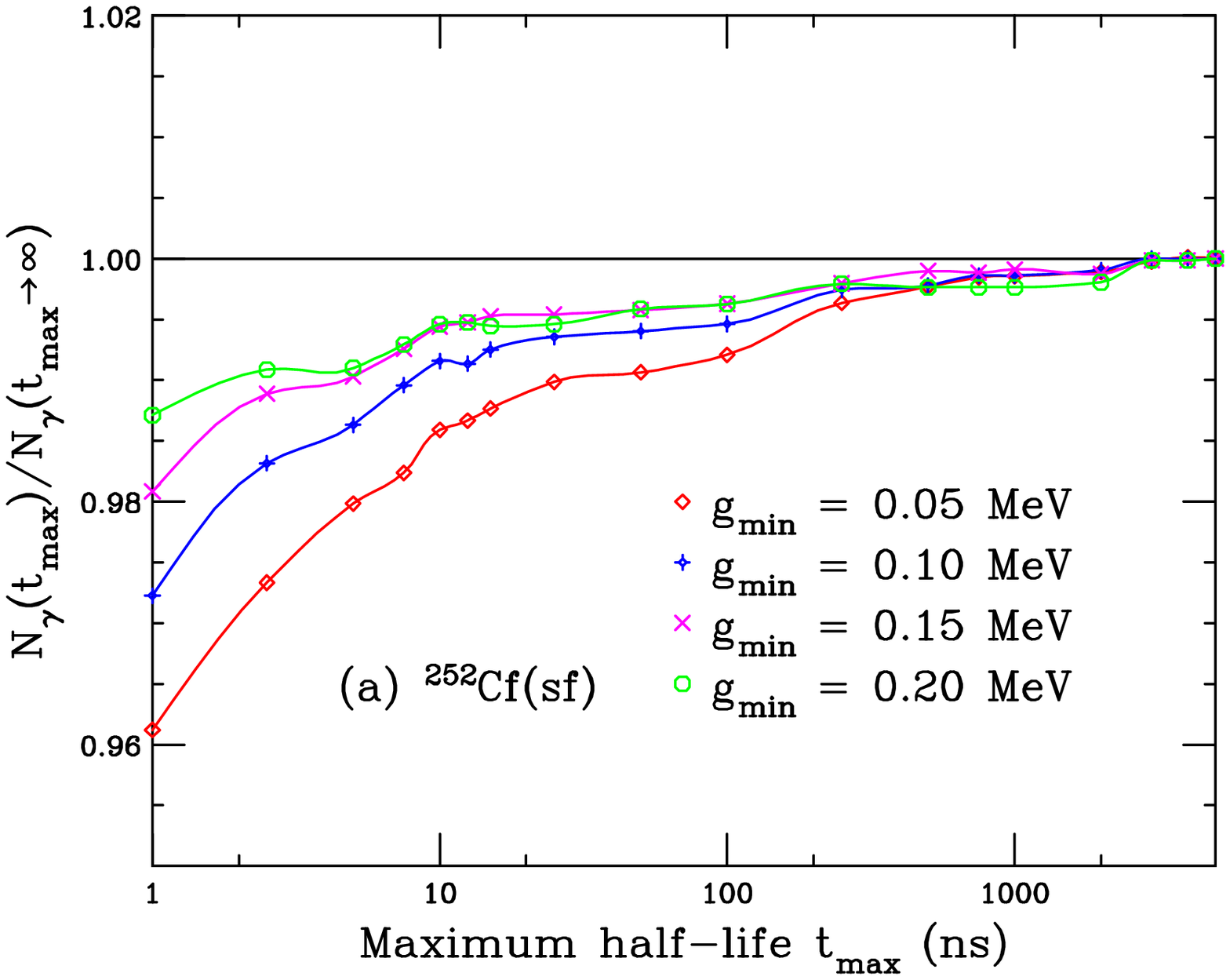}
\includegraphics[angle=0,width=0.45\textwidth]{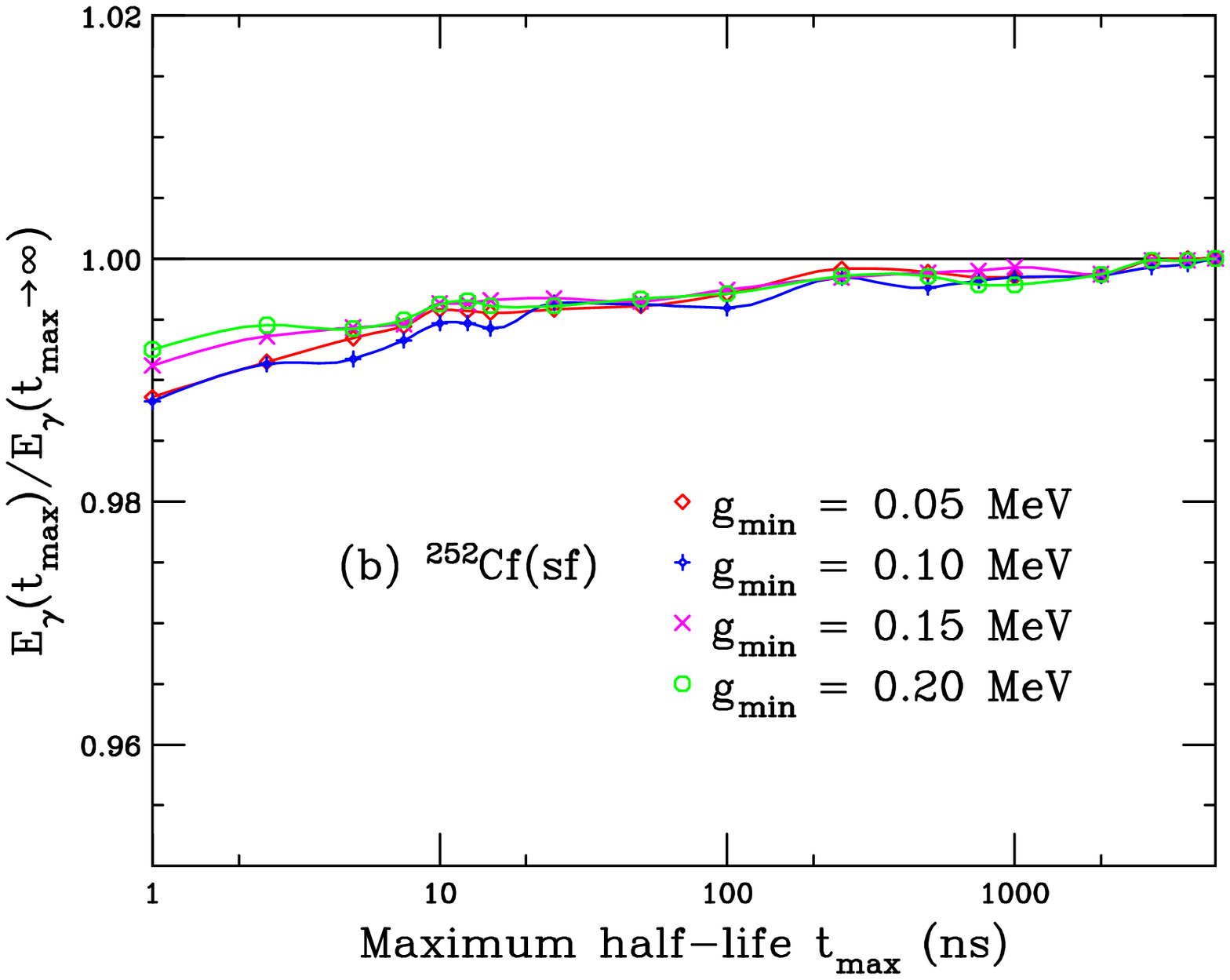}
\caption[]{(Color online)
  The total photon energy (a) and the photon multiplicity (b)
  as functions of the maximum level half-life $t_{\rm max}$,
  relative to the results for $t_{\rm max} \rightarrow \infty$,
  for $^{252}$Cf(sf) using $g_{\rm min} = 0.05$, 0.10, 0.15, and 0.20 MeV.
}\label{fig:tmax_gmin_dep_Cf}
\end{figure}

\begin{figure}[tbh]
\includegraphics[angle=0,width=0.45\textwidth]{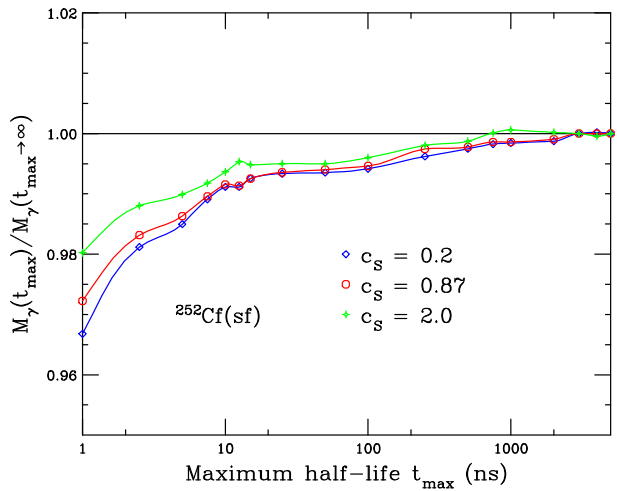}
\caption[]{(Color online)
  The total energy per photon, $E_\gamma/N_\gamma$, 
  as a function of the maximum level half-life $t_{\rm max}$,
  relative to the results for $t_{\rm max} \rightarrow \infty$,
  for $^{252}$Cf(sf) using $c_S = 0.2$, 0.87 (the default value), and 2.0.
}\label{fig:tmax_cS_dep_Cf}
\end{figure}

Our results differ somewhat from those obtained with the $\mathtt{CGMF}$ code
\cite{Talou_late}.
There are a number of ways these differences could arise.  
First, while both treatments start from the same fission fragment yields
 and the same average total kinetic energies \cite{Hambsch_Cf},
\code\ assumes that the charge distribution has a normal form
(with the experimentally measured charge variance),
whereas $\mathtt{CGMF}$ invokes Wahl systematics 
which takes account of odd-even effects.
Second, \code\ does not use the measured width of the TKE distribution
but generates it from the thermal fluctuations in the excitation energy
(controlled by the $c_T$ parameter).
Third, \code\ employs the single-valued parameter $x$ for the sharing of excitation energy
between the fragments, while $\mathtt{CGMF}$ adjusts the fragment temperature 
point-by-point to better reproduce $\nu(A)$ \cite{TalouPRC83}.

Fourth, 
$\mathtt{CGMF}$ uses the Hauser-Feshbach treatment for the fragment decays,
while \code\ uses a Weisskopf-Ewing spectrum for neutron emission
followed by the photon cascade as described in Sect.\ \ref{code}.
Finally, and perhaps most important,
the resulting ratio depends on how the RIPL-3 lines are implemented:
because the tables are rather incomplete 
a significant degree of modeling is required
to complement the measured information, especially on branching ratios,
and the two codes employ different methods for that 
(For the CGMF treatment, see Refs.\ \cite{BeckerPRC87,StetcuPRC90}.)
Of course, 
the results also depend on the specific value of $g_{\rm min}$ employed.

To show how these ratios could change with inputs, 
we look first at $N_\gamma(t_{\rm max})$ and $E_\gamma(t_{\rm max})$
for different values of $g_{\rm min}$,
as shown in Fig.~\ref{fig:tmax_gmin_dep_Cf}.  
We focus on $^{252}$Cf(sf) and
choose relatively low values of $g_{\rm min}$, from 0.05 to 0.20~MeV.
Again the largest effect is on the multiplicity, shown in
Fig.~\ref{fig:tmax_gmin_dep_Cf}(a), and for $t_{\rm max} \leq 100$~ns.
The greatest difference is between $g_{\rm min} = 0.05$~MeV and 0.10~MeV.
Higher values of $g_{\rm min}$ have a smaller effect because the discrete levels tend
to emit rather soft photons.  This is evident from the cumulative total
photon energy shown in Fig.~\ref{fig:tmax_gmin_dep_Cf}(b)
which is almost independent of $g_{\rm min}$.

Figure~\ref{fig:tmax_cS_dep_Cf} shows the cumulative multiplicity $N_\gamma(t_{\rm max})$ 
for several values of $c_S$.  
The value $c_S = 0.87$ is the best fit value from the $^{252}$Cf(sf) fit \cite{Andrew}, 
and is the same as the results shown in 
Figs.~\ref{fig:tmaxdep_CfUPu}(a) for $^{252}$Cf and in
Fig.~\ref{fig:tmax_gmin_dep_Cf}(a) for $g_{\rm min} = 0.10$~MeV.
The other two values, $c_S = 0.2$ and $c_S = 2$ are the upper and lower limits 
used in the calculations shown in the previous section.  
The dependence on $c_S$ is weaker than that on $g_{\rm min}$.  
Indeed, it is sufficiently weak to make it unnecessary to show the $c_S$ dependence 
of the cumulative total photon energy $E_\gamma(t_{\rm max})$.

Interestingly, the change in the cumulative multiplicity ratio is largest for
the lowest $c_S$ value which represents the lowest rotational energy,
while the effect is reduced for $c_S = 2$.  
Perhaps it is
less likely that the long-lived isomeric states are being populated
when the initial angular momentum is higher (see Fig.~\ref{fig:Adep_spins_cS}).

\section{Comparison to data}

We now turn to a comparison of the default \code\ results which use 
$c_S = 0.87$ as determined from the fit to $^{252}$Cf(sf) data \cite{Andrew}. 
We use $g_{\rm min}=0.1$ MeV and $t_{\rm max}=10$ ns 
unless otherwise specified.

\subsection{$^{252}$Cf(sf)}

We will compare the \code\ calculations to several previous experiments.  
Those by Nifenecker {\it et al.}\ \cite{Nifeneckergamma} and Nardi {\it et al.}\
\cite{Nardigamma} took data on photon energy as a function of fragment
mass and total kinetic energy.  Photon multiplicities as functions of fragment
mass were measured by Pleasonton {\it et al.}\ \cite{PleasontonCfgamma} and
Johansson {\it et al.}\ \cite{Johansson}.  
All these experiments were completed before the mid 1970s.

\begin{figure}[tbh]
\includegraphics[angle=0,width=0.45\textwidth]{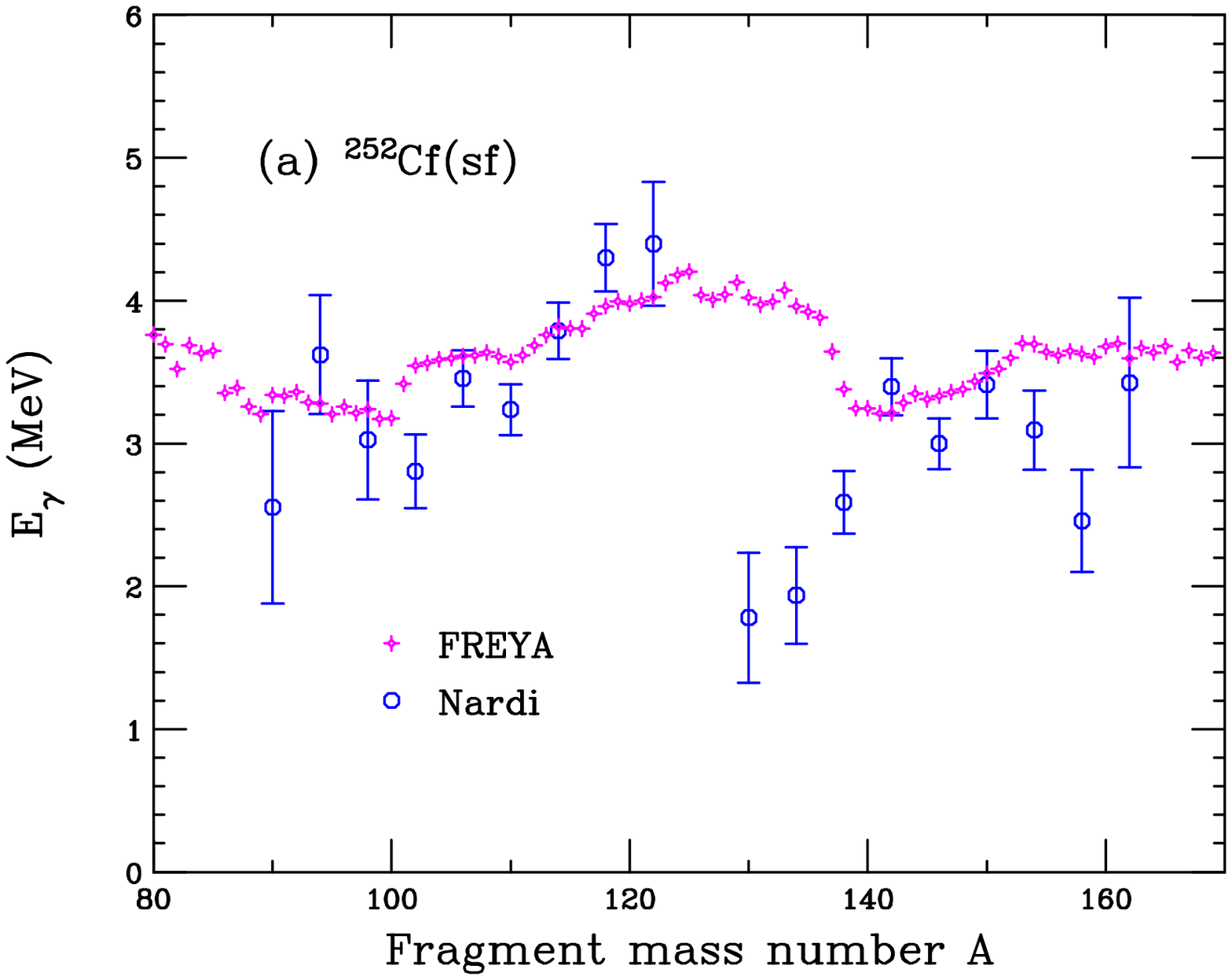}
\includegraphics[angle=0,width=0.45\textwidth]{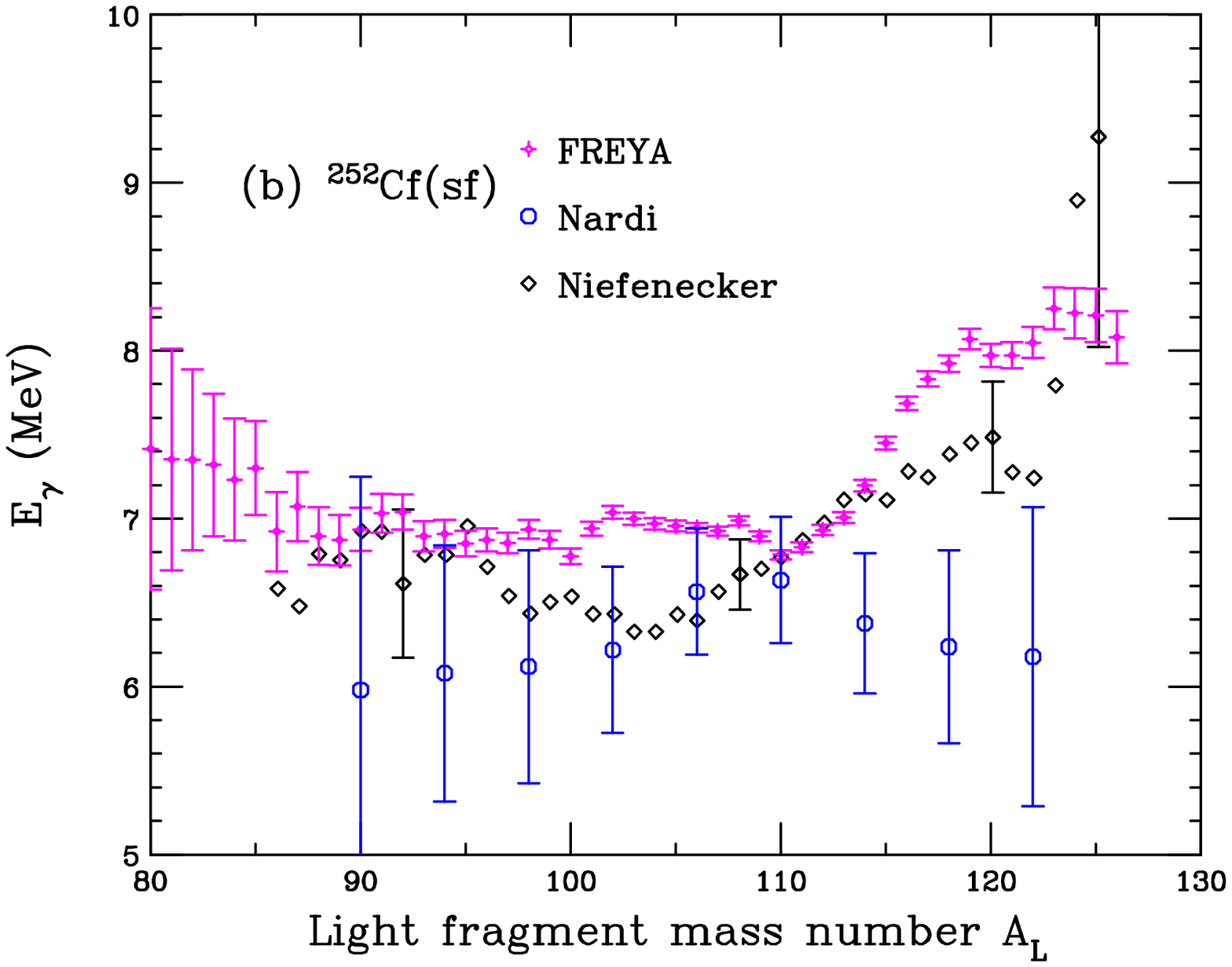}
\caption[]{(Color online) For $^{252}$Cf(sf),
  the calculated total photon energy $E_\gamma$
  as a function of the mass number $A$ of the emitting fragment (a)
  and the combined photon energy from both fragments 
  as a function of the mass number of the light fragment $A_L$ (b)
  compared to data from Nifenecker \protect\cite{Nifeneckergamma} (b)
  and from Nardi \protect\cite{Nardigamma} (a) and (b).
  The calculation used one million events;
  the associated sampling errors are shown in (b).
}
\label{fig:Adep_Cf_data}
\end{figure}

More recent experiments have not yet correlated photon production 
with fragment mass or kinetic energy.  
Billnert {\it et al.}\ \cite{BillnertCfgamma} measured the
prompt fission photon spectrum at IRMM in Belgium.
The DANCE experiment at the Los Alamos Neutron
Science Center reported the photon multiplicity
distribution \cite{Czyzhgamma} while the LiBerACE experiment at LBNL studied
neutron-photon correlations by measuring the photon multiplicity distribution
for two or four neutrons emitted \cite{LiBerACE}.

Nifenecker {\em et al.}\ \cite{Nifeneckergamma} placed the $^{252}$Cf source
and the fragment detectors in the center of a spherical gadolinium-loaded 
liquid scintillator tank one meter in diameter.   
The neutrons were distinguished from photons by timing: the
photon pulse came first, followed several microseconds later by neutrons.
The pre-evaporation mass and kinetic energy of each fragment was deduced
from the number of neutrons emitted.  
Since they could not determine which fragment emitted the photons,
they reported the average photon energy from both fragments 
as a function of the light fragment mass $A_{\rm L}$ and total fragment kinetic
energy.

\begin{figure}[tbh]
\includegraphics[angle=0,width=0.45\textwidth]{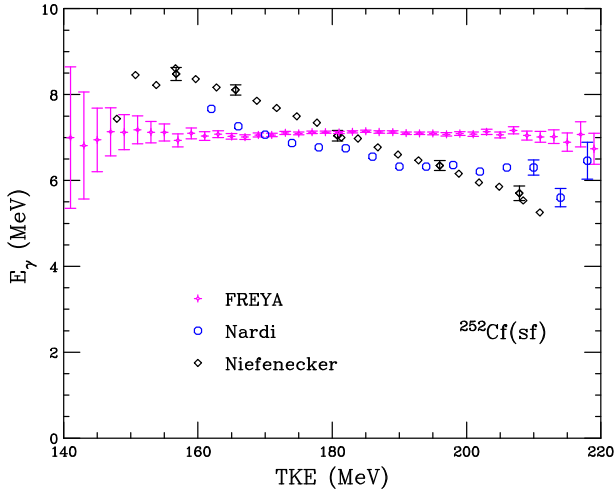}
\caption[]{(Color online) For $^{252}$Cf(sf),
  the calculated total photon energy 
  as a function of total fragment kinetic energy  
  is compared to data from Nifenecker \protect\cite{Nifeneckergamma}
  and Nardi \protect\cite{Nardigamma}.
}
\label{fig:TKE_Cf_data}
\end{figure}

Nardi {\it et al.}\ \cite{Nardigamma} used
a thin $^{252}$Cf source placed inside a vacuum chamber with fragment 
detectors on both sides, also in the chamber.  
The photons were detected with plastic scintillators that were placed 
60 cm from the source, behind the fragment detectors and outside the chamber.  
They separated photons from neutrons using time-of-flight techniques.  Because 
their geometric acceptance was small, they were able to 
measure the total energy release due to photon emission from individual
fragments, which was not possible in Nifenecker's $4\pi$ geometry.   
Thus Nardi {\it et al.}\ could report $E_\gamma$ for each individual fragment
mass, while Nifenecker {\it et al.}\ could report only the total $E_\gamma$
emitted from both fragments combined.

These results are shown in Fig.~\ref{fig:Adep_Cf_data}.  
The data as a function of $A$ from Nardi {\it et al.}\ \cite{Nardigamma}
are shown in Fig.~\ref{fig:Adep_Cf_data}(a).  The agreement of the
calculations with the data is generally very good, especially given the
uncertainties on the data.  The exceptions are the two heavy fragments 
closest to a symmetric mass split.  
In Fig.~\ref{fig:Adep_Cf_data}(b), we show the combined
total photon energy from the two fragments as a function of the light fragment
mass, $A_L$.  
The only uncertainties on the Nifenecker points 
are those representing a typical full-width half-maximum of each $A_L$, 
shown at $A_L = 92$, 108, 120, and 125.  
If these representative uncertainties are considered at all $A_L$, 
the Nifenecker data is in relatively good agreement with both the
\code\ calculations and the Nardi data.  The large uncertainties on the Nardi
data come from summing the uncertainties on $A_L$ and $A_H$
when folding the data from panel (a) over to obtain the total mean $E_\gamma$
per fission event rather than $E_\gamma$ per fragment.

As is also apparent from Fig.~\ref{fig:Adep_Cf_data}(b),
the uncertainties in the \code\ calculation are the largest
where the yields are the lowest, namely near symmetry ($120 < A_L < 126$) and, 
particularly, in the tails of the distributions ($A_L < 100$).
Increasing the number of events above one million would of course 
reduce the uncertainties correspondingly, but the trend will remain unchanged.
We note that the rise in $E_\gamma$ in the calculation for $A_L > 115$
corresponds to the rather abrupt decrease in $E_\gamma$ between $A \sim 132$
and $A \sim 140$ shown in Fig.~\ref{fig:Adep_Cf_data}(a).  
Given the relatively large uncertainties on the Nardi data 
and the implied uncertainties on the Nifenecker data (where shown), 
it would be useful to repeat these measurements with more modern detectors.

Figure~\ref{fig:TKE_Cf_data} compares the TKE dependencies obtained by
Nifenecker \cite{Nifeneckergamma} and Nardi \cite{Nardigamma} to the \code\
results.
Again, representative uncertainties are shown for the Nifenecker data 
at several values of TKE (157, 166, 181, 196 and 208 MeV).  
Aside from the smallest values of TKE,
${\rm TKE}<157$~MeV, these data decrease linearly with TKE.  This behavior is
similar to the decrease seen for neutrons, $\nu$(TKE), in other experiments.
The Nardi data, on the other hand, exhibit a slower decrease that plateaus
for ${\rm TKE}>160$~MeV.  The overall average photon energy seems to be smaller
for the Nardi measurement, as can also be observed through the comparison 
as a function of light fragment mass in Fig.~\ref{fig:Adep_Cf_data}(b).
Note that the width of the TKE distribution is rather broad, allowing for
significant photon emission up to ${\rm TKE} = 220$~MeV.

\begin{figure}[tbh]
\includegraphics[angle=0,width=0.45\textwidth]{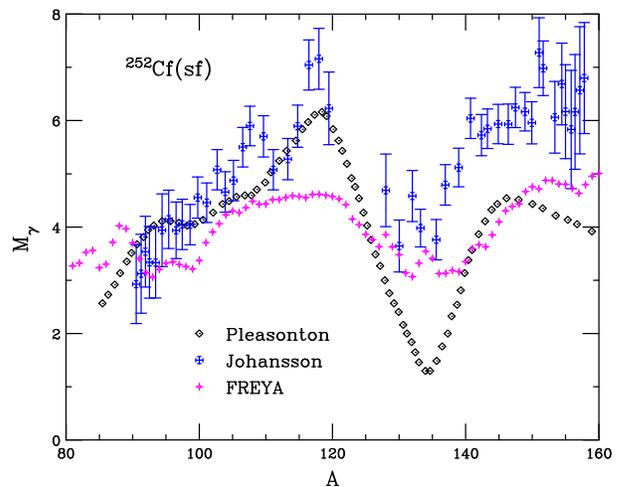}
\caption[]{(Color online) The photon multiplicity as a function of fragment mass
  calculated for $^{252}$Cf(sf) are compared to data from
  Pleasonton \protect\cite{PleasontonCfgamma} and
  Johansson \protect\cite{Johansson}.
}\label{fig:A_Cf_data}
\end{figure}

The calculations suggest that, at least for $^{252}$Cf(sf), $E_\gamma$ is
effectively independent of TKE.  We have used $g_{\rm min} = 0.1$~MeV and
$t_{\rm max} = 10$~ns in the \code\ calculations.  As shown in the previous
section, $E_\gamma$ depends only weakly on $g_{\rm min}$, 
especially relative to the total multiplicity, $N_\gamma$.
It also shows a weaker dependence on $t_{\rm max}$ than $N_\gamma$, 
see Fig.~\ref{fig:tmax_gmin_dep_Cf}.

Pleasonton {\it et al.}\ performed several experiments studying photon emission
in thermal neutron-induced fission on $^{233,235}$U and $^{239}$Pu
\cite{PleasontonUgamma,Pleasonton3UPugamma} 
as well as $^{252}$Cf(sf) \cite{PleasontonCfgamma}.  
All four experiments were performed
at Oak Ridge in the early 1970s.  The setup included two surface barrier
detectors to measure fragments and a sodium iodide detector to measure prompt
fission photons ($t_{\rm max} = 5$~ns) with energies greater than 0.122 MeV.
Data were taken in two different modes, a two-parameter mode to record only
fragment masses and energies and a four-parameter mode in which time of flight
was used to record the difference in arrival times between photons from the
two fragments.  A combined analysis of the data from the two- and four-parameter
mode runs allowed separation of the photon yields into those from light and
heavy fragments, yielding the photon energy and multiplicity as a function
of fragment mass. The neutron-induced fission data used neutrons from the
ORNL reactor while, for the Cf measurements, the apparatus remained in position
but the neutron beam was turned off.  The Pleasonton Cf data, shown in
Fig.~\ref{fig:A_Cf_data}, is digitized from Ref.~\cite{PleasontonCfgamma} 
where it was presented as a curve without uncertainties.

\begin{figure}[tbh]
\includegraphics[angle=0,width=0.45\textwidth]{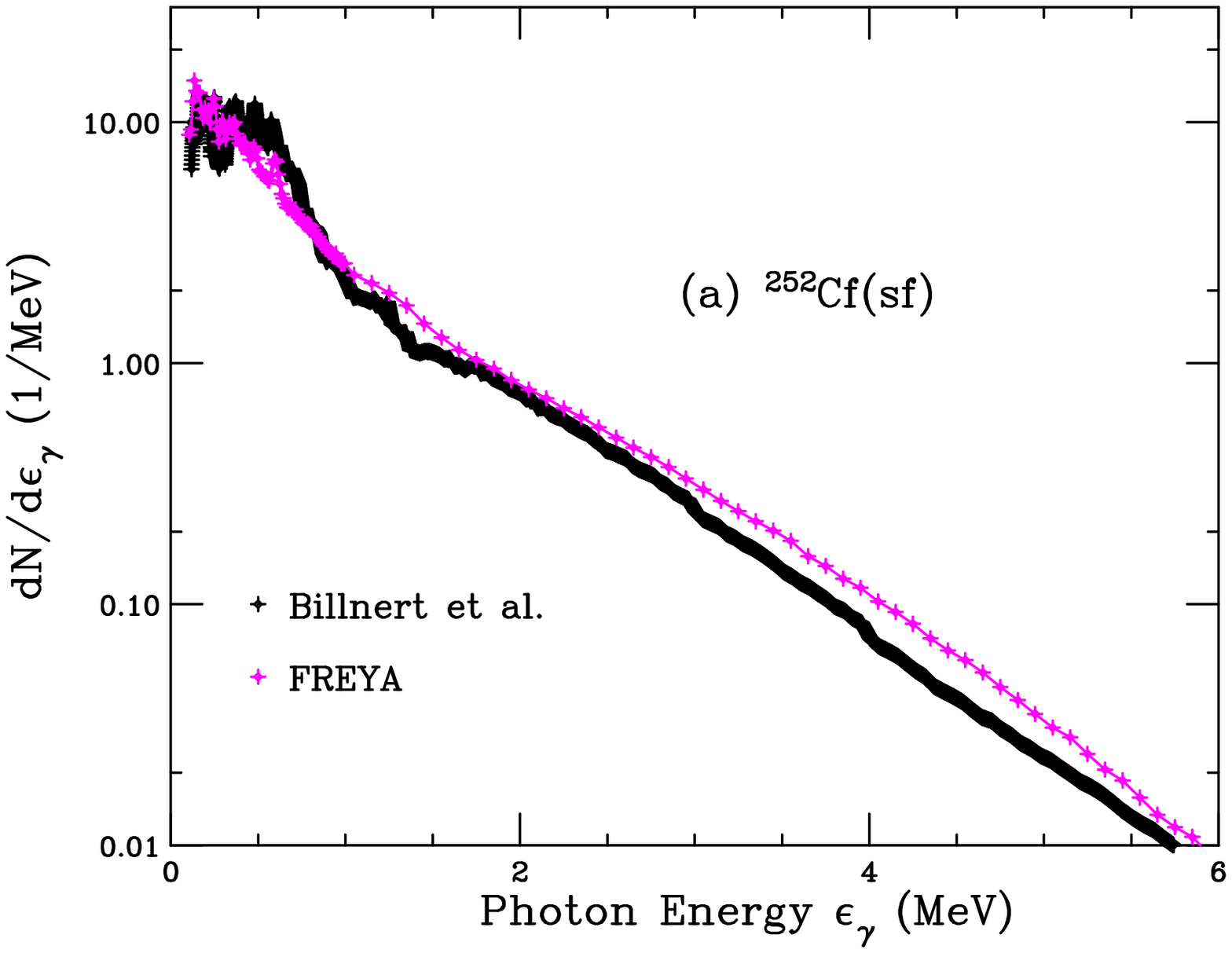}
\includegraphics[angle=0,width=0.45\textwidth]{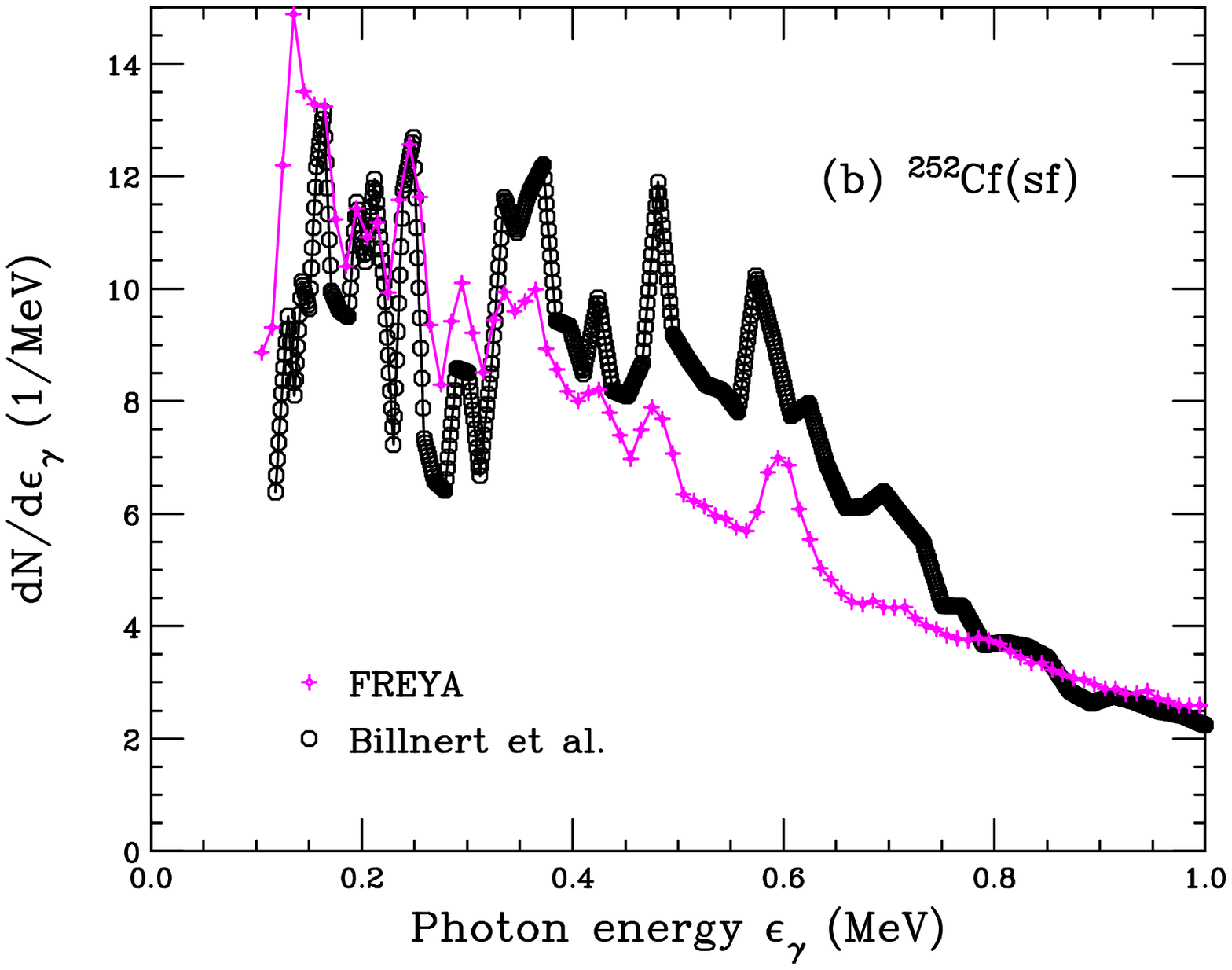}
\caption[]{(Color online) The photon energy spectrum calculated for
  $^{252}$Cf(sf) compared to data from Billnert {\it et al.}\
  \protect\cite{BillnertCfgamma}.  The total photon spectrum is shown in (a)
  while the spectrum for energies less than 1 MeV are shown in (b).
  The spectra are normalized to the fission photon multiplicity.
}\label{fig:spectra_Cf_data}
\end{figure}

The data taken by Johansson in 1964 \cite{Johansson} used a $^{252}$Cf source
with two fragment detectors placed inside a thin walled, evacuated aluminum
chamber.
Photons were detected within 1~ns of emission using a sodium-iodide crystal.
Photons from individual fragments were separated using a lead collimator: by
alternating the position of the collimator results were taken for each fragment
and the results from the two collimator positions summed.  From spectral data
for different fragment masses, Fig.~8 of Ref.~\cite{Johansson}, 
it would appear that $g_{\rm min} \sim 0.2$~MeV for the measurement 
but no explicit photon energy cutoff is given.  
While the photon multiplicity from the light fragment
reported in this experiment is compatible with that of Pleasonton, 
the Pleasonton multiplicity is considerably lower for the heavy fragment.
However, it is difficult to say how much the results differ 
without knowing the uncertainties on the Pleasonton data.  
If the energy cutoff is higher for Johansson than for Pleasonton, 
one might expect that the overall photon multiplicity would be lowered, 
similar to the \code\ results shown in Fig.~\ref{fig:Adep_Mg_gmin}.

We note that Johansson published a similar result a year later that was
focused on delayed photons with emission times of $10-100$ ns 
\cite{JohanssonII}.
The delayed emission is more sensitive to long-lived isomers 
and thus to the transition lines in the RIPL database.  There are prominent
contributions to the delayed photon multiplicity at $A = 92$, 95, 110, 130,
134, and 148 which would fill in some of the dips in the prompt photon
multiplicity, making the total multiplicity from the combined Johansson data
sets closer to a sawtooth, as observed by John {\it et al.}\ \cite{John}.

The \code\ results are similar to but somewhat below the multiplicity of
the light fragment as given by the two data sets.  Also, while the
calculations underestimate the Johansson data, they are in agreement
with the Pleasonton results for $A > 140$.  The \code\ results are furthest
off from the Pleasonton data close to symmetry, 
$115 < A < 140$, where the uncertainties are likely large.  
While our results are not a fit to the average photon energy and
multiplicity, our averages, $N_\gamma = 8.18$,
\SKIP{
multiplicity, our averages, $\langle N_\gamma \rangle = 8.47$,
$\langle E_\gamma \rangle = 7.07$~MeV, and
$\langle E_\gamma/N_\gamma \rangle = 0.864$~MeV, are compatible with those of
Pleasonton \cite{PleasontonCfgamma}:
$\langle N_\gamma \rangle = 8.32\pm 0.40$,
$\langle E_\gamma \rangle = 7.06\pm 0.35$~MeV, and
$\langle E_\gamma/N_\gamma \rangle = 0.84 \pm 0.06$~MeV.  
}
$E_\gamma = 7.11$~MeV, and
$E_\gamma/N_\gamma = 0.84$~MeV, are compatible with those of
Pleasonton \cite{PleasontonCfgamma}:
$N_\gamma = 8.32\pm 0.40$,
$E_\gamma = 7.06\pm 0.35$~MeV, and
$E_\gamma/N_\gamma = 0.84 \pm 0.06$~MeV.  
We use the same values of $g_{\rm min}$ and $t_{\rm max}$ as Pleasonton,
$g_{\rm min} = 0.12$~MeV and $t_{\rm max} = 5$~ns.
Since we do not have the exact $g_{\rm min}$ for Johansson, we use that of
Pleasonton for the comparison.

Figure~\ref{fig:spectra_Cf_data} shows the prompt fission photon spectrum for 
$^{252}$Cf(sf) measured by Billnert {\it et al.}\ \cite{BillnertCfgamma}.  They
have embarked on a campaign to make modern measurements of photon decay heat
generated during fission, in particular for isotopes relevant for reactors.  To
do this, they first made measurements of the $^{252}$Cf(sf) photon spectrum.
They used two different detectors: lanthanum bromide, for timing and energy
resolution, and cesium bromide, because of the absence of intrinsic photon
activity in this material.  The energy and timing resolution for these
detector materials is better than that of the sodium iodide-based detectors
used in previous measurements.  They were able to reduce the uncertainties
of their measurement considerably relative to previous results
\cite{BillnertCfgamma}:
\SKIP{
$\langle N_\gamma \rangle = 8.30 \pm 0.08$;
$\langle E_\gamma \rangle = 6.64 \pm 0.08$~MeV; and
$\langle E_\gamma/N_\gamma \rangle = 0.80 \pm 0.01$~MeV,} 
$N_\gamma = 8.30 \pm 0.08$;
$E_\gamma = 6.64 \pm 0.08$~MeV; and
$E_\gamma/N_\gamma = 0.80 \pm 0.01$~MeV,  
including relative to that
of Pleasonton, mentioned above.  The results for the two different detector
materials agree.  In this case, we use the same energy cutoff,
$g_{\rm min} = 0.1$~MeV, and time window, $t_{\rm max} = 1.5$~ns, in the \code\
calculations as in the measurement. 

The calculated photon spectrum is normalized to the total calculated
multiplicity, $N_\gamma = 8.37$, obtained for the $g_{\rm min}$ and $t_{\rm max}$
used by Billnert {\it et al.}.  The overall agreement, shown in
Fig.~\ref{fig:spectra_Cf_data}(a) is very good.
(Note that the uncertainties on the measurement, shown for the lanthanum bromide
detectors, are not included in the plot.  With these included, the apparent
agreement
would improve.  Incorporating the GDR into \code\ provides the harder spectrum
for high energy continuum emission, as exhibited in the data.

Note that our average calculated photon energy,
$\langle E_\gamma \rangle = 7.09$~MeV, is higher that that measured by
Billnert {\it et al.}.
Thus even though our photon
multiplicity is within the uncertainties of the data, the higher photon energy
from \code\  results in a higher average energy per photon
$E_\gamma/N_\gamma = 0.85$~MeV.

\begin{figure}[t]
\includegraphics[angle=0,width=0.45\textwidth]{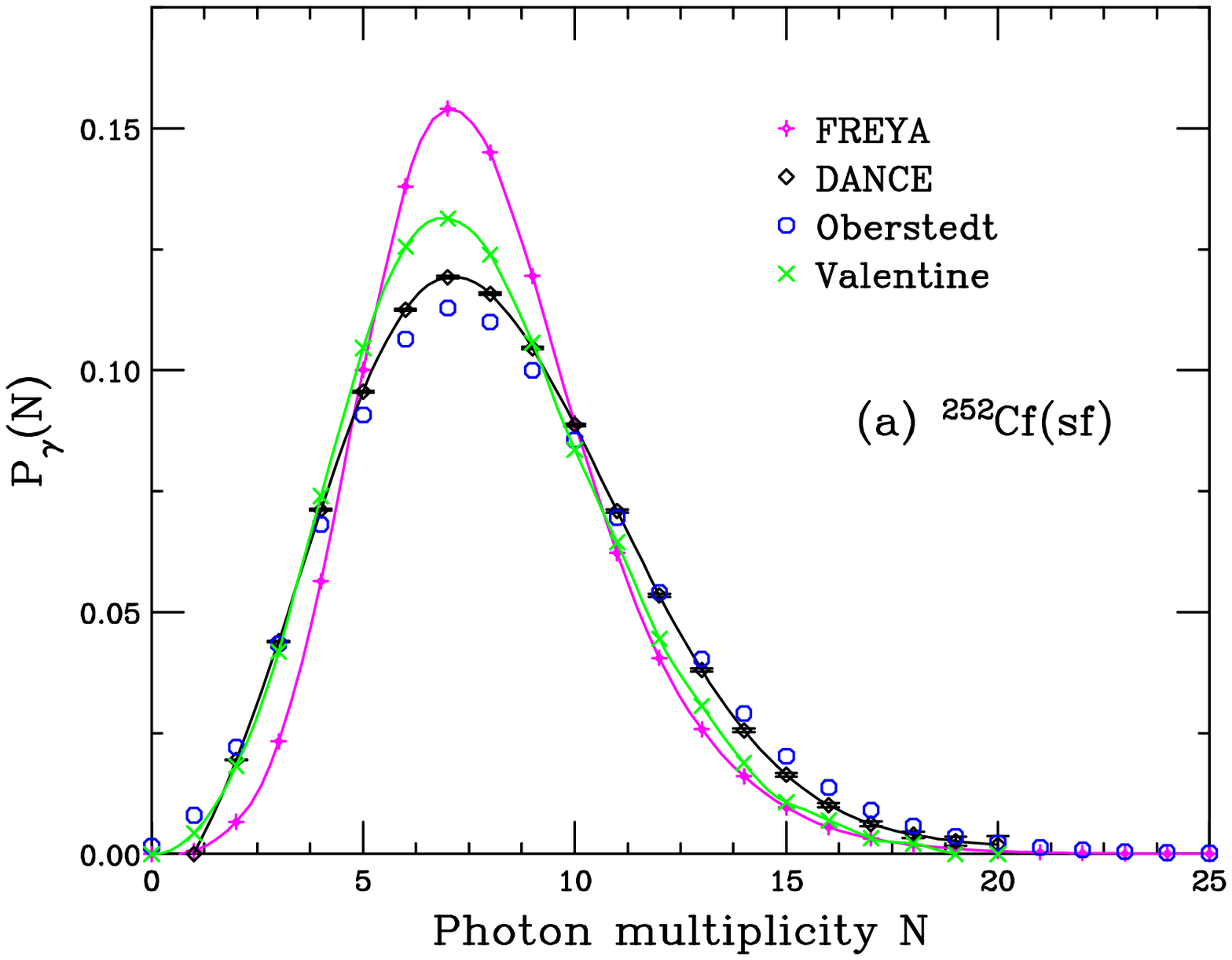}
\includegraphics[angle=0,width=0.45\textwidth]{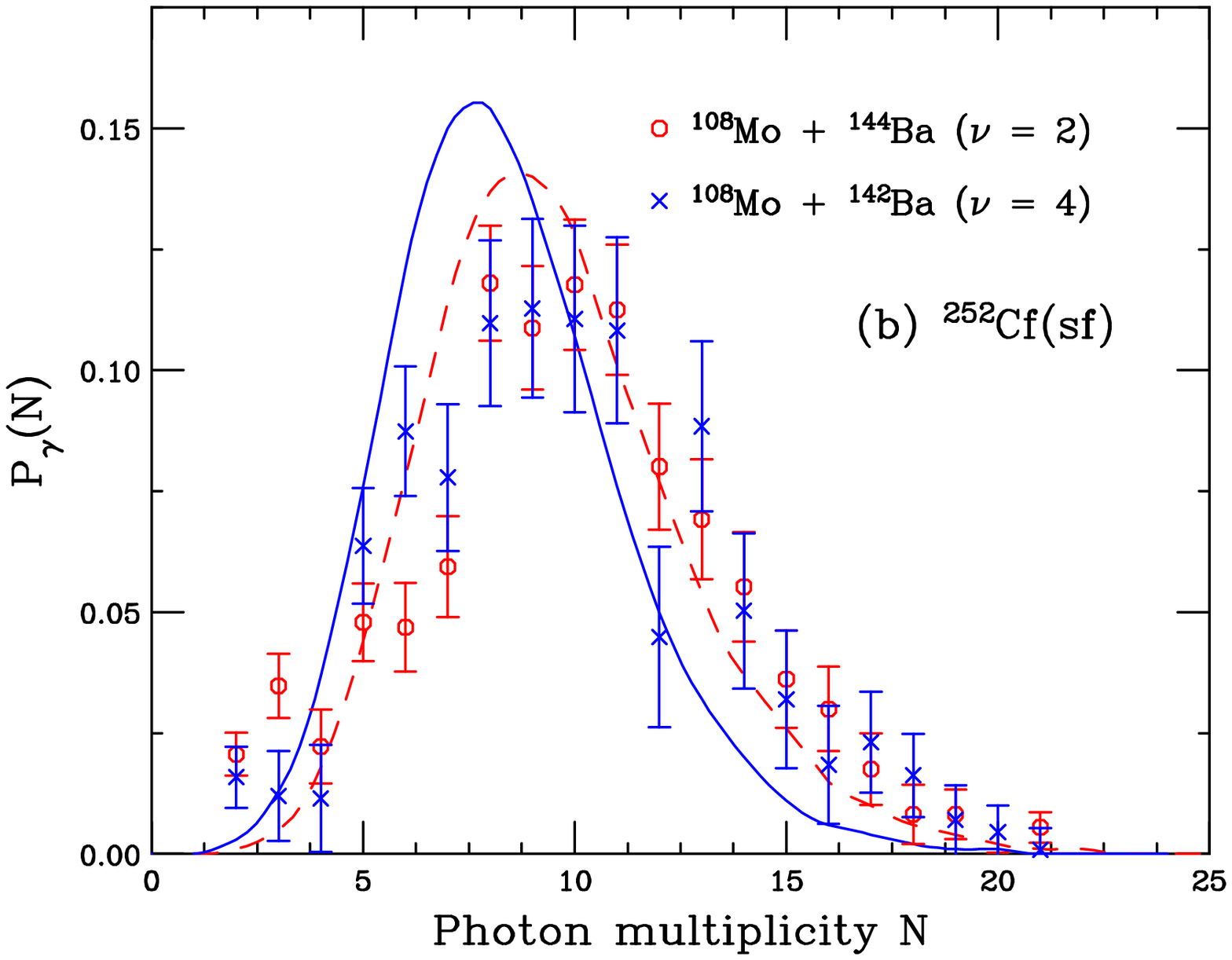}
\caption[]{(Color online) (a) The photon multiplicity distribution
  $P(N_\gamma)$ calculated for $^{252}$Cf(sf) 
  is compared to data from DANCE \protect\cite{Czyzhgamma}
  as well as data from Oberstedt {\em et al.}\ \cite{OberstedtPRC92}
  and the Valentine evaluation \cite{Valentine}.
  (b) The LiBerACE \protect\cite{LiBerACE} photon multiplicity distributions
  resulting from two and four neutrons emitted compared to the \code\ results
  for the same quantities.
}\label{fig:mult_Cf_data}
\end{figure}

Figure~\ref{fig:spectra_Cf_data}(b) focuses on the low-energy part 
of the photon spectrum, $E_\gamma < 1$~MeV, where the RIPL-3
transitions play the most important role.  
While the magnitudes of the peaks from the data and
from \code\ do not precisely match, the locations match quite well.
The differences in the strength of the peaks are likely due 
to the rather rough manner of substituting branching ratios
that are not included in the RIPL-3 tabulation.
We note that without our model
refinements, the calculations would underestimate the photon spectrum at high
energy and would not exhibit any structure at low photon energies.

Finally, in Fig.~\ref{fig:mult_Cf_data},
we compare \code\ to measured photon multiplicity distributions.

The prompt photon energy and the prompt photon multiplicity distribution 
were measured \cite{Czyzhgamma}
with the highly-segmented $4\pi$ photon calorimeter of the
Detector for Advanced Neutron Capture Experiments (DANCE) \cite{DANCE}
combined with a compact gas-filled parallel-plate avalanche counter
\cite{Wuinpress}.  The energy and multiplicity distributions were 
unfolded to produce the first experimental measurement of $P_\gamma(N)$
in spontaneous fission \cite{Czyzhgamma}.  The unfolded multiplicity
distribution
makes it possible to study the moments of the photon multiplicity distribution,
similar to studies making use of 
factorial moments of the neutron multiplicity distribution.

The distribution from DANCE, shown in Fig.~\ref{fig:mult_Cf_data}(a),
has $\langle N_\gamma \rangle = 8.14 \pm 0.40$ for
$g_{\rm min} = 0.14$~MeV.  While the average multiplicity calculated with \code\
for this same $g_{\rm min}$ is 7.94, within the uncertainty of the data, 
the \code\ distribution is significantly narrower than the data 
(the dispersion of the calculated distribution is 2.81, 
while that of the data is 3.35).
The more recent results of Oberstedt {\em et al.}\ \cite{OberstedtPRC92}
is in rather good agreement with the DANCE data.
The earlier evaluation by Valentine \cite{Valentine} is somewhat narrower
than the more recent measurement, in better agreement with \code.

We note that although the introduction of thermal fluctuations 
in the excitation energy (through the parameter $c_T$)
may affect the moments of the neutron multiplicity distribution,
these fluctuations do not affect the width of 
the photon multiplicity distribution because a significant fraction 
of the excitation energy has already been carried away through 
neutron evaporation before photon emission begins.
However, the photon multiplicity distribution is affected somewhat
by the degree of angular momentum carried away 
during the statistical part of the photon emission cascade.

The Livermore-Berkeley Array for Collaborative Experiments (LiBerACE) used
$^{252}$Cf(sf) to study photon multiplicity relative to neutron emission
\cite{LiBerACE}.  
They surrounded the $^{252}$Cf source with high-purity germanium detectors 
enclosed in bismuth-germanate detectors.
The geometry of the detector array provided good solid angle coverage.
Room background, as well as photons from cosmic rays, 
were subtracted by counting photons with no source present.

The LiBerACE Collaboration exploited the observation of discrete energy 
photons coming from known transitions in identified fission products, after 
neutron emission, to study neutron-photon correlations.  
They hoped to determine whether neutron and photon emission was 
positively or negative correlated.
Energy and momentum-conserving calculations of neutron and photon emission
in fission \cite{RVJR_spont,LemaireGamma}, such as \code,
predicts an anti-correlation between photons and neutrons, 
{\it i.\,e.}\ the average photon multiplicity would decrease 
with increasing neutron multiplicity.
On the other hand, Nifenecker {\it et al.}\ \cite{Nifeneckergamma} suggested 
that there was a positive correlation between neutron multiplicity and
photon energy.

They focused on two pairs of deformed even-even product nuclei, 
$^{106}$Mo/$^{144}$Ba and $^{106}$Mo/$^{142}$Ba,
which are associated with the emission of two or four neutrons, 
respectively.
They compared the photon multiplicity distributions from these product pairs.
If the emission is anti-correlated, a backward shift in the centroid 
of the photon multiplicity distribution should be observed
when comparing the first pair with the second pair,
{\em i.\,e.}\ comparing two-neutron emission with four-neutron emission,
and vice versa if there is a positive correlation.
As seen in Fig.~\ref{fig:mult_Cf_data}(b), there is no observable difference in 
the location of the centroid for the selected Mo/Ba ratios within
their significant statistical uncertainties, 
suggesting the absence of a correlation between neutron and photon emission.

\code\ results for the photon multiplicity distribution with two and four
neutrons emitted from all fragment pairs
  are also shown in Fig.~\ref{fig:mult_Cf_data}(b).  The
calculations use the experimental $g_{\rm min}$ value of 0.1~MeV and the time
window with $t_{\rm max} = 2$~$\mu$s.  There is a clear shift in the
calculations to lower photon multiplicity for the emission of four neutrons.
We note, however, that the \code\ results shown here are based on 
all fragment pairs, not just the two Mo/Ba pairs employed in the measurement.

\subsection{$^{235}$U($n_{\rm th}$,f)}
\label{sec:235U}

Here we compare the \code\ results to the Pleasonton measurements 
of photon energy and multiplicity as a function of fragment mass 
and total kinetic energy \cite{PleasontonUgamma} 
as well as the multiplicity measurement as a
function of fragment mass by Albinsson \cite{Albinsson}.
The recent spectral measurement by Oberstedt {\em et al.}\
\cite{BillnertUgamma} is also included.
We have not modified $c_S$ for neutron-induced fission but use the value 
determined from the preliminary fit to the $^{252}$Cf(sf) data, $c_S=0.87$.
The parameters $x$, $c_T$ and $d$TKE are tuned to $^{235}$U($n_{\rm th}$,f) neutron
data employing $c_S = 0.87$ while the value of $e_0$ is assumed to be the
same for all fissioning nuclei, both spontaneous and neutron-indiced.

The detector setup for the Pleasonton experiment was the same as for the
$^{252}$Cf(sf) measurement described in the previous section except that,
for this measurement, as well as for other measurements, shown in the
appendix for $^{233}$U($n_{\rm th}$,f) and $^{239}$Pu($n_{\rm th}$,f)
\cite{Pleasonton3UPugamma}, the data were taken with thermal neutrons from
the Oak Ridge National Laboratory research reactor.  The target, in this case
a 99.44\% pure thin deposit of $^{235}$U$_3$O$_8$ on a backing, was placed at
a 45$^\circ$ angle to the neutrons from the reactor.  The fragment detectors
were placed at 45$^\circ$ angles on either side of the target, 90$^\circ$ away
from the neutron direction and in line with the sodium iodide crystal to
measure photons from fission events.  As described previously, the experiment
was run in two different modes, the two parameter mode to study fragment masses
and kinetic energies and the four parameter mode for timing to separate the
photons from the light and heavy fragments.  They collected 306K events in the
four-parameter mode and 852K events in the two-parameter mode.
In such cases, the statistics from the four-parameter mode
sets the level of statistics for the data.

\begin{figure}[tbh]
\includegraphics[angle=0,width=0.45\textwidth]{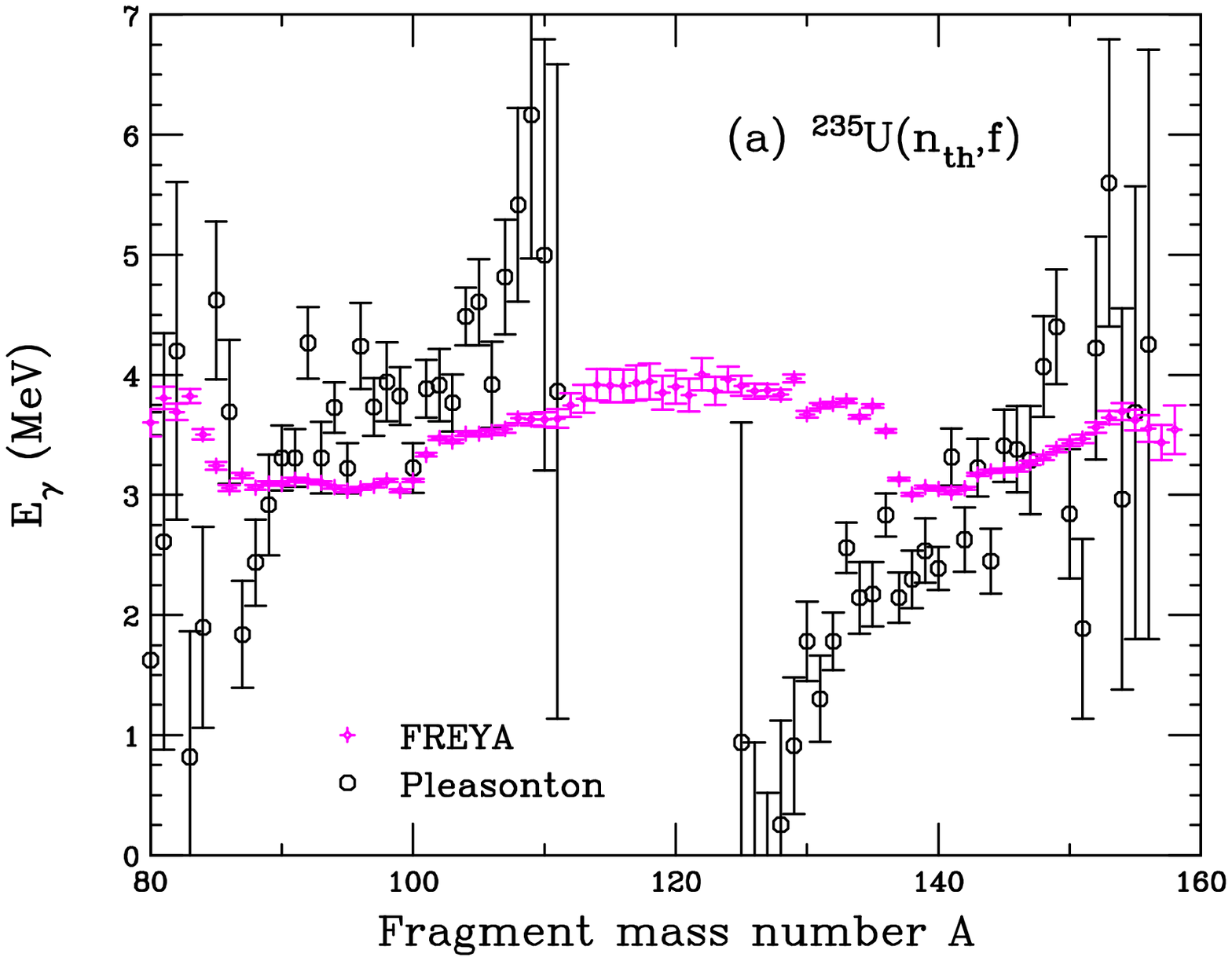}
\includegraphics[angle=0,width=0.45\textwidth]{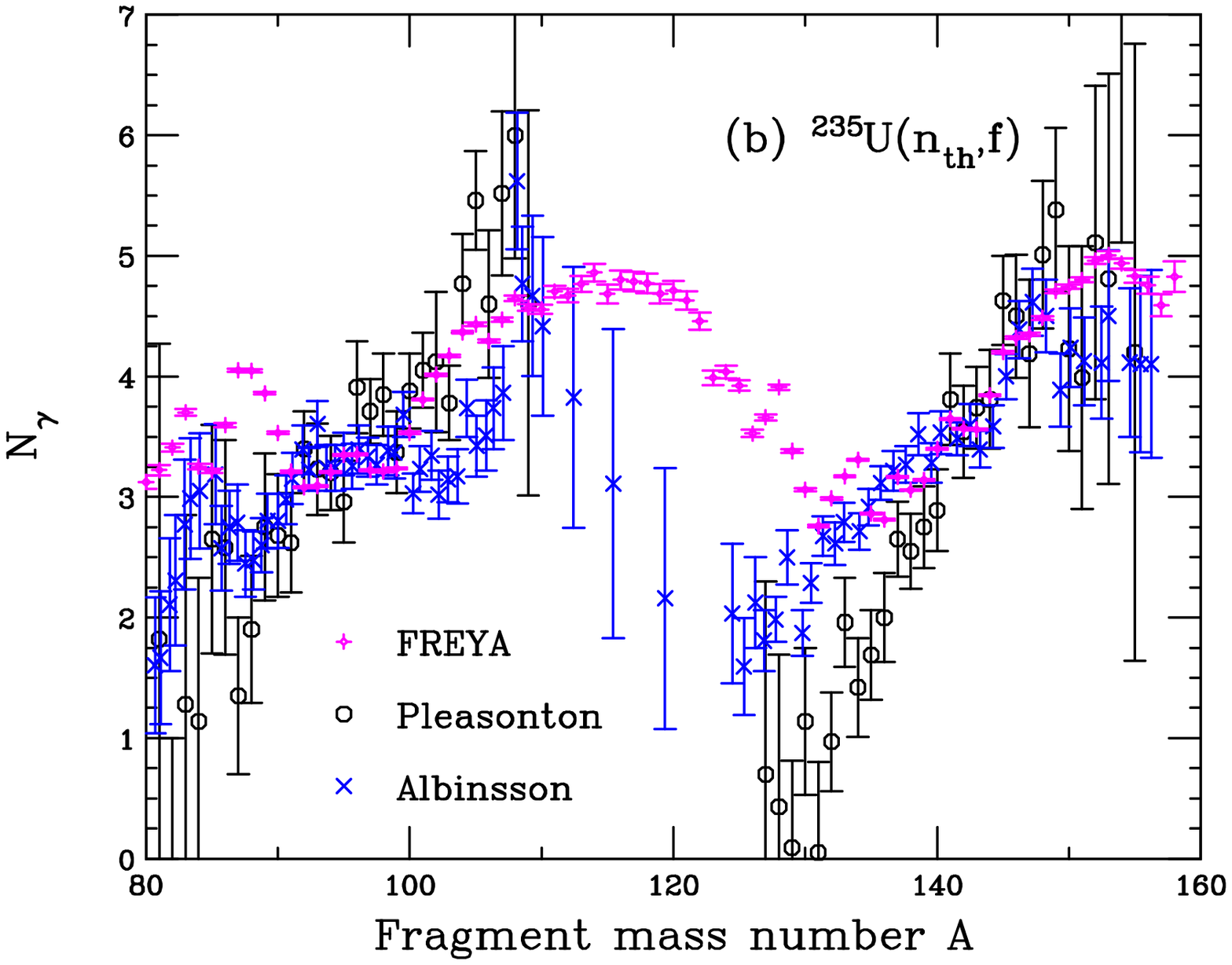}
\caption[]{(Color online) The total photon energy
  (a) and the photon multiplicity (b) as a function of fragment mass
  calculated for $^{235}$U($n_{\rm th}$,f) are compared to
  data from Pleasonton \protect\cite{PleasontonUgamma} (a) and (b) and
  data from Albinsson \protect\cite{Albinsson} (b).  
}\label{fig:Eg_Mg_vsA_U}
\end{figure}

The \code\ results are compared to the Pleasonton data on photon energy 
as a function of fragment mass in Fig.~\ref{fig:Eg_Mg_vsA_U}(a).  
The uncertainties in the
data are rather large and there is considerable point-to-point scatter.  
The trend of the data appears to be an approximately linear increase in
$E_\gamma$
with $A$ in both the light and heavy fragment regions.  The region near
symmetry, $110 < A < 125$, is effectively devoid of data due to the small
fragment yields in this region.  The tails of the fragment distributions,
$A < 90$ and $A > 145$, also having small fragment yields, exhibit large
uncertainties in $E_\gamma$ as well.  Thus the data are also effectively
consistent with being independent of $A$ within one standard deviation.
The \code\ calculations are nearly independent of $A$ and are also within
one standard deviation of the data for $A < 110$ and $A > 130$.
The dependence of $E_\gamma$ on $A$ in \code\ has not changed significantly with
the model improvements.

Note that the statistical uncertainties on the calculation are shown as well.
Since our results are based on 1M events with several photons emitted per event,
the \code\ results have significantly higher statistics than the four-parameter 
mode of the experiment that recorded the photon data.  
The largest statistical uncertainties in the calculation 
are in the symmetric fission region.

\begin{figure}[tbh]
\includegraphics[angle=0,width=0.45\textwidth]{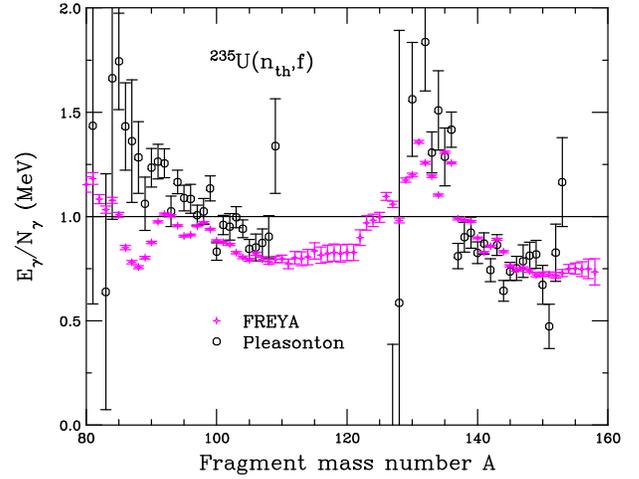}
\caption[]{(Color online) The energy per photon as a function of fragment mass
  calculated for $^{235}$U($n_{\rm th}$,f) is compared to data from
  Pleasonton \protect\cite{PleasontonUgamma}.
}\label{fig:EgoMg_vsA_U}
\end{figure}

In addition to the Pleasonton data on photon multiplicity, results from
Albinsson and Lindow \cite{Albinsson} are also shown in
Fig.~\ref{fig:Eg_Mg_vsA_U}(b).  The measurement by Albinsson and Lindow used an
experimental set-up similar to Pleasonton, with a collimated neutron beam from
the Studsvik R2 reactor.  However, similar to the Johansson measurement
\cite{Johansson}, they used a movable lead collimator to track fragments from
individual fragments and for timing purposes.  Their results are similar to
those from Pleasonton although the multiplicity trends lower for Albinsson
for $A > 100$ and higher for $125 < A < 140$.  We note that the calculation
shown is done using the values of $g_{\rm min}$ and $t_{\rm max}$ suitable for
the Pleasonton experiment.  These values are, however, compatible with those
used in the measurement by Albinsson and Lindow.
Given the large uncertainties in the two data sets, the agreement with the
\code\ calculation is good.  Comparison with
Fig.~\ref{fig:Adep_Eg_Mg_GDR_RIPL}(b) shows that the model refinements makes
the shape of $N_\gamma(A)$ more compatible with the data.

\begin{figure}[bh]
\includegraphics[angle=0,width=0.45\textwidth]{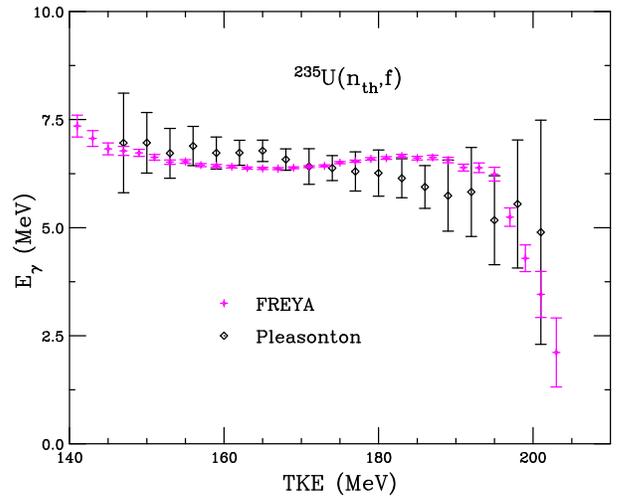}
\caption[]{(Color online) The total photon energy as a function of total
  fragment kinetic energy
  calculated for $^{235}$U($n_{\rm th}$,f) is compared to data from
  Pleasonton \protect\cite{PleasontonUgamma}.
}\label{fig:TKE_U_data}
\end{figure}

Figure~\ref{fig:EgoMg_vsA_U} compares the energy per photon from Pleasonton
with \code.  Overall, the agreement is quite good.  The \code\
calculation reproduces the peak in $E_\gamma/N_\gamma$ near $A\approx130$.
We note that only including the model improvements produces the peak in
$E_\gamma/N_\gamma$ at $A \sim 130$, see Fig.~\ref{fig:Adep_EgoMg_GDR_RIPL}.
While there are some differences for $A < 90$, due to a small enhancement in 
the photon multiplicity for $A\approx88$, producing a dip in $E_\gamma/N_\gamma$
at this value of $A$, the uncertainties are large enough for the two to still
be reasonably compatible.
The results for the mean photon multiplicity, total photon energy,
and energy per photon,
\SKIP{
$\langle N_\gamma \rangle = 6.82$;
$\langle E_\gamma \rangle = 6.47$~MeV; and
$\langle E_\gamma/N_\gamma \rangle = 0.95$~MeV,
agree well with those of Pleasonton \cite{PleasontonUgamma},
$\langle N_\gamma \rangle = 6.51 \pm 0.30$;
$\langle E_\gamma \rangle = 6.43 \pm 0.30$~MeV; and
$\langle E_\gamma/N_\gamma \rangle = 0.99 \pm 0.07$~MeV,
}
$N_\gamma = 6.82$;
$E_\gamma = 6.47$~MeV; and
$E_\gamma/N_\gamma = 0.95$~MeV,
agree well with those of Pleasonton \cite{PleasontonUgamma},
$N_\gamma = 6.51 \pm 0.30$;
$E_\gamma = 6.43 \pm 0.30$~MeV; and
$E_\gamma/N_\gamma = 0.99 \pm 0.07$~MeV,
and are also compatible with the earlier measurement of Verbinski
{\em et al.}\ \cite{Verbinski}.

Our results are also compared to the photon energy as a function of total
kinetic energy in Fig.~\ref{fig:TKE_U_data}.  Contrary to the $^{252}$Cf(sf)
calculation shown in Fig.~\ref{fig:TKE_Cf_data}, the \code\ calculation for
$^{235}$U($n_{\rm th}$,f) is not independent of TKE.  Although the curvature
appears to be somewhat different than that of the data, the results agree
within the uncertainties. Note also the falloff of the calculation for
${\rm TKE}>190$~MeV, due to the narrower TKE distribution for
$^{235}$U($n_{\rm th}$,f) relative to $^{252}$Cf(sf).  In this case the upper
bound of TKE is 205~MeV.

\begin{figure}[tbh]
\includegraphics[angle=0,width=0.45\textwidth]{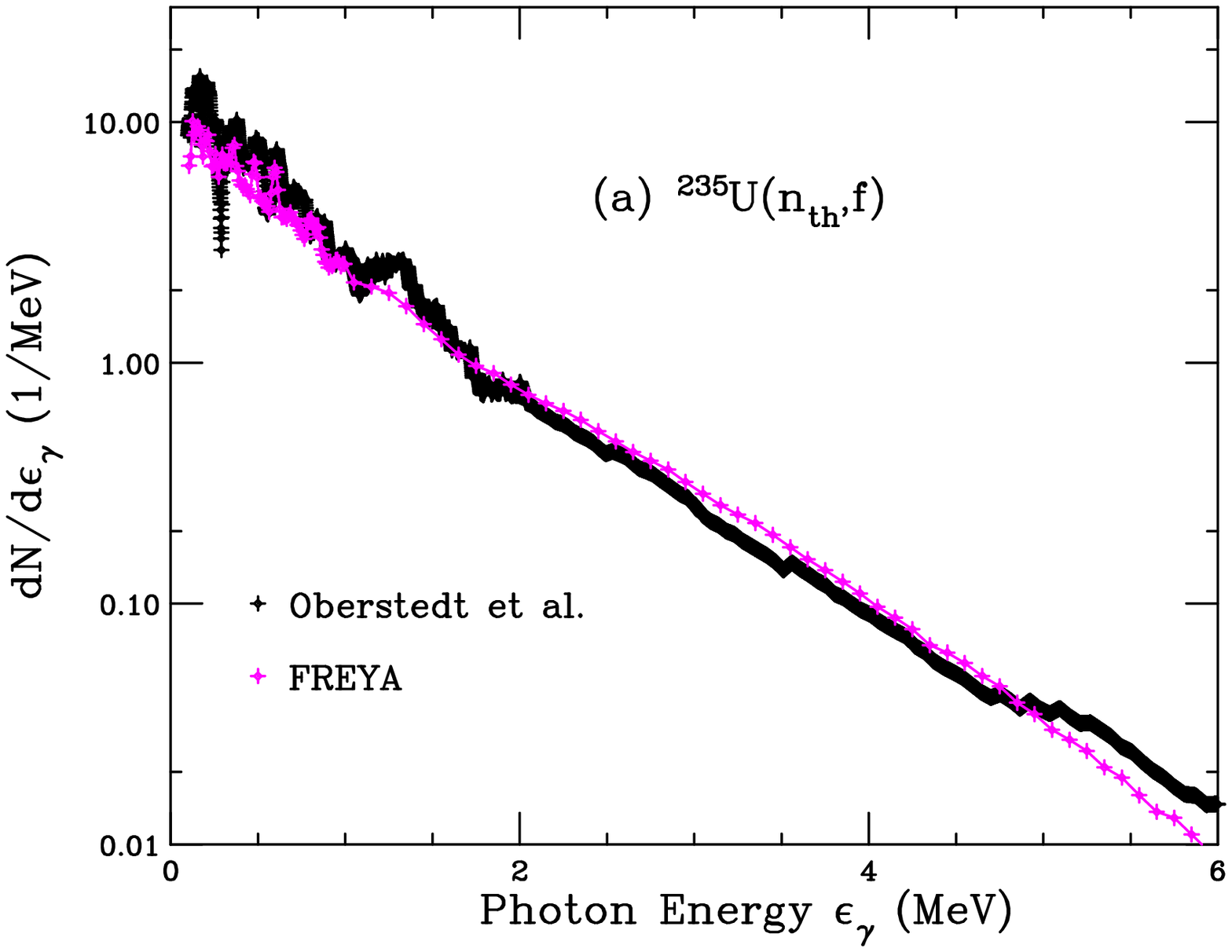}
\includegraphics[angle=0,width=0.45\textwidth]{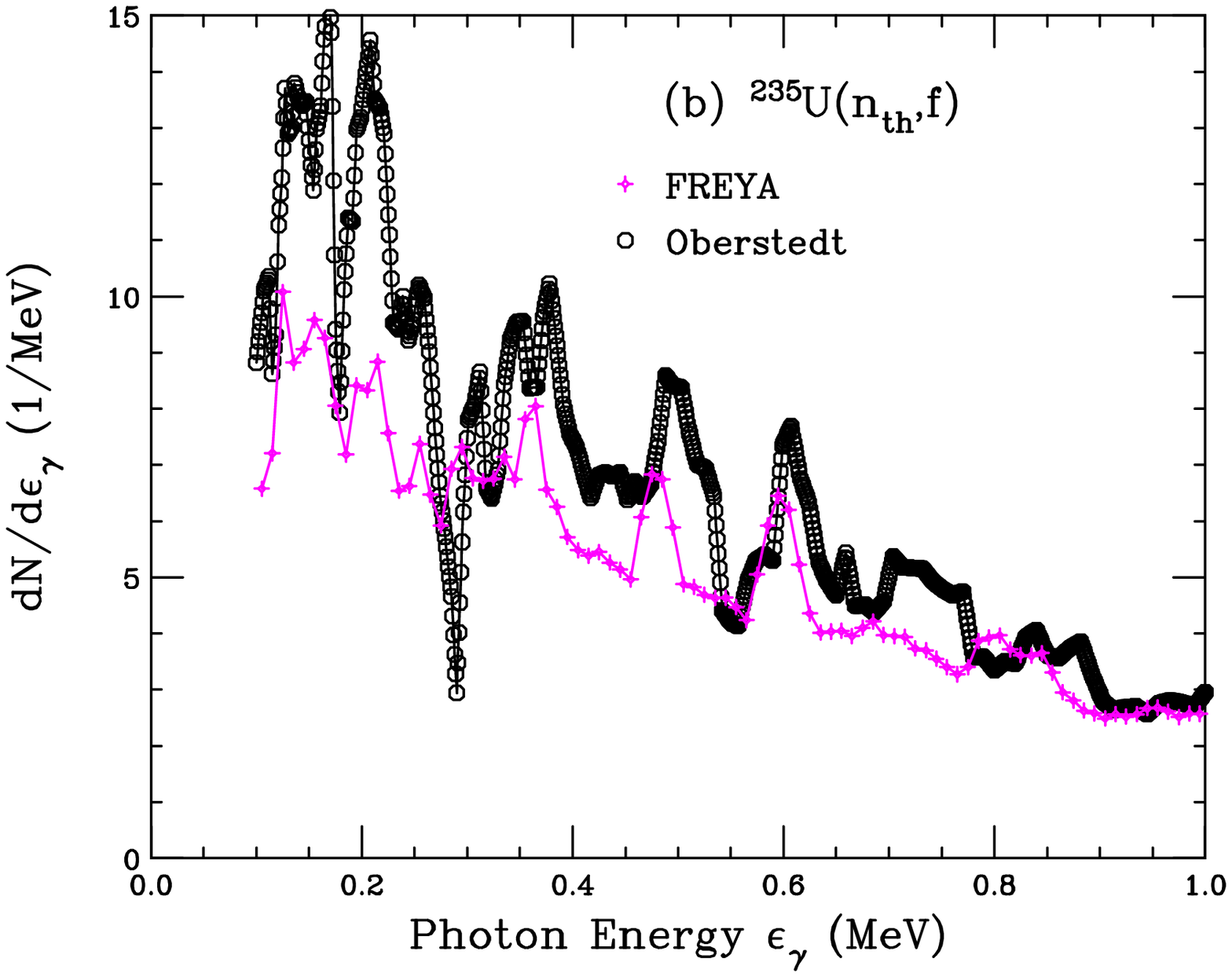}
\caption[]{(Color online) The photon energy spectrum calculated for
  $^{235}$U($n_{\rm th}$,f) compared to data from Oberstedt {\it et al.}\
  \protect\cite{BillnertUgamma}.  The total photon spectrum is shown in (a)
  while the spectrum for energies less than 1 MeV are shown in (b).
  The spectra are normalized to the fission photon multiplicity.
}\label{fig:spectra_U_data}
\end{figure}

Finally, we compare our calculated prompt fission photon spectrum to the
results of Oberstedt {\it et al.}\ \cite{BillnertUgamma}.  
This measurement is a continuation of the work of Billnert {\it et al.}\ 
\cite{BillnertCfgamma} for $^{252}$Cf(sf) 
shown in Fig.~\ref{fig:spectra_Cf_data} with the same detectors and values 
of $g_{\rm min}$ and $t_{\rm max}$ as in that work.  
In this case the measured high-energy slope of the photon energy spectrum 
is in good agreement with the \code\ calculation,
even without including the experimental uncertainties.  Without our model
refinements, the calculations would underestimate the photon spectrum at high
energy and would not exhibit any structure at low photon energies.
The peaks observed in the low-energy part of the photon spectrum, 
shown in Fig.~\ref{fig:spectra_U_data}(b), 
also agree well with the \code\ calculation.
\SKIP{Note, however, this time the peaks in the data are higher relative to our
calculation than for the Cf result, likely the main contribution to the
relatively large discrepancy between our calculated multiplicity and the data
using the same $g_{\rm min}$ and $t_{\rm max}$.}
Oberstedt {\it et al.}\ measured
\SKIP{
$\langle N_\gamma \rangle = 8.19 \pm 0.11$,
$\langle E_\gamma \rangle = 6.92 \pm 0.09$~MeV, and
$\langle E_\gamma/N_\gamma \rangle = 0.85 \pm 0.02$~MeV,  
while, for the same cutoffs, we find
$\langle N_\gamma \rangle = 6.93$,
$\langle E_\gamma \rangle = 6.48$~MeV, and
$\langle E_\gamma/N_\gamma \rangle = 0.93$~MeV.  The somewhat lower
}
$N_\gamma = 8.19 \pm 0.11$,
$E_\gamma = 6.92 \pm 0.09$~MeV, and
$E_\gamma/N_\gamma = 0.85 \pm 0.02$~MeV,  
while, for the same cutoffs, we find
$N_\gamma = 6.93$,
$E_\gamma = 6.48$~MeV, and
$E_\gamma/N_\gamma = 0.93$~MeV.  The somewhat lower
values of $g_{\rm min}$ and $t_{\rm max}$ for this experiment results in the
slightly higher $N_\gamma$ calculated here 
than the calculation for the Pleasonton data.

We note that, overall, we have achieved rather good agreement with the photon
data, despite not having made any global parameter analyses to extract $c_S$
specifically for $^{235}$U($n_{\rm th}$,f).

\section{Conclusions}

We have shown that the inclusion of the GDR form factor 
and the RIPL-3 transitions into \code\ 
has improved the photon emission process, resulting in better agreement
with photon observables as compared to our previous work \cite{RVJR_gamma}.  
In particular, there is significant improvement for the photon energy spectrum 
and the $A$ dependence of the energy per photon.
However, there is still room for further improvements, 
particularly with regard to the shape of the photon multiplicity distribution.  
We will return to this in future work.

We have also studied the sensitivity of our results to the degree of rotation
imparted to the fragments during scission by means of the scale factor $c_S$
controlling the ratio between the fragment spin temperature 
and the scission temperature.
Furthermore, 
we have illustrated how the photon energy and the photon multiplicity 
measured by different experiments depend on detector characteristics 
such as the minimum energy of the detected photons, $g_{\rm min}$, 
and the time window over which the measurement is made, $t_{\rm max}$.

The calculations were made with a value of $c_S$ optimized for $^{252}$Cf(sf).
(Note that, as mentioned in Sec.~\ref{sec:235U}, the parameters $x$, $c_T$, and
$d$TKE were tuned to the neutron observables for the fixed value $c_S = 0.87$.)
Even so, the agreement of our results with photon measurements 
in neutron-induced fission  is also quite good.
Further improvement can likely be obtained by performing 
a global parameter optimization including $c_S$, a subject of future work. 

We expect that the quality of the agreement of the \code\ calculations
with experimental data will continue to improve as further model refinements
are made and more measurements become available for both optimization and
predictive capability, for example, fission induced by (d,$p$) scattering,
see Ref.~\cite{Oslo}.

\section*{Acknowledgments}
We thank P.\ Talou for helpful discussions.
This work was supported by the Office of Nuclear Physics
in the U.S.\ Department of Energy's Office of Science under Contracts No.\ 
DE-AC52-07NA27344 (RV) and DE-AC02-05CH11231 (JR),
 as well as by the U.S.\ Department of Energy's
National Nuclear Security Administration

\appendix
\section{$^{233}$U($n_{\rm th}$,f) and $^{239}$Pu($n_{\rm th}$,f)}

In this appendix we compare our results with those measured by Pleasonton
for $^{233}$U($n_{\rm th}$,f) and $^{239}$Pu($n_{\rm th}$f,)
\cite{Pleasonton3UPugamma}.  The experimental setup is the same as for the
previously described Pleasonton measurements of $^{235}$U($n_{\rm th}$,f).
The values of $g_{\rm min}$ and $t_{\rm max}$ were the same as those for the
measurement.  In both cases, $c_S = 0.87$, obtained for $^{252}$Cf(sf),
is used in the calculations.  The values of $x$, $c_T$ and $d$TKE were fixed
to neutron observables.  We are working toward global analyses, including
fitting $c_S$ to the data, in the future.  Nonetheless, we can already check
whether or not the calculated trends are reasonable.

\subsection{$^{233}$U($n_{\rm th}$,f)}

\begin{figure}[h]
\includegraphics[angle=0,width=0.45\textwidth]{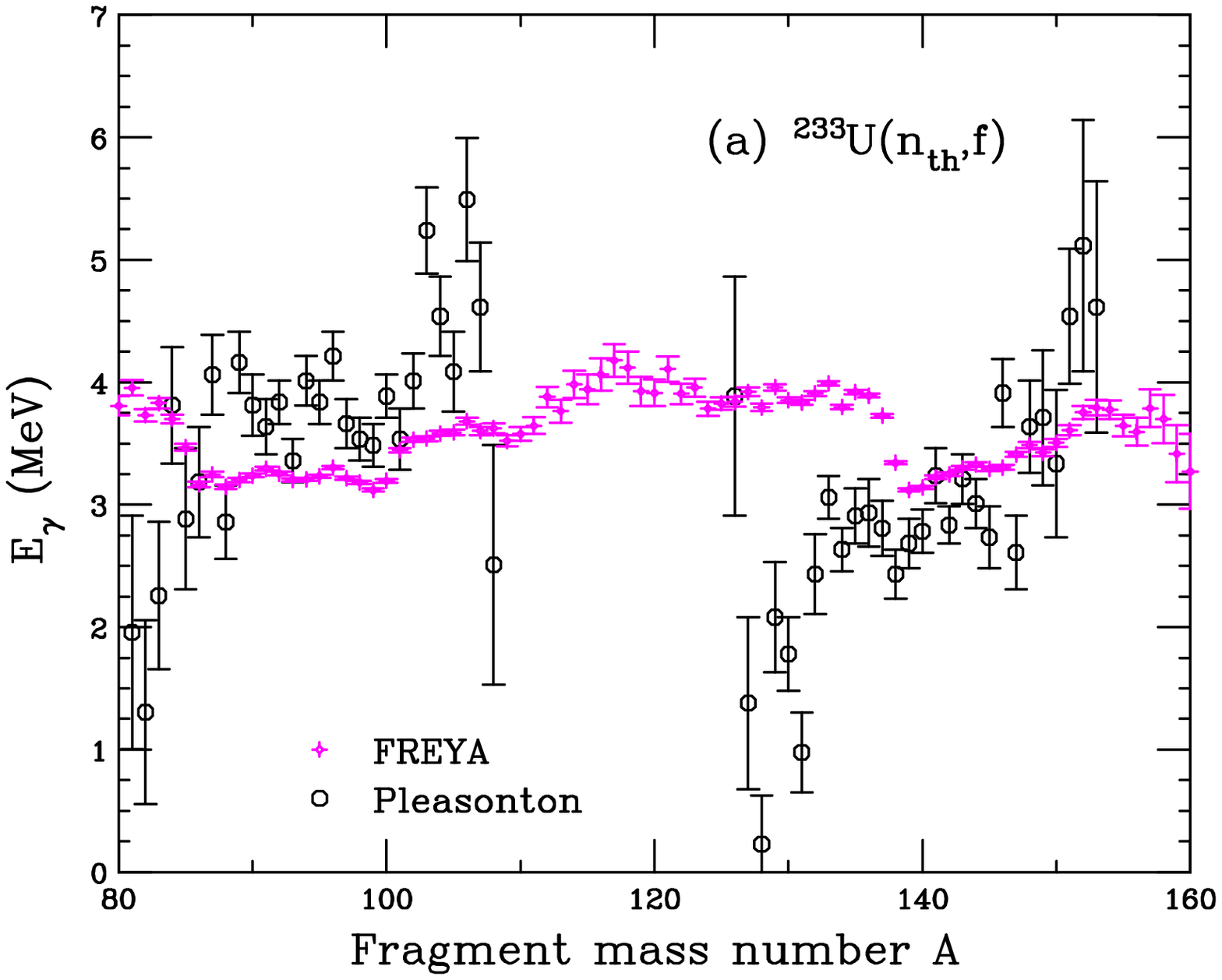}
\includegraphics[angle=0,width=0.45\textwidth]{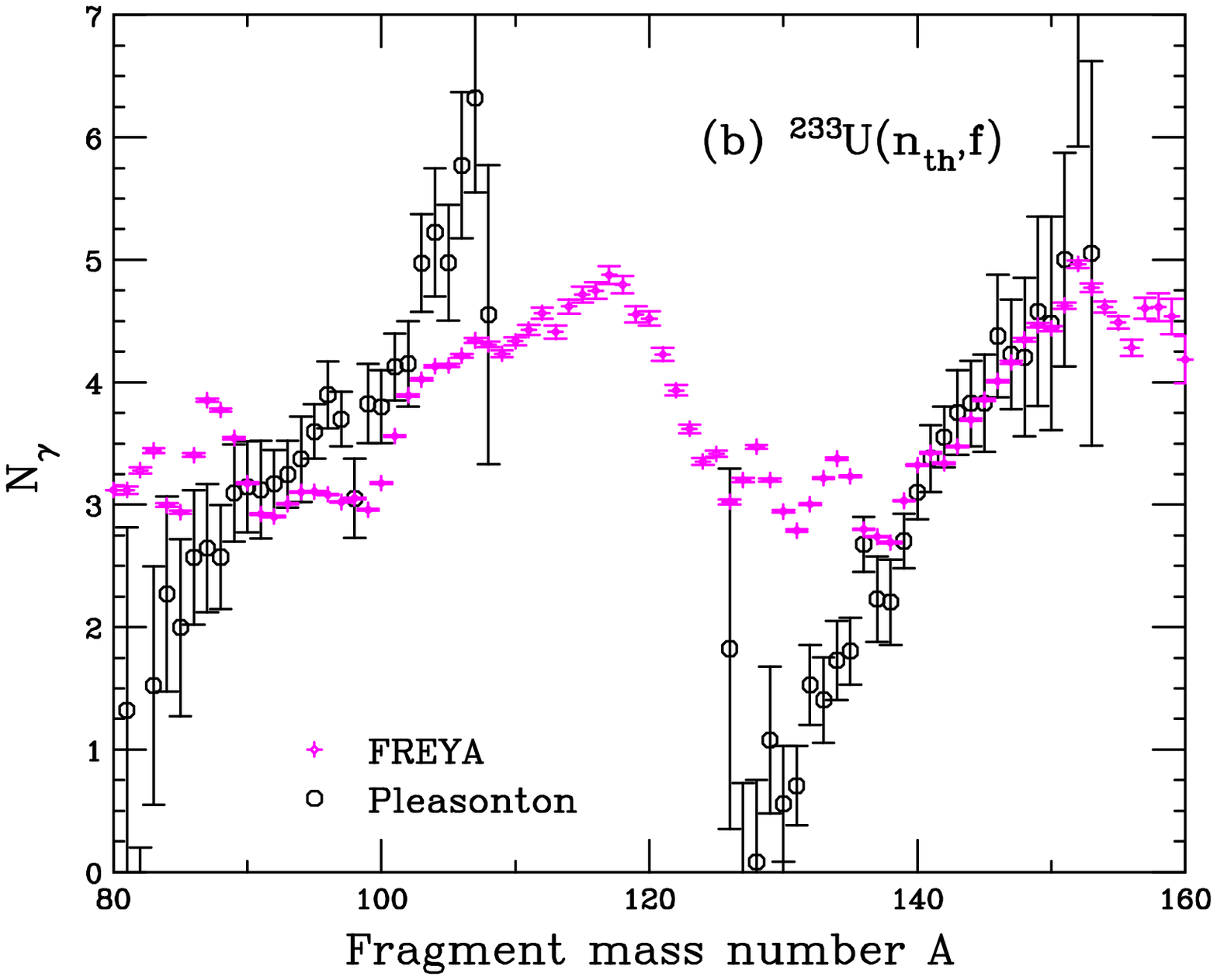}
\includegraphics[angle=0,width=0.45\textwidth]{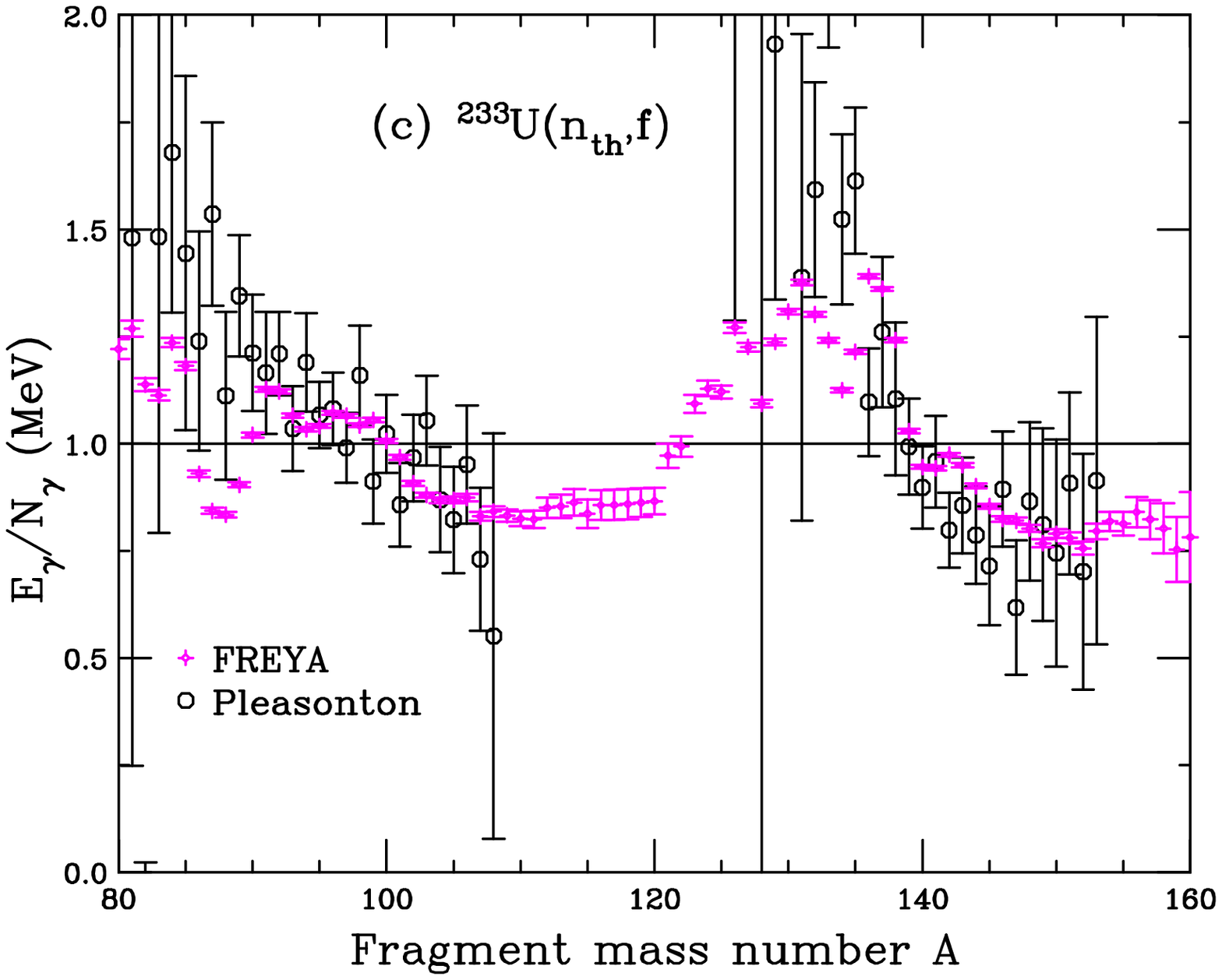}
\caption[]{(Color online) The total photon energy (a),
  the photon multiplicity (b), and
  the energy per photon (c) as functions of fragment mass number
  calculated for $^{233}$U($n_{\rm th}$,f) are compared to data from
  Pleasonton \protect\cite{Pleasonton3UPugamma}.
}\label{fig:Eg_Mg_vsA_U3}
\end{figure}

For $^{233}$U($n_{\rm th}$,f), a 99.9\% pure $^{233}$U$_3$O$_8$ target was
placed on a carbon film;
920K events were collected in the two-parameter mode 
and 350K events in the four-parameter mode.

The results for the total photon energy, photon multiplicity, and energy
per photon are shown in Fig.~\ref{fig:Eg_Mg_vsA_U3}.  The trends of the
calculations and the data are similar.  Here the total photon energy appears
to have a somewhat weaker dependence on fragment mass $A$ than did the
$^{235}$U measurement in Fig.~\ref{fig:Eg_Mg_vsA_U}(a).  

The measured average multiplicities and energies are
$\langle N_\gamma \rangle = 6.31 \pm 0.30$;
$\langle E_\gamma \rangle = 6.69 \pm 0.30$~MeV; and
$\langle E_\gamma/N_\gamma \rangle = 1.06 \pm 0.07$~MeV.
Our calculations give
$\langle N_\gamma \rangle = 6.66$;
$\langle E_\gamma \rangle = 6.87$~MeV; and
$\langle E_\gamma/N_\gamma \rangle = 1.03$~MeV, in very good overall agreement
with the data.

\begin{figure}[tbh]
\includegraphics[angle=0,width=0.45\textwidth]{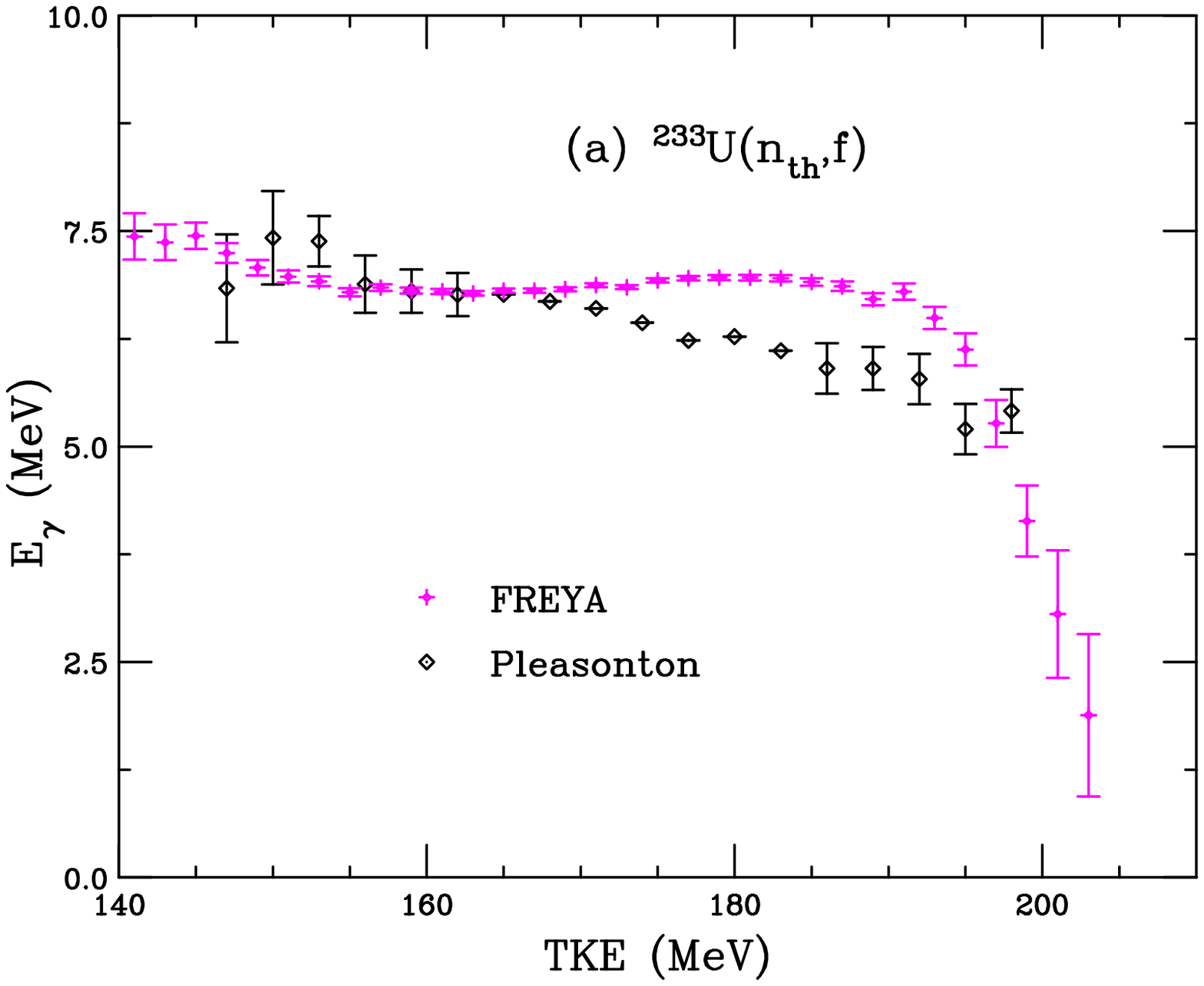}
\includegraphics[angle=0,width=0.45\textwidth]{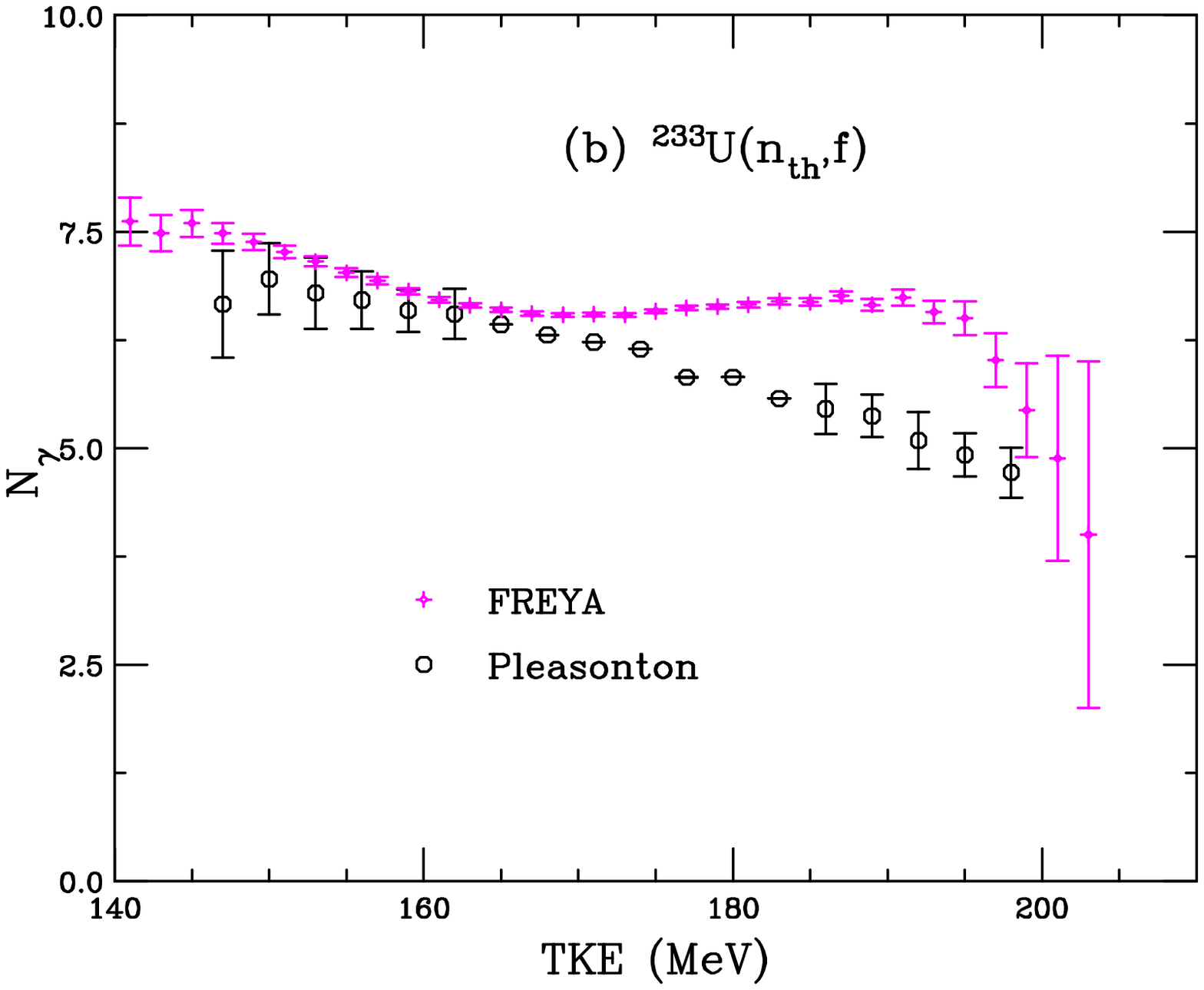}
\caption[]{(Color online) The total photon energy as a function of total
  fragment kinetic energy
  calculated for $^{233}$U($n_{\rm th}$,f) is compared to data from
  Pleasonton \protect\cite{Pleasonton3UPugamma}.
}\label{fig:TKE_U3_data}
\end{figure}

The total photon energy and multiplicity are shown as functions of the total
fragment kinetic energy in Fig.~\ref{fig:TKE_U3_data}.  The calculated shape
is similar to that obtained for $^{235}$U.  Here, however, the measured
uncertainties shown are smaller than those for $^{235}$U, giving a clearer
suggestion of small decrease in both $E_\gamma$ and $N_\gamma$ with TKE.  The
calculations agree with the data for ${\rm TKE}<170$~MeV.

\subsection{$^{239}$Pu($n_{\rm th}$,f)}

For $^{239}$Pu($n_{\rm th}$,f), a 99.11\% pure $^{239}$PuO$_2$ target was
deposited on a carbon film;
641K events were collected in the two-parameter mode 
and 209K events in the four-parameter mode.

\begin{figure}[tbh]
\includegraphics[angle=0,width=0.45\textwidth]{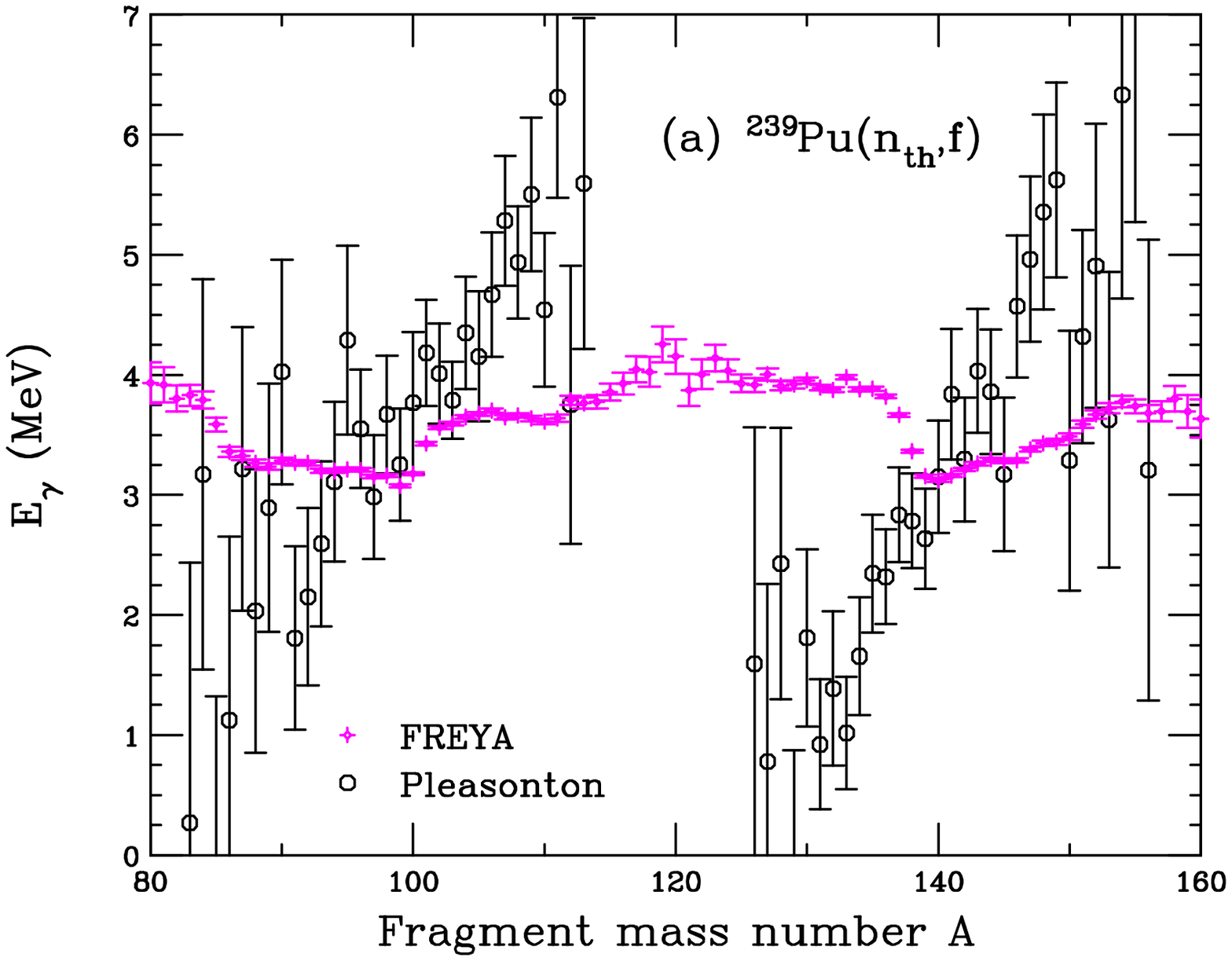}
\includegraphics[angle=0,width=0.45\textwidth]{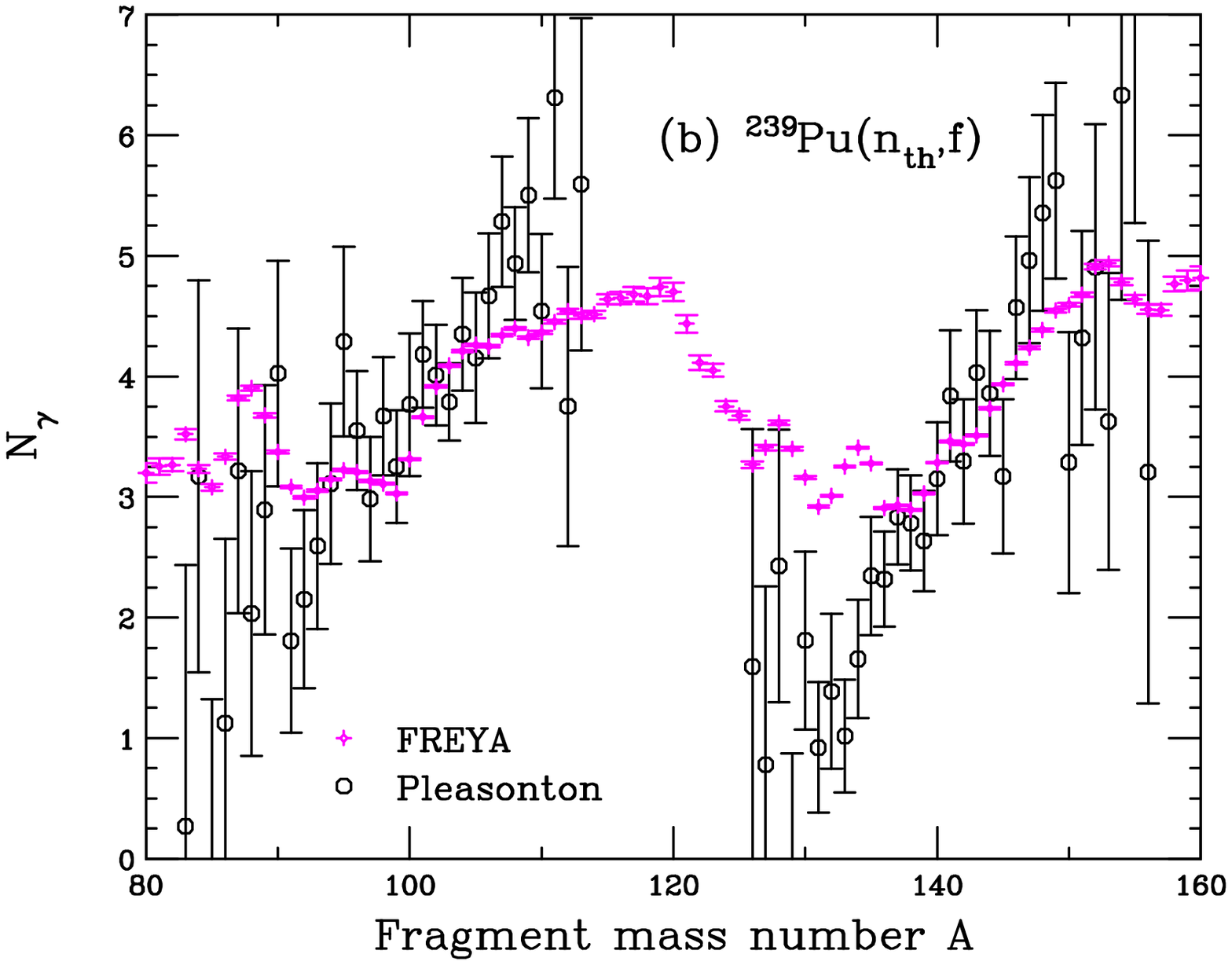}
\includegraphics[angle=0,width=0.45\textwidth]{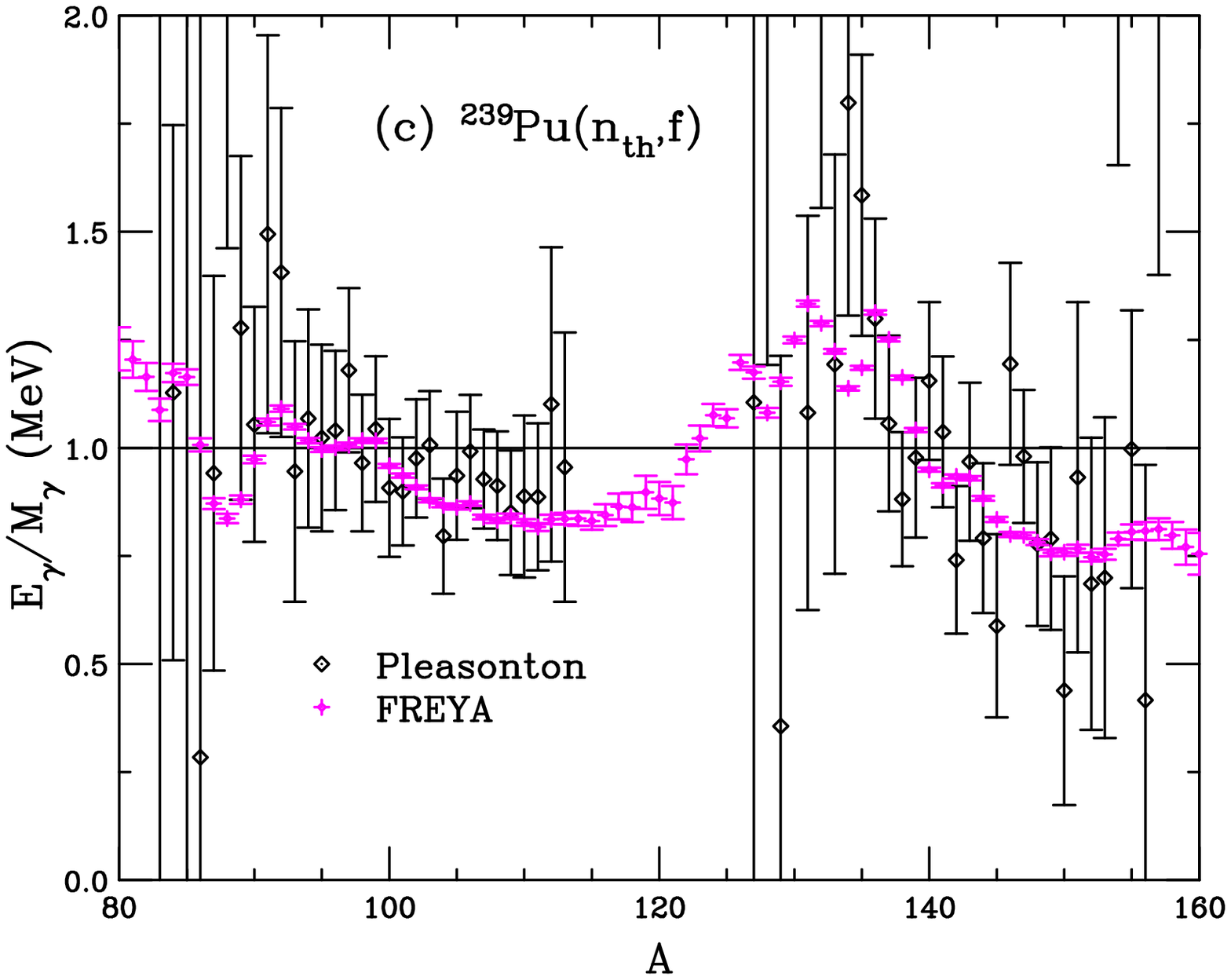}
\caption[]{(Color online) The total photon energy
  (a) and the photon multiplicity (b) as a function of fragment mass
  calculated for $^{239}$Pu($n_{\rm th}$,f) are compared to data from
  Pleasonton \protect\cite{Pleasonton3UPugamma}.
}\label{fig:Eg_Mg_vsA_Pu}
\end{figure}

The results for the total photon energy, photon multiplicity, and energy
per photon are shown in Fig.~\ref{fig:Eg_Mg_vsA_Pu}.  The trends of the
calculations and the data are similar and the agreement is generally good.
However, the low statistics of this data set make it difficult to
conclude anything substantial.

The measured average multiplicities and energies are
$\langle N_\gamma \rangle = 6.88 \pm 0.35$;
$\langle E_\gamma \rangle = 6.73 \pm 0.35$~MeV; and
$\langle E_\gamma/N_\gamma \rangle = 0.98 \pm 0.07$~MeV.
Our calculations give
$\langle N_\gamma \rangle = 7.19$;
$\langle E_\gamma \rangle = 6.95$~MeV; and
$\langle E_\gamma/N_\gamma \rangle = 0.98$~MeV.  While our calculated averages
are somewhat higher than the data, they are still within the uncertainties.

\begin{figure}[tbh]
\includegraphics[angle=0,width=0.45\textwidth]{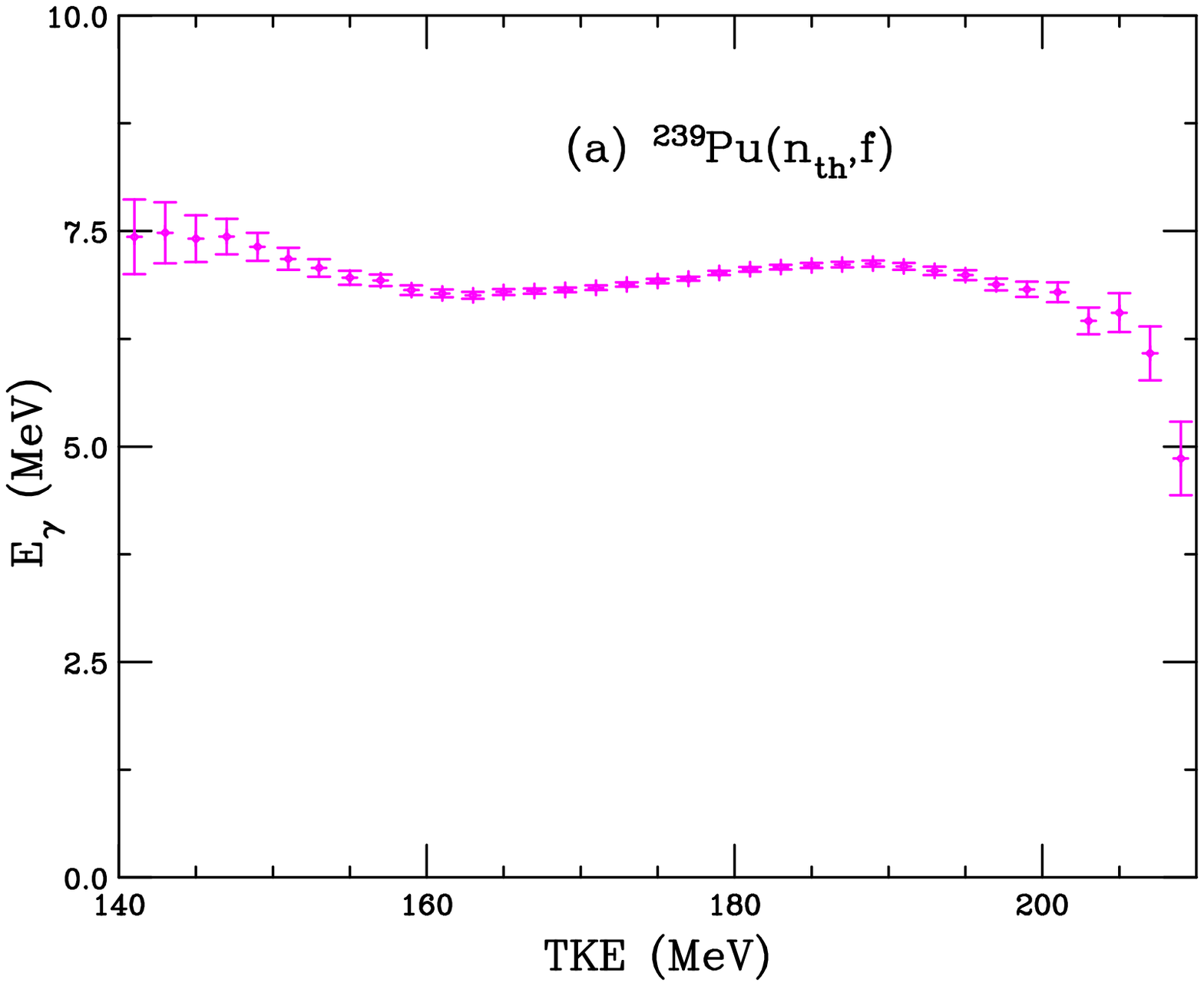}
\includegraphics[angle=0,width=0.45\textwidth]{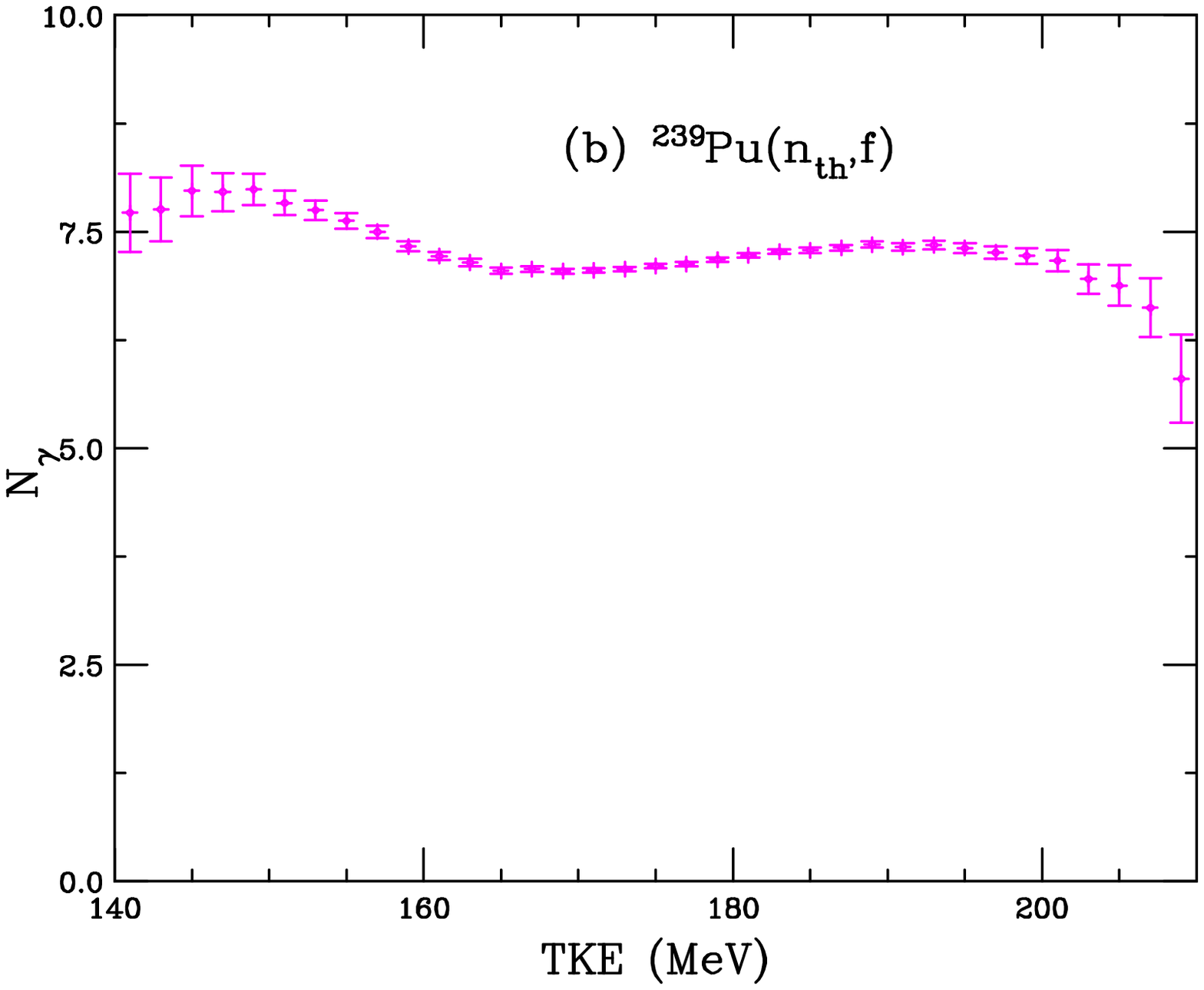}
\caption[]{(Color online) The total photon energy as a function of total
  fragment kinetic energy calculated for $^{239}$Pu($n_{\rm th}$,f).
}\label{fig:TKE_Pu_data}
\end{figure}

For completeness, the photon energy and multiplicity are shown as functions of
TKE in Fig.~\ref{fig:TKE_Pu_data}.  The results are similar to those for the
other neutron-induced fission calculations shown above.  
No data were available for comparison for this case.  Note that the
calculations extend to higher values of TKE than for $^{233}$U($n_{\rm th}$,f) and
$^{235}$U($n_{\rm th}$,f) because the tail of the TKE distribution extends to
higher TKE for $^{239}$Pu($n_{\rm th}$,f), up to 215~MeV.


\newpage

\end{document}